\documentclass[fleqn,usenatbib]{mnras}

\usepackage{caption}
\usepackage{amsmath}
\usepackage{amssymb}
\usepackage{multicol}
\usepackage{rotating}
\usepackage{lscape}
\usepackage{graphicx}
\usepackage{color}
\usepackage{placeins}
\usepackage{xcolor}
\usepackage{float}
\usepackage{verbatim}
\usepackage{bbold}
\usepackage{lipsum,adjustbox}
\usepackage{subcaption}

\usepackage{tikz}
\usetikzlibrary{matrix, fit, positioning,arrows.meta,arrows, ,decorations.pathreplacing,shapes.geometric,backgrounds}

\tikzstyle{startstop}=[rectangle, rounded corners, minimum width=2.7cm, minimum height= 1cm, text centered, draw=black, fill=red!20, text width = 2cm]
\tikzstyle{startstop_nb}=[rectangle, rounded corners, minimum width=2.5cm, minimum height=0.8cm, text centered, draw=black, fill=gray!20, text width = 2cm]
\tikzstyle{io}=[trapezium, trapezium left angle = 70, trapezium right angle =110,minimum width=2cm, minimum height= 1cm, text centered, draw=black, fill=blue!30]
\tikzstyle{io_short}=[rectangle,rounded corners, minimum width=0.2cm, minimum height= 1cm, text centered, draw=black, fill=blue!5, line width = 0.05cm]
\tikzstyle{process}=[rectangle, rounded corners,minimum width=3cm, minimum height= 0.7cm, text centered, draw=black, fill=orange!30]
\tikzstyle{process_DUMMY2}=[rectangle, minimum width=2.7cm, minimum height= 1cm, text centered, draw=white, fill=white!0, text width = 3cm]
\tikzstyle{process_DUMMY}=[rectangle, minimum width=0cm, minimum height= 0cm, draw=white, fill=white!0]
\tikzstyle{process_SDUMMY}=[circle, minimum width=0cm, minimum height= 0cm, draw=white, fill=white!0]

\tikzstyle{process_NO_BOX}=[circle, minimum width=0.5cm, minimum height= 0.5cm, text centered, draw=gray, fill=white!0]
\tikzstyle{startstop2}=[rectangle, rounded corners, minimum width=2.7cm, minimum height= 1cm, text centered, draw=black, fill=gray!20, text width = 3cm]

\tikzstyle{startstop3}=[rectangle, rounded corners, minimum width=2.7cm, minimum height= 1cm, text centered, draw=black, fill=white, text width = 3cm]

\tikzstyle{process_TITLE}=[rectangle, rounded corners, minimum width=1cm, minimum height= 0.4cm, text centered, draw=black, fill=gray!30]
\tikzstyle{nomen}=[rectangle, rounded corners, minimum width=4cm, minimum height= 1cm,  draw=black, fill=yellow!10]
\tikzstyle{process_short}=[rectangle, rounded corners,minimum width=1cm, minimum height=0.5cm, text centered, draw=black, fill=orange!20]
\tikzstyle{decision}=[diamond, rounded corners,minimum width=2cm, minimum height= 0.7cm, text centered, draw=black, fill=green!20]
\tikzstyle{question}=[diamond,rounded corners, minimum width=2.5cm, minimum height= 0.7cm, text centered, draw=black, fill=blue!20]
\tikzstyle{results}=[rectangle,rounded corners, minimum width=2.5cm, minimum height= 0.7cm, text centered, draw=black, fill=blue!20]
\tikzstyle{decisionfwd}=[diamond, rounded corners,minimum width=2cm, minimum height= 0.7cm, text centered, draw=black, fill=green!20]
\tikzstyle{decision_short}=[diamond, rounded corners,minimum width=1cm, minimum height= 1cm, text centered, draw=black, fill=green!20]
\tikzstyle{decision_short_input}=[diamond, minimum width=0.8cm, minimum height=0.05cm, text centered, draw=black, fill=gray!20]
\tikzstyle{arrow}=[thick, ->, >= stealth]
\tikzstyle{arrow_d}=[dotted, ->, >= stealth]
\tikzstyle{arrow_new}=[dotted, ->, >= stealth]
\tikzstyle{arrow_pi}=[dashed, ->, >= stealth, line width=0.03cm]
\tikzstyle{arrow_nl}=[dashed, -, >= stealth, line width=0.03cm]
\tikzstyle{arrow_nls}=[thick, -, >= stealth, line width=0.03cm]
\tikzstyle{arrow_new}=[dotted, ->,>= stealth, color=red, line width = 0.02cm]
\tikzstyle{arrow_pib}=[dashed, ->, >= stealth, line width=0.03cm, color=blue]
\tikzstyle{arrow_nlb}=[dashed, -, >= stealth, line width=0.03cm, color=blue]

\newcommand{\beq}{\begin{equation}}
\newcommand{\eeq}{\end{equation}}
\newcommand{\beqa}{\begin{eqnarray}}
\newcommand{\eeqa}{\end{eqnarray}}

\definecolor{gray}{gray}{0.55}

\font\BF=cmmib10

\def\v{{\hbox{\BF v}}}
\def\u{{\hbox{$u_z$}}}

\newcommand{\kmsmpc}{\mbox{km\,s$^{-1}$\,Mpc$^{-1}$}}
\newcommand{\mbi}[1]{\mbox{\boldmath$#1$}}

\newcommand{\mat}[1]{\mbox{\rm\bf #1}}
\newcommand{\lsim}{\mbox{${\,\hbox{\hbox{$ < $}\kern -0.8em \lower 1.0ex\hbox{$\sim$}}\,}$}}
\newcommand{\gsim}{\mbox{${\,\hbox{\hbox{$ > $}\kern -0.8em \lower 1.0ex\hbox{$\sim$}}\,}$}}

\newcommand{\dd}{{\rm d}}

\def\beqn{\vspace{2mm}
\begin{eqnarray}}
\def\eeqn{\vspace{2mm}
\end{eqnarray}}

 \def\la{\langle}

\newcommand{\be}{\begin{equation}}
\newcommand{\ee}{\end{equation}}
\newcommand{\ba}{\begin{eqnarray}}
\newcommand{\ea}{\end{eqnarray}}
\newcommand{\brr}{\begin{array}}

\newcommand{\err}{\end{array}}
\newcommand{\bc}{\begin{center}}
\newcommand{\ec}{\end{center}}

\title[Cosmic density from light-cone data]{\texttt{COSMIC BIRTH}: Efficient Bayesian Inference of the Evolving Cosmic Web from Galaxy Surveys}

\author[F.-S.~Kitaura et al.]{Francisco-Shu~Kitaura$^{1,2}$\thanks{E-mail: \href{mailto:fkitaura@iac.es}{fkitaura@iac.es}}, Metin Ata$^{3}$, Sergio A. Rodr\'{\i}guez-Torres$^{1,2}$,  \and  M{\'o}nica Hern{\'a}ndez-S{\'a}nchez$^{1,2}$, A. Balaguera-Antol\'{\i}nez$^{1,2}$ and Gustavo Yepes$^{4,5}$
\\ \\
$^1$Instituto de Astrof\'{\i}sica de Canarias (IAC), Calle V\'{\i}a Lactea s/n, 38200, La Laguna, Tenerife, Spain \\ 
$^2$Departamento de Astrof\'{\i}sica, Universidad de La Laguna (ULL), E-38206, La Laguna, Tenerife, Spain\\ 
$^3$Kavli IPMU (WPI), UTIAS, The University of Tokyo, Kashiwa, Chiba 277-8583, Japan\\
$^{4}$Departamento de F\'isica Te\'{o}rica, M\'{o}dulo 8, Facultad de Ciencias, Universidad Aut\'{o}noma de Madrid, 28049 Madrid, Spain\\
$^{5}$CIAFF, Facultad de Ciencias, Universidad Aut\'{o}noma de Madrid, 28049 Madrid, Spain}

\date{Accepted XXX. Received YYY; in original form ZZZ}

\pubyear{2019}

\begin{document}
\label{firstpage}
\pagerange{\pageref{firstpage}--\pageref{lastpage}}
\maketitle

% Abstract of the paper
\begin{abstract}
We present {\tt COSMIC BIRTH}: {\tt COSM}ological {\tt I}nitial {\tt C}onditions from {\tt B}ayesian {\tt I}nference {\tt R}econstructions with {\tt TH}eoretical models: an algorithm to reconstruct the primordial and evolved cosmic density fields from galaxy surveys on the light-cone. 
The displacement and peculiar velocity fields are obtained from forward modelling at different redshift snapshots given some initial cosmic density field within a Gibbs-sampling scheme.
This allows us to map galaxies, observed in a light-cone, to a single high redshift and hereby provide tracers and the corresponding survey completeness in Lagrangian space including tetrahedral tessellation  mapping.
These Lagrangian tracers in turn permit us to efficiently obtain the primordial density field, making the {\tt COSMIC BIRTH} code general to any structure formation model. Our tests are restricted for the time being to Augmented Lagrangian Perturbation theory.
%The {\tt COSMIC BIRTH} code includes a novel higher order Hamiltonian Monte Carlo sampling, including a non-diagonal mass matrix, to obtain fast independent samples, greatly reducing the correlation length.
We show how to robustly compute the non-linear Lagrangian bias from clustering measurements in a numerical way, enabling us to get unbiased dark matter field reconstructions at initial cosmic times.
We also show that we can {\color{black}accurately} recover the information of the dark matter field from the galaxy distribution based on a detailed simulation.
Novel key ingredients to this approach are a higher-order Hamiltonian sampling technique and a non-diagonal Hamiltonian mass-matrix. This technique could be used to study the Eulerian galaxy bias from galaxy surveys and could become an ideal baryon acoustic reconstruction technique.
In summary, this method represents a general reconstruction technique, including in a self-consistent way a survey mask, non-linear and non-local bias and redshift space distortions, with an efficiency about {\color{black}10} times superior to previous comparable methods.
\end{abstract}


%%%%%%%%%%%%%%%%%%%%%%%%%%%%%%%%%%%%%%%%%%%%%%%%%%

%%%%%%%%%%%%%%%%% BODY OF PAPER %%%%%%%%%%%%%%%%%%

% sections
% what/why
% results

\section{Introduction}
\label{sec:intro}

The observed accelerated expansion of the Universe \citep[][]{1998AJ....116.1009R,1999ApJ...517..565P} poses some of the most intriguing questions in modern cosmology: what is the origin of such a dynamical state? \citep[see e.g.][]{2008Natur.451..541G}; {\color{black} is the so-called dark energy component responsible for it?, and what is its nature?} In the recent years, a number of wide-field galaxy surveys have been designed in order to answer these fundamental questions, such as eBOSS \cite[][]{2016AJ....151...44D}, Euclid \cite{Euclid}, DESI \cite[][]{DESI}, 4MOST \citep{4most}, WFIRST \citep[][]{2019arXiv190205569A} and LSST \citep[][]{2009arXiv0912.0201L}.
Additionally, pencil-beam surveys with smaller footprints but deeper and more abundant target sampling, e.g. VIPERS \citep[]{2014A&A...566A.108G} and PFS \citep[]{10.1093/pasj/pst019}, and also dense and deep phootometric surveys, such as  DES \cite{DES} and J-PAS \citep{jpas} contribute to understand the cosmic evolution of large-scale structures. The acquisition of observational data, as generated by these surveys, is potentially reaching the precision requirements to be able to tackle not only the above mentioned questions, but indeed more profound ones, such as the validity of General Relativity on the largest cosmological scales \citep[see e.g.][ and references therein]{2019LRR....22....1I}. To that end, the tools envisaged to perform data analysis need to keep track with the observational campaigns in order to be able to exploit the data to its maximal information content.

Recent measurements of the statistical properties of the spatial distribution of galaxies \citep[see e.g.][]{Alam:2016hwk} or quasars \citep[see e.g.][]{Ata:2017dya} are currently paving the road for careful analyses aiming at shedding light into the different physical processes involved in the formation of large-scale structures. 
The identification of some of such properties dates back to pioneering papers \cite[see][]{1989RvMP...61..185S,bond1996,platen2011}, with the realization that galaxies follow an intricate pattern, the \emph{the cosmic-web}. A large variety of tools have been envisaged to study it \citep[see e.g.][and references therein]{2018MNRAS.473.1195L}. Another particular feature in the spatial distribution of galaxies is the baryon acoustic oscillations (BAOs). Its detection \citep[see e.g.][]{2005ApJ...633..560E,2007MNRAS.381.1053P} represents the establishment of cosmological standard ruler, and a sensitive probe for the equation of state of the dark energy \citep[e.g.][]{2003ApJ...598..720S,2003ApJ...594..665B,2015PhRvD..92l3516A,2017NatAs...1..627Z}.

Extracting the content of cosmological information encoded in the BAO is not a trivial task. At early cosmological times, such information was entirely contained in the two-point correlation function (or its Fourier counterpart, the power spectrum), which fully {\color{black}characterizes} linear Gaussian over-density fields. However, as cosmic density fields evolved, non-linear evolution not only induced shifts in the position of the acoustic peak \citep[e.g.][]{2008PhRvD..77b3533C}, but also dragged part of the information of the BAO signature from the two-point to higher order statistics \citep[see e.g.][]{2015PhRvD..92l3522S}, thus making mandatory the assessment of the galaxy three-point correlation function or the galaxy bispectrum \citep[see e.g.][]{2017MNRAS.469.1738S,2017MNRAS.465.1757G}. 

In order to obtain a clean the detection of the BAO in the spatial distribution of galaxies (or quasars) in the two--point statistics, it is now standard to apply the concept of \emph{reconstruction}: a technique which takes the spatial galaxy distribution back in time to a higher redshift, in which cosmic density fields are closer to linear \citep[e.g.][]{2007ApJ...664..675E,2012MNRAS.427.2132P}. However, a number of systematic uncertainties based on technical aspects of the observation strategy,  such as the survey mask, or the radial selection function, together with other observational uncertainties with a physical background, such as galaxy bias, or redshift space distortions, have to be taken into account in reconstruction studies.
A Bayesian approach represents a natural framework to deal with these systematic uncertainties, in which a posterior distribution function relates the linear density field to the observational data \citep{1995ApJ...449..446Z,2008MNRAS.389..497K}.

There are additional arguments to rely on this type of statistical approach.
While mapping the linear to the non-linear density field has a clear physical foundation governed by gravity in an expanding background Universe, its inverse mapping is not trivial. The phase-space information is reduced to the spatial distribution at late cosmic times in a galaxy survey. Shell-crossing has already set in, and the trajectories of the tracers of the large-scale structure are not uniquely defined. 
To solve this problem, forward methods (sampling the posterior distribution function of the primordial density field, given some galaxy survey data) have been proposed in the literature \citep{2013MNRAS.432..894J,2013ApJ...772...63W,2014ApJ...794...94W,2019MNRAS.488.2573B,2019A&A...625A..64J}. See also the corresponding cosmic-web analysis based on forward modelling \citep{2014MNRAS.445..988N,2015JCAP...06..015L}.
However, these methods require to sample the initial density field in Lagrangian coordinates as a function of the final density field in the Eulerian frame. This is not only computational very expensive, but has also the drawback of adjusting the sampling procedure to the particular (i.e. particle mesh or Lagrangian perturbation theory) forward structure formation model. 
One of the main disadvantages of these methods is the computational cost.

To increase the computational efficiency, an effective bias prescription at the field level is used. This introduces a stochastic bias component, which requires a detailed likelihood modelling \citep[][]{2015MNRAS.446.4250A,2019JCAP...01..042S, 2015JCAP...07..030M}.
The correlation lengths between the iterations sampling the posterior distribution as reported in these papers are of the order of 1000 and producing tens of thousands of iterations, in which each time a gravity solver is applied yields only tens of independent samples. For this reason, other approaches have been proposed, in which, instead of sampling the full posterior, the maximum a-posteriori is computed \citep[][]{2019JCAP...10..035H,2010MNRAS.403..589K,2019arXiv190104454S} .
Nonetheless, sampling the posterior distribution function has several advantages as a variety of compatible solutions can be obtained with the posterior assessment of confidence regions and therefore realistic error bars and in general, covariance matrices.

In this work we propose an alternative approach. Grounded in the philosophy of previous works such as  \citet{1999MNRAS.308..763M} and  \citet{2013MNRAS.429L..84K}, our proposal goes one step further and introduces additional developments aimed at retrieving the distribution of tracers in Lagrangian space. In particular, we implement a nested Gibbs- and Hamiltonian sampler, in which the final (i.e, the observed) galaxy distribution in redshift space is translated to real-space at initial high-redshift coordinates defined on the light-cone. The required peculiar velocities and displacements are iteratively obtained from the primordial density field at a single high redshift with forward modelling.

%Hence, the reconstructed density fluctuations are very small within a Lagrangian framework at early cosmic times with negligible shell-crossing for which a Lognormal-Poisson model is very accurate. An arbitrary structure formation model can be applied to connect the initial field with the final galaxy distribution.

%One key ingredient is the transformation of the response operator including the survey geometry and the radial selection function to high resdhifts with the reconstructed displacement field by using phase-space mapping. 
%In this way we obtain accurate estimates of the completeness of galaxy distribution transformed to high redshifts. 

Our approach is particularly efficient due to a novel higher-order leapfrog algorithm applied at the solution of Hamilton equations (core of the Hamiltonian sampling technique). We have implemented an explicit and time-reversible symplectic integrator, widely used to solve quantum field theoretical phenomena of many-body fermionic systems \citep[see e.g.][]{PhysRevLett.63.9}. In a companion paper (Hern\'andez-S\'anchez et al., in preparation) we show that this increases the computational efficiency by factor of about $20$. Furthermore, we have implemented a non-diagonal Hamiltonian mass which includes the response operator to further increase the speed of the Hamiltonian Monte Carlo sampler. 

In order to test our method, we use a galaxy mock catalog based on an $N$-body simulation, which reproduces clustering on the light-cone as measured from the CMASS sample \citep[][]{2016MNRAS.460.1173R}. Our results demonstrate that we can obtain unbiased linear primordial density fields up to $k \sim 0.4$ $h$ Mpc$^{-1}$.  
This method promises to be especially suited for the reconstruction of baryon acoustic oscillations. Moreover, we have achieved that the reconstruction depends only on cosmological parameters, solving for dependencies on internal parameters associated to the resolution of the mesh. Another advantage of working with tracers based on the galaxy distribution is that they retain the small scale clustering information, when taking them to higher redshifts. The displacements and velocities are obtained in Lagrangian space, while aliasing and shot-noise are accurately corrected for through a Bayesian posterior sampling. Our method uses an iterative scheme which uses only differences of particle positions and their peculiar velocities, thus being flexible to be implemented with any arbitrary gravity solver.

The method presented here shows a way of taking into account nonlinear and non-local bias in the reconstruction process. This is particularly important, as the models used so far in the literature \citep[e.g.,][]{2014MNRAS.439L..21K,2014MNRAS.441..646N,2015MNRAS.450.1836K} as well as those used within Bayesian reconstruction algorithms \citep[e.g.][]{2019A&A...625A..64J} are based on bias descriptions which have been shown to be quite inaccurate for low-mass tracers \citep[see e.g.][]{2019arXiv191013164P}. While those analytic effective Eulerian bias models can represent the distribution of LRGs to a good accuracy \citep[e..g][]{2016MNRAS.456.4156K}, the approach presented in this work can be  particularly relevant for surveys based on ELGs and bright galaxies.

The outline of this paper is as follows. In \S~\ref{sec:method} we discuss the theoretical framework and the main motivations of our reconstruction approach. Section \ref{sec:birth} describes the main ingredients and operations performed within the \texttt{COSMIC BIRTH} approach. In \S~\ref{sec:application} we present the validation of the method. We end with conclusions.

During this work we are dealing with different estimates of redshift, viz, the cosmological redshift (in Lagrangian $z_{q}$ or Eulerian coordinates $z_r$), and that affected by peculiar velocities $z_s$ (in Eulerian coordinates). If nothing else indicated than $z$, this corresponds to $z_r$, i.e., the true redshift at which a galaxy resides. In table \ref{table:symbols} we have summarized some of the symbols used in the paper.

\begin{figure}
\hspace{0.5cm}
    \includegraphics[width=8.cm]{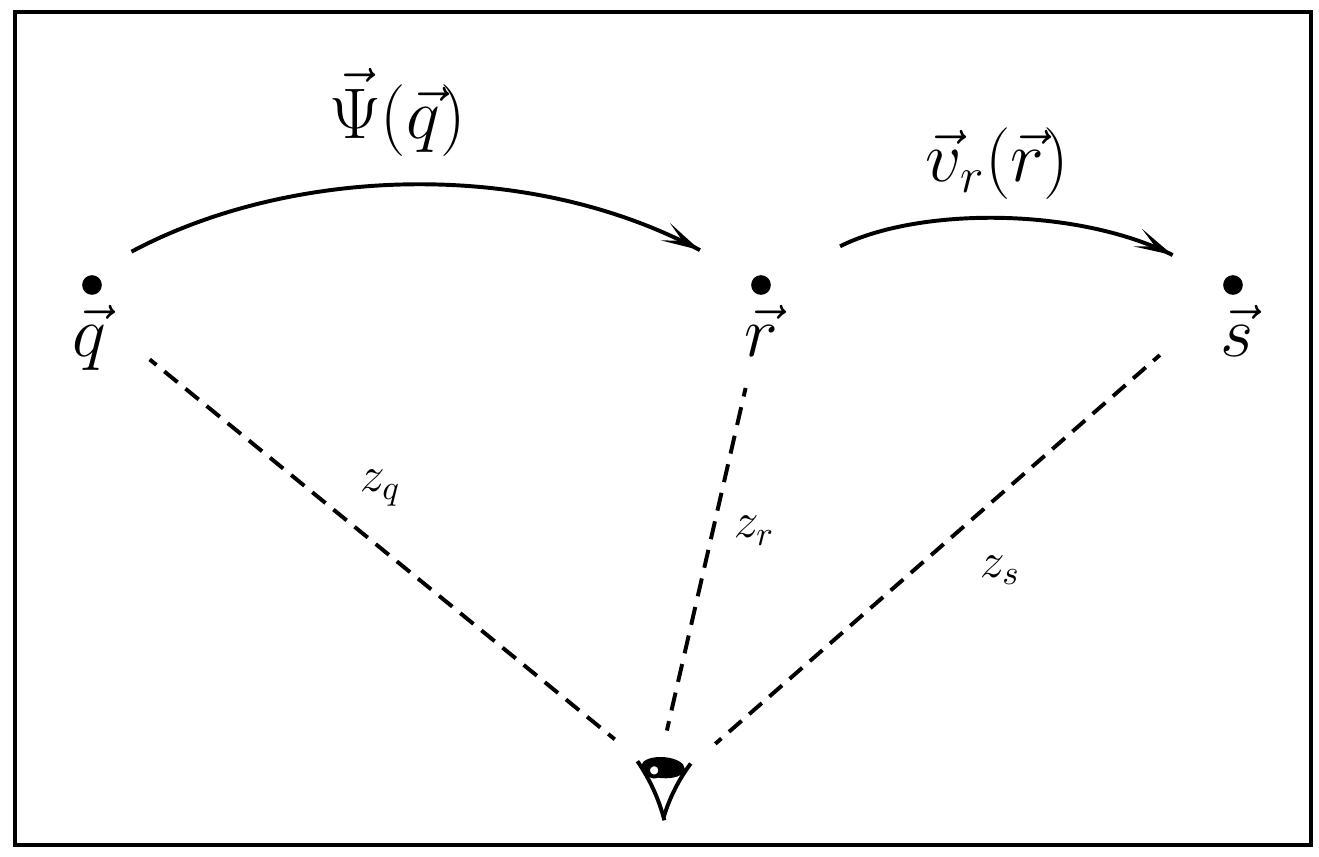}
   \caption{\label{fig:lagtoeul} Sketch representing the mapping from Lagrangian coordinates $\mbi q$ to 
Eulerian in redshift space $\mbi s$. The fist step consists in the 
mapping $\mbi q$ $\to$ $\mbi r$ mediated by the displacement field $\mbi 
\Psi(\mbi q)$. The second step, mediated by the peculiar velocity field 
(projected along the line of sight $\mbi v_r(\mbi r)$) takes us to $\mbi 
s$. The displacement field requires prior knowledge of Lagrangian space 
$\mbi q$, and the peculiar velocity field requires  prior knowledge of 
real-space $\mbi r$.}

\end{figure}

\begin{table}
\centering
\begin{tabular}{c l }
\hline
\textcolor{blue}{Symbol}      &       \textcolor{blue}{Description} \\
  \hline
$\mbi q$ & Position vector in Lagrangian coordinates \\
$\mbi r$ & Position vector in Eulerian coords. (real space) \\
$\mbi s$ & Position vector in Eulerian coords. (red-shift space)\\
$\Psi(\mbi q)$ & Displacement field\\
$z_{q}$ & Redshift at the Eulerian coords. \\
$z_{r}$ & Redshift at the Lagrangian coords. (real space)\\
$z_{s}$ & Redshift at the Lagrangian coords. (redshift space)\\
$\mathcal{B}$ & Bias description \\
$\mat R$ & Response function \\
\texttt{DMDF} & Dark matter density field. \\
\texttt{ALPT} & Augmented Lagrangian Perturbation Theory\\
$D(z)$ & Growth factor at cosmological redshift $z$\\
$\mathcal{M}$ &  Model\\

\hline \end{tabular}
\caption{Different symbols used in the text.}
\label{table:symbols}
\end{table}

% ==============================================================================================

\section{Description of the problem}
\label{sec:method}
In this section we describe the theoretical context in which the    \texttt{COSMIC BIRTH} approach resides.

\subsection{The phase-space mapping problem}
The data, as obtained from galaxy survey, represents a distribution of tracers (galaxies) in Eulerian redshift-space. In general, there is no velocity information, except for the Local Universe \citep[see][]{Courtois_2013, lanikea, sorce2014}. This implies on one side that an incomplete picture of the phase-space information is available, and on the other, that highly non-linear evolution (e.g. shell-crossing) is ubiquitously present. 

Inferring the real-space Lagrangian coordinates $\mbi q$ from Eulerian coordinates in redshift space, $\mbi s$, is a complex task, as one needs already prior knowledge of the Lagrangian coordinates for the displacement field $\mbi\Psi(\mbi q)$ and of the real-space coordinates for the peculiar velocity field along the line of sight $\mbi v_r(\mbi q)$\footnote{Note that the velocity field is a function of the real-space Eulerian coordinates $\mbi r$, which is in turn a function of the Lagrangian coordinates $\mbi q$, i.e., $\mbi v(\mbi r)=\mbi v(\mbi r(\mbi q))=\mbi v(\mbi q$).}, according to  
\be
\label{eq:lagtotoeul}
\mbi q=\mbi s-\mbi\Psi(\mbi q)-\mbi v_r (\mbi q)\,,
\ee
as depicted in Fig.~\ref{fig:lagtoeul}. In order to illustrate the complexity of the problem, let us use the Zel'dovich approximation \citep{1970A&A.....5...84Z,RevModPhys.61.185,2014MNRAS.439.3630W}.
At early enough times in the evolution of perturbations in the dark matter density field, (i.e. before shell-crossing), their dynamics is well described by a potential flow with a displacement field given by $\mbi\Psi(\mbi q,z)=D(z)\,\boldsymbol{\nabla}\Phi(\mbi q)$, where $D(z)$ is the growth factor of linear perturbations \citep[computed throughout this work as  in][]{1977MNRAS.179..351H}, $z$ the cosmological redshift and $\Phi$ is proportional to the initial gravitational potential of the perturbations satisfying Poisson's equation $\nabla^2\Phi(\mbi q)=\delta(\mbi q)$. The position and velocity of a test-particle at a fixed cosmological redshift can be written respectively as
\ba
\label{eq:zeld}
\mbi r(z) & = & \mbi q+ \mbi\Psi(\mbi q,z)=\mbi q + D(z)\,\boldsymbol{\nabla}\Phi(\mbi q),\\
\mbi v(z) & = & \frac{\dd \mbi r(z)}{\dd \tau}=\frac{\dd \mbi \Psi(\mbi q,z)}{\dd \tau}=\dot{D}(z)\,\boldsymbol{\nabla}\Phi(\mbi q)\nonumber\,,
\ea
where $\tau$ is the corresponding conformal time. The tracers defined by Eq.~(\ref{eq:zeld}) occupy a three-dimensional sub-manifold
of the entire six-dimensional phase-space according to the time-dependent mapping:
\be
\label{eq:subman}
%\mbi q \mapsto \left(\mbi q+ \mbi\Psi(\mbi q,z), \mbi v(\mbi q,z)\right)=\left(\mbi q+ D(z)\,\boldsymbol{\nabla}\Phi(\mbi q), \dot{D}(z)\,\boldsymbol{\nabla}\Phi(\mbi q)\right)\,.
\mbi q \mapsto \left(\mbi q+ D(z)\,\boldsymbol{\nabla}\Phi(\mbi q), \dot{D}(z)\,\boldsymbol{\nabla}\Phi(\mbi q)\right)\,.
\ee 
From such mapping it becomes clear that the knowledge of the initial density field $\delta(\mbi q)$ determines all posterior evolution. This is valid (without loss of generality) beyond the Zel'dovich approximation, in which case Eq.~(\ref{eq:zeld}) becomes more complex \citep[e.g.][]{2013JCAP...06..036T,doi:10.1093/mnrasl/slt101,2016MNRAS.463.2273F}.
The map between $\mbi r(z)$ and $\mbi q$ (e.g., as in Eq.~(\ref{eq:zeld})) is uniquely defined (bijective) until more than one stream of dark matter exists at one spatial location (shell crossing). Irregardless of the complexity of solving the phase-space collisionless fluid equations back in time, the problem becomes irreversible having only information on the (redshift-space) positions of galaxies without knowing their peculiar motions.
%easier: no unique solution of the EUR-LAG mapping due to lack of peculiar velocity information

\subsection{Previous velocity and displacement field reconstructions}

In the basic reconstruction scheme widely used for baryon acoustic oscillation reconstruction \citep[][]{2007ApJ...664..675E,2012MNRAS.427.2132P}, the displacement and peculiar velocity fields are obtained from smoothing the galaxy field in Eulerian redshift space 
$\Psi(\mbi s)=\Psi(K\otimes \delta_{\rm g}(\mbi s))$ leading to Lagrangian coordinates expressed as
\be
\label{eq:lagtotoeulwrong}
\mbi q=\mbi s-\mbi\Psi(\mbi s)-\mbi v_r (\mbi s)\,.
\ee
This can be improved with an iterative method envisaged to effectively solve Eq.~(\ref{eq:lagtotoeul}) (see \citet[][]{hada18} and previous works developing this technique e.g. \citet{1991ApJ...372..380Y,1999MNRAS.308..763M,2012MNRAS.420.1809W,2012MNRAS.425.2443K,2013MNRAS.429L..84K,2016MNRAS.457L.113K}).
Some strategies are based on seeking a unique optimal solution \citep[e.g.][]{1989ApJ...344L..53P,2000MNRAS.313..587N,2003MNRAS.346..501B,2018PhRvD..97b3505S,2019MNRAS.484.3818S}.
As mentioned in the introduction, we are interested in forward modelling approaches within a Bayesian framework, which yield an ensemble of solutions compatible with the observations.
Let us present the problem in a more formal context below leading to our chosen strategy.
For an overview of other works pioneering this field we refer to the introduction given in \citet[][]{2019MNRAS.488.2573B}.
We should stress here that methods like the ones explored in \citet{1999MNRAS.308..763M,hada18} do not use a Bayesian formalism, which permits to correct for the shot noise of the discrete galaxy distribution, and the survey mask and radial selection function in the reconstruction process. A Gaussian smoothing is applied in these methods limiting the reconstruction power towards small scales. We will present here the method, which permits us to deal with Lagrangian tracers within a Bayesian formalism.

%The large scale bias is usually measured in redshift bins (and galaxy populations according to various properties).

\section{The \texttt{COSMIC BIRTH} approach}\label{sec:birth}

%The methodology of the approach presented in this study is based on the same principles as the Kigen code \citep{2013MNRAS.429L..84K}, aiming at sampling the initial cosmic density field conditional to the observed galaxy distribution. This is achieved within a hierarchical Bayesian framework, using a nested Gibbs and Hamiltonian sampling scheme. Let us summarize this in two main crucial Gibbs-sampling  steps (additional steps will be discussed further below):

The methodology of the \texttt{COSMIC BIRTH} approach aims at sampling the initial cosmic density field, conditional to the observed galaxy distribution on the light-cone. To this end, it relies on an iterative Gibbs-sampling method, \citep[see e.g.][]{2008MNRAS.389..497K,2012MNRAS.420...61K}. 
The work presented here extends this approach to similarly sample displacement fields together with the peculiar velocities, as presented in \citet{2016MNRAS.457L.113K}. This is similar to the method presented by \citet{2013MNRAS.429L..84K}, with some exceptions which we will discuss below. Moreover, novel methods to sample the non-linear and non-local bias and the response function $\textbf{R}$ in Lagrangian space are presented. {\color{black} The response function stands in this context for the survey geometry and radial selection function as further explained below (see Eq. \ref{eq:response}).} 
Let us discuss each step in detail below .
%However, a number of differences  that within this study we will consider only one tracer at Lagrangian coordinates associated to each galaxy, as opposed to a larger number. Another difference comes from the way Eq. \label{eq:lagtotoeul} is solved.

\begin{figure*}
\hspace{0.5cm}
\begin{tikzpicture}[font=\ttfamily\small,node distance=1.45cm]
\hspace{0cm}

\node(start1)[startstop]{\huge{} \{$\vec{s}^{\ \rm o}$\}};
\node(start2)[startstop, below of= start1, yshift=-0.4cm]{\huge $R(\vec{s})$};
\node(start3)[startstop, below of= start2, yshift=-0.4cm]{\huge $\mathcal{B}(\vec{s})$};
\node(start4)[startstop, below of= start3, yshift=-0.4cm]{\huge ...};
\node(PROD)[startstop2, above of= start1, yshift=0.4cm, xshift=0.0cm]{\large INPUT: \\ Eulerian \\ Redshift-space};

\node(TR1_1)[decision_short, right of = start2, xshift=1.8cm, yshift=-0.93cm]{\large$\vec{v}(\delta(\vec{q}))$};

\draw[arrow](start1)--(TR1_1);
\draw[arrow](start2)--(TR1_1);
\draw[arrow](start3)--(TR1_1);
\draw[arrow](start4)--(TR1_1);

\node(PROD2)[startstop3, right of= PROD, yshift=-0.2cm, xshift=4.0cm]{\large Eulerian \\ Real-space};
\node(M1)[process_short, right of =start1, xshift = 4cm]{\Large \{$\vec{r}^{\hspace{0.1cm}\rm o}$\}};
\node(M2)[process_short, right of =start2, xshift = 4cm]{\Large $R(\vec{r})$};
\node(M3)[process_short, right of =start3, xshift = 4cm]{\Large $\mathcal{B}(\vec{r})$};
\node(M4)[process_short, right of =start4, xshift = 4cm]{\Large ...};

\draw[arrow](TR1_1)--(M1);
\draw[arrow](TR1_1)--(M2);
\draw[arrow](TR1_1)--(M3);
\draw[arrow](TR1_1)--(M4);

\node(TR1_2)[decision_short, right of = TR1_1, xshift=3cm]{\large$\vec{\Psi}(\delta(\vec{q}))$};

\draw[arrow](M1)--(TR1_2);
\draw[arrow](M2)--(TR1_2);
\draw[arrow](M3)--(TR1_2);
\draw[arrow](M4)--(TR1_2);

\node(PROD3)[startstop3, right of= PROD2, yshift=-0.0cm, xshift=3.0cm]{\large Lagrangian \\ Real-space};
\node(L1)[process_short, right of =M1, xshift = 3cm]{\Large \{$\vec{q}^{\hspace{0.1cm}\rm o}$\}};
\node(L2)[process_short, right of =M2, xshift = 3cm]{\Large $R(\vec{q})$};
\node(L3)[process_short, right of =M3, xshift = 3cm]{\Large $\mathcal{B}(\vec{q})$};
\node(L4)[process_short, right of =M4, xshift = 3cm]{\Large ...};

\draw[arrow](TR1_2)--(L1);
\draw[arrow](TR1_2)--(L2);
\draw[arrow](TR1_2)--(L3);
\draw[arrow](TR1_2)--(L4);

\node(RS)[results, right of = TR1_2, xshift=3.5cm]{\huge $\delta(\vec{q})$};
\node(ham)[process_DUMMY2, above of = RS, yshift=1.5cm]{\Large Hamiltonian \\ sampling};
\draw[arrow](RS)--(ham);

\draw[arrow](L1)--(RS);
\draw[arrow](L2)--(RS);
\draw[arrow](L3)--(RS);
\draw[arrow](L4)--(RS);

\node(DUMMYL)[process_DUMMY, below of = RS, yshift=-2.6cm]{}; 
\node(DUMMY1)[process_DUMMY, below of = TR1_1, yshift=-2.6cm]{}; 
\node(DUMMY2)[process_DUMMY, below of = TR1_2, yshift=-2.6cm]{}; 

\draw[arrow_nlb](RS)--(DUMMYL);
\draw[arrow_pib](DUMMY1)--(TR1_1);
\draw[arrow_pib](DUMMY2)--(TR1_2);
\draw[arrow_nlb](DUMMY1)--(DUMMYL);

\node(corner)[process_DUMMY, right of = start4, xshift=0.2cm, yshift=-0.8cm]{};
\draw[dotted, thick] (start1.north) + (-1.7, 0.3) rectangle (corner.east) ;

\node(corner2)[process_DUMMY, right of = M4, xshift=-0.7cm, yshift=-0.8cm]{};
\draw[dotted, thick] (start1.north) + (4.6, 0.3) rectangle (corner2.east) ;

\node(corner3)[process_DUMMY, right of = L4, xshift=-0.7cm, yshift=-0.8cm]{};
\draw[dotted, thick] (start1.north) + (9.0, 0.3) rectangle (corner3.east) ;

\draw[decorate,decoration={brace,amplitude=15pt,mirror}] 
    (-1.7,-7.3) coordinate (t_k_unten) -- (13.9,-7.3) coordinate (t_k_opt_unten); 
\node at (6,-8.2){\Large Gibbs sampling};

\end{tikzpicture}
%\vspace{-1.2cm}
\caption{\label{fig:flowchart} Flowchart depicting the Gibbs-sampling scheme of $\tt BIRTH$ to sample the density field $\delta(\mbi q)$ at initial cosmic times. The input data reside in Eulerian redshift-space and can be transformed to real-space with a given peculiar velocity field $\mbi v$. Once the data have been transformed to Lagrangian space with the displacement field $\mbi \Psi$, we can sample the corresponding density field with higher-order Hamiltonian-sampling. We note, that this scheme can be extended to account for more variables of the model (power spectrum, growth rate, etc).}
\end{figure*}
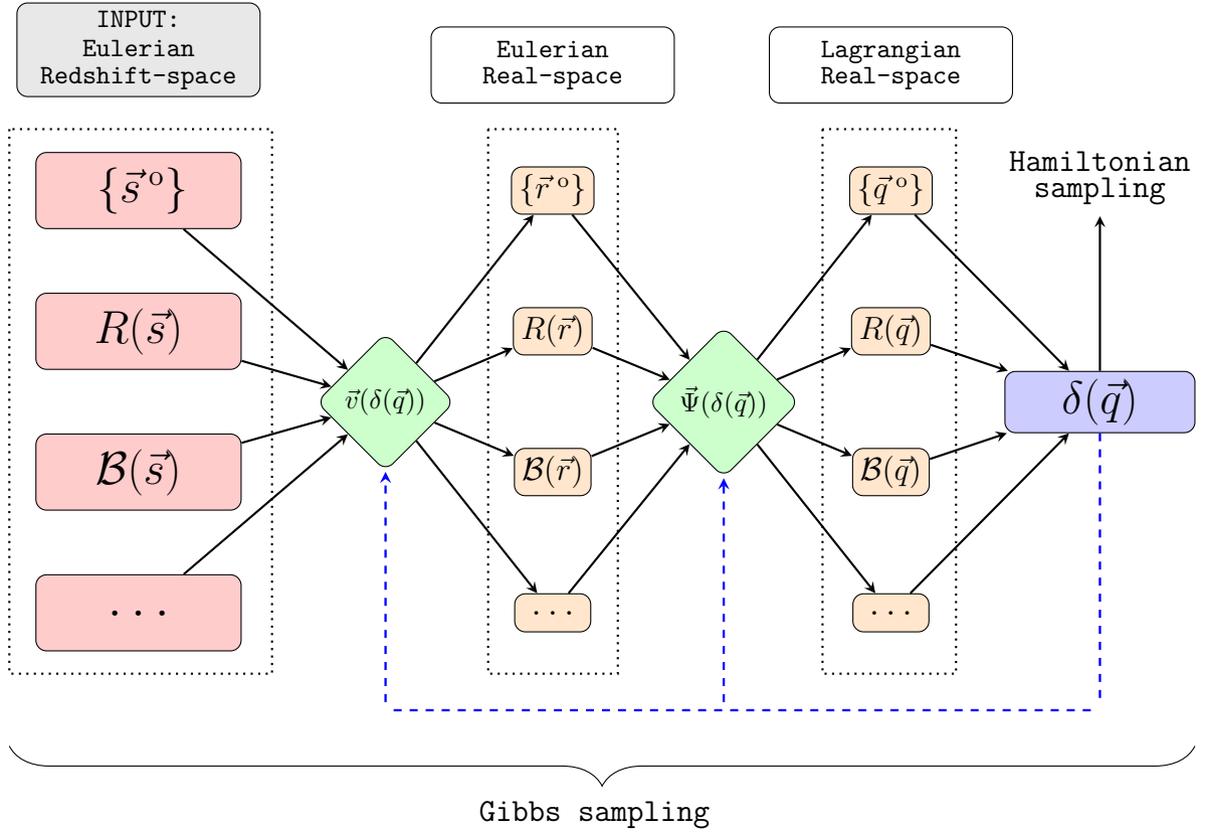

\begin{figure*}
\hspace{-1.cm}
   \includegraphics[width=15.cm]{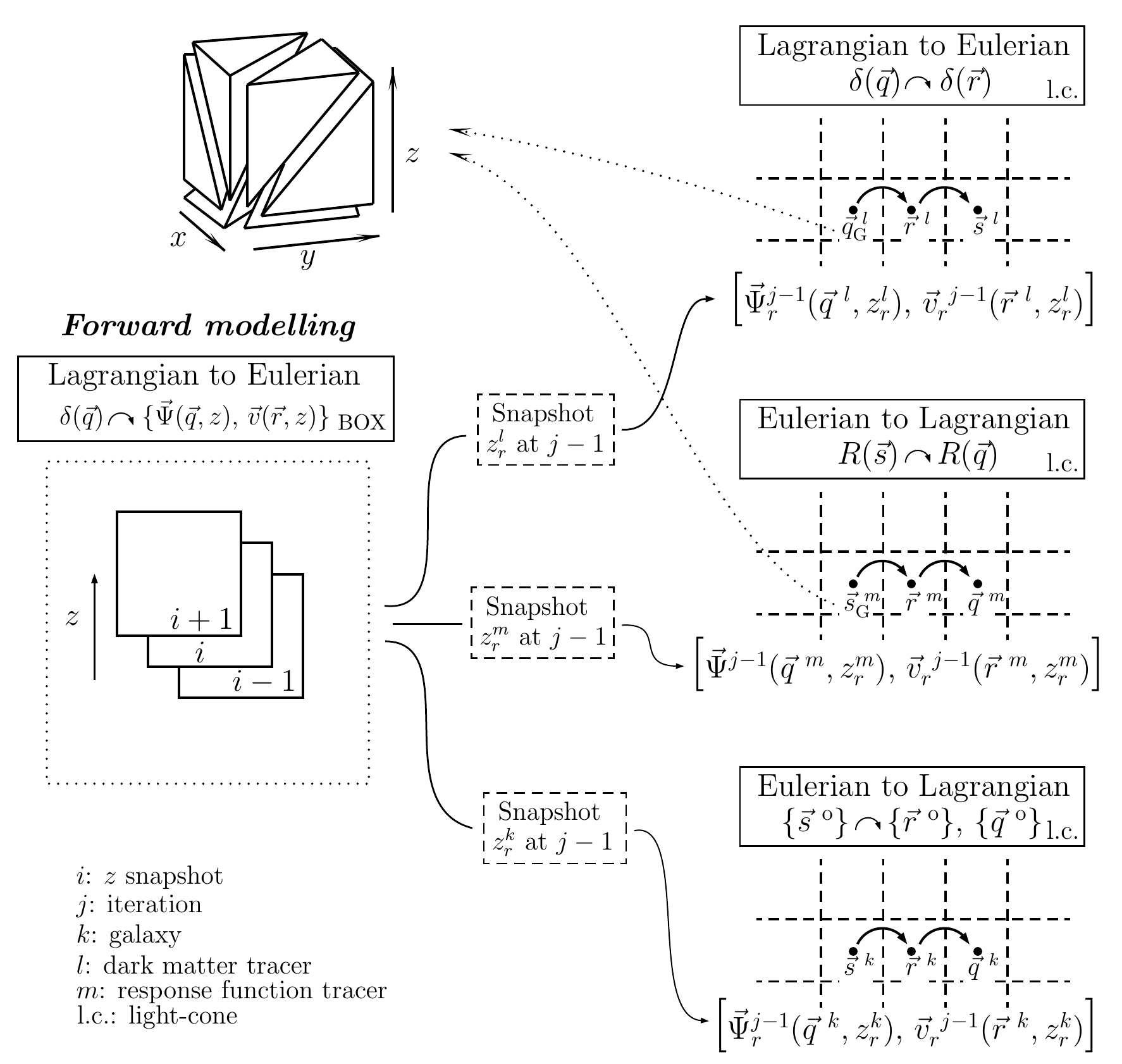}
  \caption{\label{fig:phasespace} This sketch illustrates the various Lagrangian to Eulerian and vice-versa mappings performed within {\tt COSMIC BIRTH}. 
  At the bottom right it is illustrated how galaxies are mapped to real Lagrangian space looking up the information from a forward simulation based on the previous iteration yielding outputs at different redshift snapshots ($\dots$, $i-1$, $i$,  $i+1$, $\dots$), as shown on the left. The light-cone dark matter field calculation from Lagrangian to Eulerian space and the response function mapping from Eulerian to Lagrangian space happen in the same loop, each of one looking up the corresponding redshift snapshot information, when going through the cells of the mesh. For both these two cases a {\color{black} Lagrangian tetrahedral tessellation} is applied. To this end the corresponding tracers are defined at the cell center and thus denoted with a subscript G.
  }
\end{figure*}

\subsection{The Gibbs-sampling scheme}

%To give an overview over the sampling scheme, let us introduce first anumber of useful quantities and definitions. 
We are interested in the matter over-density field $\delta(\mbi q)$ evaluated at Lagrangian coordinates $\mbi q$ and defined on a regular cubical mesh with $N_{\rm c}$ cells. The data is represented by the three dimensional galaxy distribution, as observed in Eulerian redshift space, with coordinates $\{\mbi s^{\rm o}\}$. Given the peculiar velocities $\mbi v$, we are in position to infer their corresponding real space coordinates $\{\mbi r^{\rm o}\}$. Furthermore, knowing the displacements $\mbi \Psi$ connecting the initial cosmic times with the final ones, we can compute their corresponding Lagrangian coordinates $\{\mbi q^{\rm o}\}$, as discussed in the previous section.
The galaxy number counts on the mesh $\mbi N_{\rm g}$  can be related to the matter density field according to our likelihood model and a bias description $\mathcal{B}$ presented in \S \ref{sec:bias_rel}. Furthermore, we use a response function $\mat R$, which accounts for the survey geometry, or angular completeness, and for the radial selection function.  
In particular, the joint PDF of all the above mentioned variables can be sampled within a Gibbs-sampling scheme based on  the corresponding conditional PDFs:
\ba
\label{eq:inference}
\delta(\mbi q)&\curvearrowleft& \mathcal P_{\delta}\left(\delta_{\mbi q}|{\{\mbi q^{\rm o}\}}, {\mat R}_{\mbi q},{\mat C}_{\mbi q}\left(\{p_{\rm c}\}\right),\{\mathcal B_{\mbi q}\}\right)\,{,}\label{eq:sig} \nonumber \\
{\{\mbi r^{\rm o}\}}&\curvearrowleft& \mathcal P_{r}\left(\{\mbi r^{\rm o}\}|\{\mbi s^{\rm o}\},\{\mbi v^z\left(\delta_{\mbi q},f^z_\Omega\right)\},{\cal M}_{v}\right)\,,\nonumber\\
{\{\mbi q^{\rm o}\}}&\curvearrowleft& \mathcal P_{q}\left(\{\mbi q^{\rm o}\}|\{\mbi r^{\rm o}\},\{\mbi \Psi^z_{\mbi q}\},{\cal M}_{\Psi}\right)\,,\nonumber\\
{\mat R}\left(\mbi q\right)&\curvearrowleft& \mathcal P_{R}\left({\mat R}_{\mbi q}|{\mat R}_{\mbi s},\{\mbi \Psi^z_{\mbi q}\},{\cal M}_{\Psi}\right)\,,\nonumber\\
\{\mathcal B\left(\mbi q\right)\}&\curvearrowleft& \mathcal P_{B}\left(\{{\mathcal B}_{\mbi q}\}|\{{\mathcal B}_{\mbi s}\},\{\mbi \Psi^z_{\mbi q}\},{\cal M}_{\Psi}\right)
\,{,}
\ea
where the subscripts $q$ and $s$ stand for Lagrangian real space, and Eulerian redshift space coordinates, respectively. {\color{black} The curved left arrows stand for the sampling process.} The superscript $z$ stands for redshift bin. ${\cal M}_{v}$ and ${\cal M}_{\Psi}$ represent the models describing peculiar motions and displacement fields, respectively.
 The growth rate $f^z_\Omega$ will be further discussed in \S \ref{sec:LtoE}.
%In case the bias is not interpolated, a galaxy with $\mbi q$ at redshift bin $j-1$, and $\mbi s$ at redshift bin $j$ will have a large scale bias $b_j$ and not $b_{j-1}$. \aba{some of the symbols here are not defined}

The approach described above has the great advantage of being general for any structure formation model, as only the initial and final positions with their peculiar motions are needed.
In this study, we only consider the approach provided by the Augmented Lagrangian perturbation theory \texttt{ALPT} \citep[][]{doi:10.1093/mnrasl/slt101}, which is being successful describing clustering down to a few Mpc scales and has been implemented for the generation of halo mock catalogs  generation based on bias mapping methods \citep[][]{2016MNRAS.457L.113K,   2019MNRAS.483L..58B}. 

Figs.~\ref{fig:flowchart} and \ref{fig:phasespace} depict, with a flow-chart, the main steps followed within the Gibbs sampling method. In the following sub-sections we summarize the first three main crucial Gibbs-sampling steps and the corresponding assumptions (additional steps will be subsequently presented).

\subsection{Step 1: sampling the linear density field}

 The continuous primordial dark matter field is sampled assuming that the observed galaxies are identified in real Lagrangian-space at high redshift. This is done within a Bayesian framework, in which a log-normal-Poisson posterior distribution function (PDF) is assumed. The log-normal PDF stands for the prior of the dark matter distribution, while the Poisson PDF represents the likelihood describing the distribution of discrete Lagrangian space tracers \citep[e.g.][]{2010MNRAS.403..589K}. We note that the log-normal prior assumes on one side a comoving Lagrangian framework, and, on the other, that tracers can be uniquely followed, i.e. neglecting shell crossing \citep{1991MNRAS.248....1C,2012MNRAS.425.2443K}. This precisely applies for Lagrangian tracers at high redshift. Also, the logarithmic transformation of the normalised density $\log(1+\delta)$ (with $\delta=\rho/\bar{\rho}-1$) tends towards the over-density field for $|\delta|\ll1$. 
 
 On the other hand, Poissonity is a reasonable assumption for homogeneously distributed tracers at high redshift, e.g, before  gravity introduces small-scale clustering. Such small-scale clustering generates over-Poisson dispersions at the scale of the sub-volume element at which the galaxy number counts are defined, since at larger scales an inhomogeneous Poisson distribution accounts for the large scale clustering modulated by the dark matter density field \citep[e.g.][]{1980lssu.book.....P,1989ApJ...341..588S,1998MNRAS.299..207S,2014MNRAS.439L..21K,2014MNRAS.441..646N,2015MNRAS.450.1486A}.
 In any case such a deviation from Poissonity can also be included in a Bayesian framework \citep{2015MNRAS.446.4250A}.  %, which as we will show below is not necessary in our study case.
 
 {\color{black} The density estimation step is the bottle neck in our computations. }
We would like to stress that \texttt{COSMIC BIRTH} achieves a high efficiency with the development of a novel version of the Hamiltonian Monte Carlo (HMC) sampling introduced by \citet{2010MNRAS.407...29J}, and studied in detail in a companion paper (Hern\'andez-S\'anchez in preparation). For further reference to the basic implementation see also \citet[][]{2017MNRAS.467.3993A}. In particular, we have introduced, as a novel ingredient, a higher-order discretisation of Hamilton equations using subsequent second order Leapfrog operators. In this work we use the 4th order discretisation, which consists of a forward time step integration, followed by a backward one, with a third and final forward one with different time-step lengths following \citet{PhysRevLett.63.9}. 

 In addition, we introduce a strategy to deal with non-diagonal Hamiltonian mass-matrices including the survey geometry, which act as a \emph{preconditioner} to additionally speed up the algorithm (see Appendix \ref{app:hmc}).

The Hamiltonian sampler could have assigned a larger number of tracers in Lagrangian space, as available in the Eulerian light-cone, i.e., the observed galaxy distribution. This was done in  \citet{2013MNRAS.429L..84K,2012MNRAS.427L..35K,2013MNRAS.435.2065H}. Here we want to control the evolving bias on the light-cone (see \S \ref{sec:bias_rel}). Therefore we will restrict in this first paper the Lagrangian tracers to be equal in number to the Eulerian ones. We note however, that the bias does not change, if the number of Lagrangian tracers is equal for each Eulerian tracer, as the over-density field does not change. In case, one would consider a different number of tracers depending for instance on the location of the galaxy in the cosmic web this picture becomes more complicated, although not unsolvable. We leave this investigation for future work.

\begin{figure}
   \hspace{.0cm}
   \includegraphics[width=8.cm]{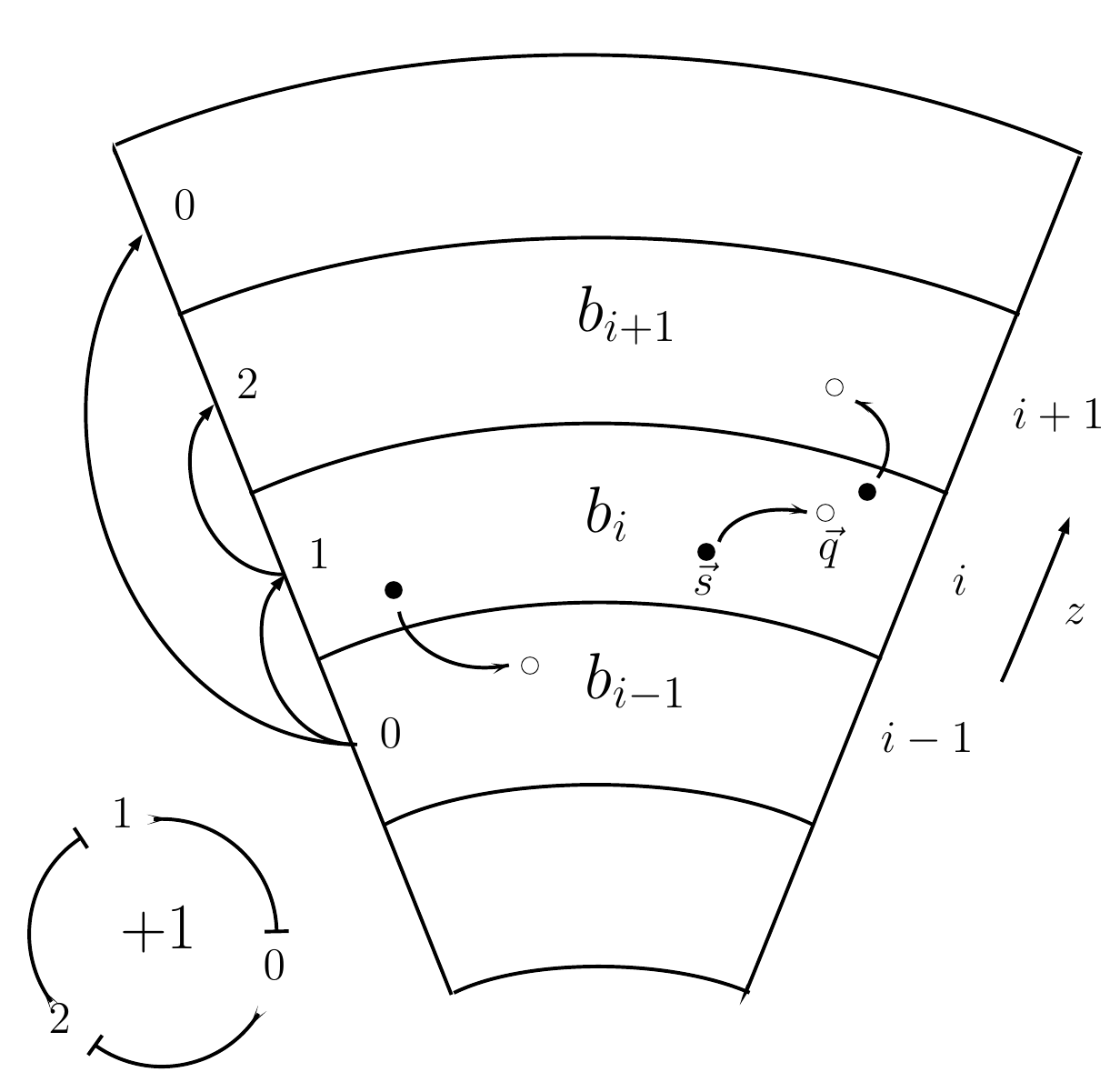}
  \caption{\label{fig:lightconeper} This is a sketch of the lightcone of a galaxy survey divided in redshift bins $\dots$, $i-1$, $i$,  $i+1$, $\dots$, with the corresponding large scale bias $\dots$, $b_{i-1}$, $b_i$,  $b_{i+1}$, $\dots$. 
  A galaxy may stay in its redshift bin when doing reconstruction $\mbi s\rightarrow\mbi q$, jump to a lower, or to a higher redshift bin. 
  {\tt COSMIC BIRTH} goes through the data in a cyclic order of even permutations: $\dots$-0-1-2-0-$\dots$, requiring the storage in RAM of the displacements and velocities of only three redshift snapshots at once.}
\end{figure}

\subsection{Steps 2 \& 3: displacements and peculiar velocity fields}
\label{sec:LtoE}

The Lagrangian tracers are sampled assumimg that the initial cosmic density field is known.  
  This can be achieved using a given structure formation model, yielding the displacement and peculiar velocity fields at different redshifts from that initial field. These in turn can be used to obtain the Lagrangian tracers which correspond to the observed tracers in Eulerian space.
  We note that this step is similar to the analysis of $N$-body simulations for which the initial conditions are known and the particles composing a halo are traced back to high redshift \citep[e.g.][]{2011MNRAS.413.1961L}.
  The additional uncertainty in our study comes from the lower resolution at which we reconstruct the initial Gaussian field, which is in general not high enough to resolve the halos hosting the observed galaxies. Also, galaxy bias  and redshift space distortions contribute to this uncertainty, as we will discuss below. 
  
  In this second step \texttt{COSMIC BIRTH} obtains the real space position for each galaxy, given its observed redshift space $s^{\rm obs}$ position (required for the first step). The latter is obtained by sampling the peculiar velocities $\{\mbi v\left(\mbi\delta,f_\Omega\right)\}$ (with the growth rate given by $f_{\Omega}\equiv d\log D(a)/d\log a$), assuming that the density field and the growth rate $f_\Omega$ are known\footnote{We note that assuming a wrong growth rate will yield an anisotropic reconstructed density field. This was recently investigated \citep[][]{2015A&A...583A..61G} by jointly sampling the anisotropic power spectrum including the growth rate and the redshift space density field.}. One can add a dispersion term to the displacement and peculiar velocity field accounting for the uncertainty. In practice this is assumed to be  Gaussian distributed, and set to low standard deviations of about 1 $h^{-1}\,{\rm Mpc}$ \citep[see][]{2016MNRAS.457L.113K,2017MNRAS.467.3993A}.

%The first step samples the linear density fields at Lagrangian coordinates defined on a mesh with $N_{\rm c}$ cells compatible with the number counts on that mesh $\mbi N_{\rm g}$  of the galaxy distribution in real space $\{\mbi r\}$.  

In a further publication we will explore in more detail the redshift-space distortions corrections achieved with \texttt{COSMIC BIRTH}.
We should also note that the solution to the Lagrangian-to-Eulerian mapping problem is not solved here in the same way to the approach presented in \citet{2013MNRAS.429L..84K}. While in the latter approach a large number of  large-scale structure tracers are displaced forward in time and then linked to observed galaxy distribution (in a likelihood comparison step), here we solve Eq.~(\ref{eq:lagtotoeul}) by evaluating the forward computed displacement and velocity fields at the Lagrangian locations of the previous iteration (see Fig.~\ref{fig:phasespace}). This is not yielding a completely self-consistent relation between the sampled initial Gaussian field and the final positions of tracers, as the Gaussian fields change from iteration to iteration given various uncertainties modelled in the Bayesian framework. Therefore, we expect some improvement at least on small scales, if the approach formulated in \citet{2013MNRAS.429L..84K} is implemented in the \texttt{COSMIC BIRTH} code. As anticipated, we will investigate this in detail in future work.

%Let us stress here that this formulation admits 

%After these probabilities reach their so-called stationary distribution, the drawn samples are representatives of the target distribution.
%In the following sections we will dig deeper into Eq.~(\ref{eq:sig}) in order to describe our sampling strategy.

\subsection{Step 4: response function in Lagrangian space}
\label{sec:response}

The problem arising from the type of approach described  in this section is that the observables are obtained in Eulerian space, while the reconstruction of the matter density field is performed in Lagrangian space, under the assumption that the data is in that space. This is still an unsolved problem for the survey geometry or the radial selection function. Previous studies \citep[][]{2013MNRAS.429L..84K} have augmented empty regions with some mock galaxies and inverse weighted the selection function to mitigate these issues. Since this operation is particularly dangerous when the selection function acquires very low values \citep[see discussion in ][]{2009MNRAS.400..183K}, such studies were restricted only to small cosmological volumes \citep{2012MNRAS.427L..35K}.

One can improve this in a number of ways. The key concept is data augmentation  {\color{black} to enable a balanced likelihood analysis} \citep[see e.g.][]{zbMATH00458721}. {\color{black} We enumerate several strategies as follows.}
\begin{enumerate}
    \item One possibility is based on the production of data exactly compensating for the incompleteness of the survey in each Gibbs sampling iteration. This can be done according to some bias model and the density field obtained in a given iteration. Marginalisation over such augmented data could be done by sampling, in each Gibbs sampling iteration, new augmented data, discarding the previous ones (the true observed data is untouched during the iterations). Note that this approach needs an accurate Eulerian bias model, which is particularly difficult to achieve \citep[][]{2019arXiv191013164P}.
    
    \item A second approach can be that of introducing a noise-component  \citep[as is done with Wiener filtering, e.g.][]{1995ApJ...449..446Z,2019JCAP...10..035H}, which depends on the completeness, being larger in less observed areas. The inconvenience of this approach is that in our particular case one would need some degree of arbitrary fine-tuning to get sensible results as the level of noise is a complex function of the completeness for which we lack a proper model \citep[e.g.][]{1999ApJ...520..413Z}.
    
    \item
    The most natural option consists of including the completeness in a response function and let the Bayesian model compensate the survey mask and radial selection function in the reconstructed initial cosmic density field \citep[see][]{1995ApJ...449..446Z,2008MNRAS.389..497K}.
    This has been done for the first time connecting Lagrangian to Eulerian space including cosmic evolution in \citet[][]{2013MNRAS.432..894J}, and later adopted by \citet[][]{2013ApJ...772...63W} and \cite{2019MNRAS.488.2573B}.
    All these approaches compute gradients of structure formation demanded within the Hamiltonian sampling. Also, an accurate description of Eulerian bias is requested, 
    without which those methods are likely to down-weight the data (see e.g. \citet[][where the likelihood is down-weighted with a factor of 0.3]{2019A&A...625A..64J}.

    \item We propose here to calculate the response function in Lagrangian space to be able to apply the standard Bayesian approach. We explain this in detail below.
    \end{enumerate}

    The response function can be straightforwardly computed for the radial selection function, as we just have to calculate it based on the reconstructed galaxy sample at Lagrangian coordinates.
    However, the angular survey mask is not trivial to compute in Lagrangian space.
    First we have to project it to the three-dimensional space as it is introduced in \cite{2009MNRAS.400..183K}.
    Then that Eulerian field has to be mapped to Lagrangian space using the reconstructed forward displacement field on the lightcone. 
    In particular, we assume that each cell center $\mbi r_{\rm g}$ represents a response function tracer. We need to keep track of the Lagrangian coordinates of each cell center in the same way we do it with the galaxies. This enables us in principle to make a mapping of the response function to Lagrangian space. 
    However, the finite number of tracers (finite resolution of the grid) yields inaccurate estimates of this mapping.

Therefore we resort to {\color{black} Lagrangian tetrahedral tessellation} \citep[see][]{shandarintetra,abel2012,2013MNRAS.434.1171H}, which makes a tessellation of the mesh into tetrahedrons and uses the positions and displacement field information to get accurate density estimates even on coarse resolutions \citep[e.g.][]{2019MNRAS.483L..58B}. See also the works by \citet[][]{2012ApJ...754..126F, neyrinck2012, neyrinck2013}.

Let us call the resulting three dimensional projected angular mask as $\mbi w_\alpha$, and refer to the corresponding the angular response function as
\be
\label{eq:response}
\mat R_\alpha=\mbi w_\alpha \mathbb{1}\,.
\ee
We note that the angular completeness does not care about real or redshift space, which affect only the radial direction. We consider thus the redshift $z_s$ as the final one of the displacement field (in the Zel'dovich approximation this would be $\Psi(q,z)=D(z_r)\Psi(q)$).

The radial selection function is computed from the $z$-distribution of large scale structure tracers normalised by the volume enclosed in shells \citep[or divided by $z^2$ as it is described in][]{2017MNRAS.467.3993A}. We do this computation in Lagrangian space, as we need all quantities defined in that space in order to perform the step 1.
Let us refer to the radial selection function part of the response function as
\be
\mat R_r=\mbi w_r \mathbb{1}\,,
\ee
where $\mbi w_r$ is the three dimensional projected spherical symmetric radial selection function.
The total response function $\mat R$ will be the product of $\mat R_\alpha$ and $\mat R_r$:
\be
\mat R=\mat R_\alpha\cdot\mat R_r\,.
\ee
We note that (iteratively) computing the radial selection function in real (Lagrangian) space prevents the  so-called "Kaiser rocket" effect \citep{1987MNRAS.227....1K,2014ApJ...788..157N}. 
The Lagrangian framework we are using has another advantage. We can define the radial selection function as described above ignoring light-cone effects, as we are performing the reconstruction at a single snapshot at high redshift. However, direct reconstructions of the dark matter density field in Eulerian space require to take into account a cosmological selection function \citep[see][]{2015A&A...583A..61G}.

\begin{figure}
   \hspace{-0.5cm}
   \includegraphics[width=8.cm]{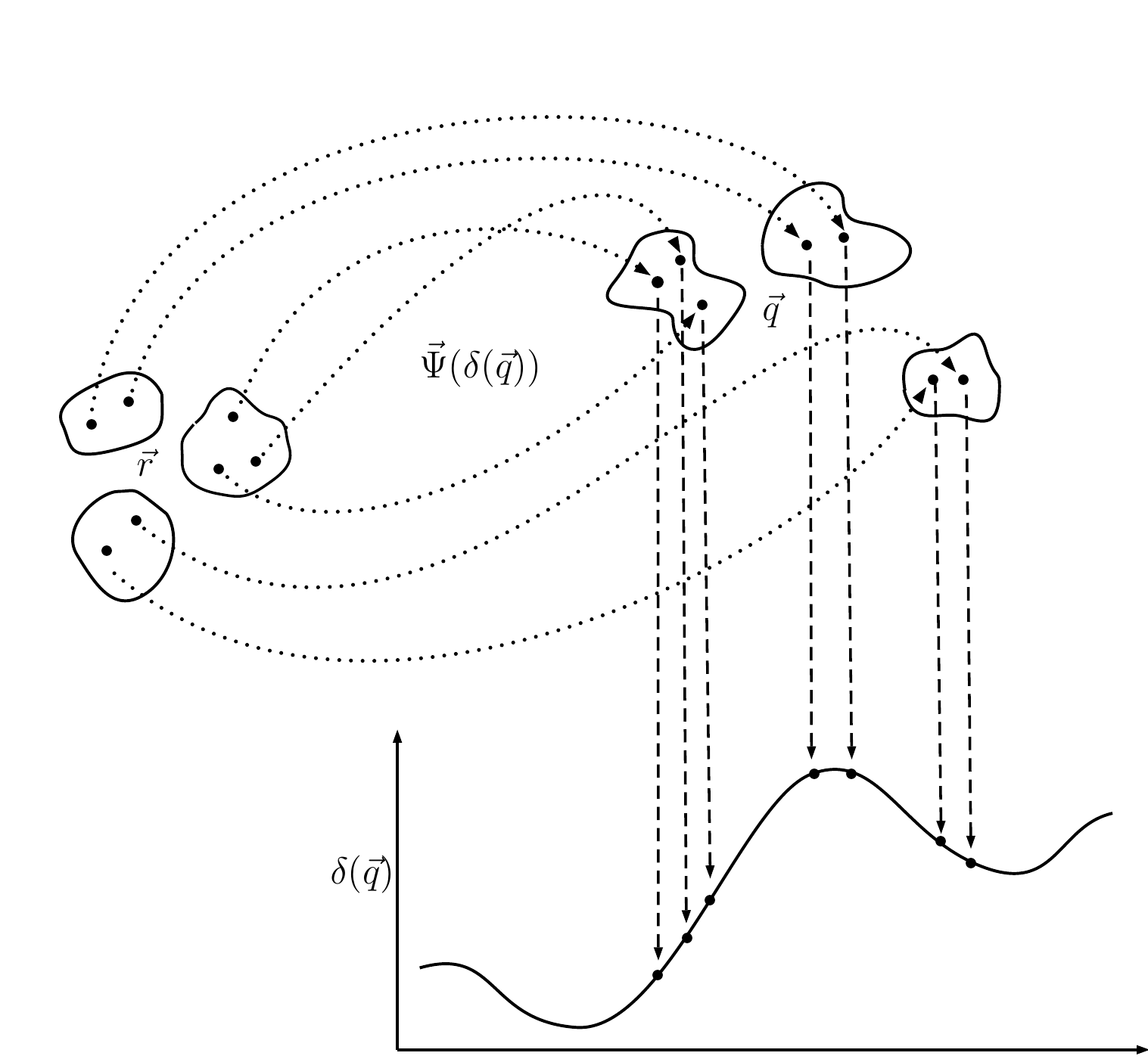}
  \caption{\label{fig:biasmodel} This sketch illustrates the relation between galaxy and halo bias at Eulerian and Lagrangian space. Galaxies (represented by dots) are depicted as tracers of halos (represented by asymmetric regions). The corresponding proto-halos are non-spherical and do not necessarily trace the peaks of the initial cosmic density field. This is accounted for in {\tt COSMIC BIRTH} when mapping the galaxies to Lagrangian space and using a bias without selecting the peaks, i.e. without threshold bias. A higher galaxy number density yields a more accurate description of the proto-halo regions. This hints towards the advantage of using multiple galaxies at the same Eulerian position instead of varying the mass, as these will be mapped differently in Lagrangian space. }
\end{figure}

\subsection{Step 5: Sampling the galaxy-dark matter bias relation}\label{sec:bias_rel}

The description of the galaxy distribution (in our context: galaxy number counts per cell mapping the observed volume onto a mesh) with respect to the large scale dark matter field (defined on the same mesh) requires effective bias models, encoding the underlying physics of galaxy formation in a non-linear, non-local functional dependence. 
The large scale bias can be measured in redshift bins (and galaxy populations according to various properties) using different probes of clustering \cite[e.g.][]{2002MNRAS.335..432V,2005PhRvD..71d3511S,2005MNRAS.356..456C,2009MNRAS.392..682C,2014MNRAS.440.1527L,2015MNRAS.451..539G,2018MNRAS.476.1050B,2019arXiv190912069P}.
The galaxy bias is in general a non-linear function of the underlying continuous dark matter field. In the attempt of modelling such a relation, a Taylor expansion has been suggested both as a function of the dark matter i) over-density field \citep{1993ApJ...413..447F}, and ii) to its logarithm  \citep[][]{1992ApJ...399L.113C}.
In fact the latter expansion corresponds, truncated to  first order, to a power-law, giving already a fair description at the two point statistics \citep[see e.g.][]{2013MNRAS.435..743D}. 
However, it has been shown that a threshold bias based on the peak split-background picture \citep[e.g.][]{1984ApJ...284L...9K} is crucial for an accurate description of the three-point statistics \cite[see e.g.][]{2014MNRAS.439L..21K,2015MNRAS.450.1836K}.
This model has been refined to have a smoother drop-off behaviour towards the low density regime by \citep[][]{2014MNRAS.441..646N} and has been successfully applied to reproduce the LRG distribution of the BOSS survey \citep[]{2016MNRAS.456.4156K}. 
The expected galaxy number counts is then given by
\be
\rho_{\rm g}(\mbi r, z)=\gamma(z)\mathcal B(\mbi r, z)\,,
\ee
with a normalisation  of
\be
\gamma(z)=\frac{\bar{N}(z)}{\langle\mathcal B(\mbi r, z)\rangle}\,,
\ee
and a non-linear deterministic bias given by
\be
\label{eq:biasthreshold}
\mathcal B(\mbi r, z)={\exp}\left[
-\left( \frac{\rho(\mbi r,z)}{\bar{\rho}(z)b_{\rho}(z)} \right)^{b_{\epsilon}(z)} \right]
\left(\frac{\rho(\mbi r,z)}{\bar{\rho}(z)}\right)^{b_{\rm p}(z)}\,,
\ee
where the density is linked to the over-density field through $\rho(\mbi r,z)=\bar{\rho}(z)(1+\delta(\mbi r,z))$.
However, this simple non-linear model lacks a proper non-local bias description \cite[see e.g.][]{2009JCAP...08..020M}, which can be modelled through the tidal field to second order \citep[see e.g.][]{ 2019MNRAS.483L..58B, 2019arXiv190606109B}.  A complete bias description also demands in principle the dependence with the initial cosmic field \citep[e.g.][]{2018PhR...733....1D} and there are current attempts to include this within a Bayesian context.

\begin{figure}
\includegraphics[width=8.cm]{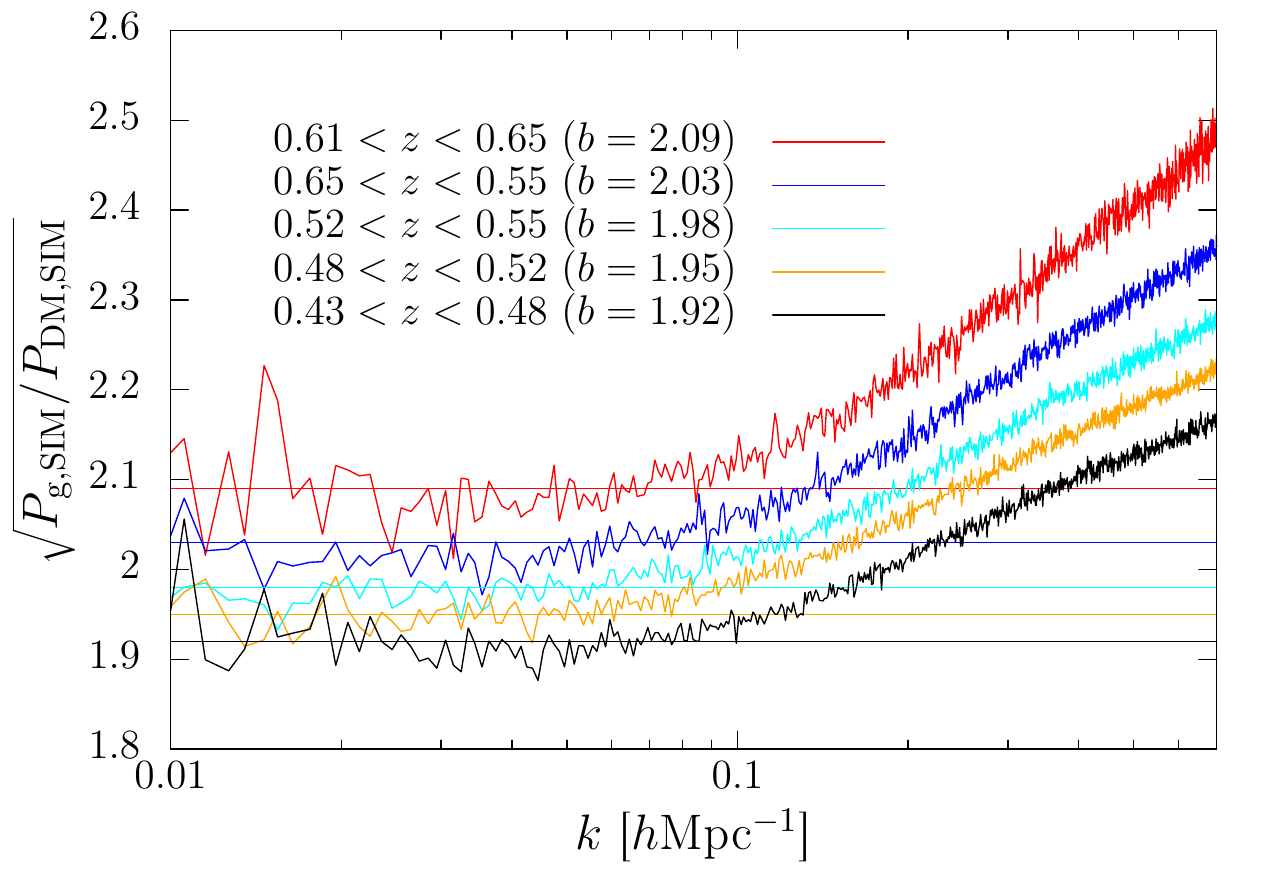}
\caption{\label{fig:bias} Large-scale bias obtained from the power spectrum ratio between the galaxy and the dark matter field at different redshift snapshots from the BigMD simulation (SHAM catalogs and dark matter particles, respectively). For illustrative purposes 5 bins are shown from the 10 to 20 used in our study. }
\end{figure}

The goal of this work is to find a practical Lagrangian bias description which can be directly derived from the observations, assuming that the tracers of the large scale structure reside in Lagrangian space. This is achieved within our Gibbs-sampling scheme through an Eulerian to Lagrangian mapping which already accounts for non-local and non-linear bias, simplifying the bias relation in Lagrangian space. This is represented in a sketch in Fig.~\ref{fig:biasmodel}. 

In the remainder of this section we will derive a complete formalism which connects the observed redshift large scale clustering over the large scale Lagrangian bias,  to a non-linear Lagrangian bias model including the dependence on the chosen mesh resolution to represent the galaxy number counts and the dark matter field.

\subsubsection{Eulerian large scale bias}

The clustering of galaxies in redshift space with respect to some fiducial cosmology provides a measure of the large scale bias.
Following \citet[][]{2017MNRAS.467.3993A},
given a redshift $z$  one can define  the ratio between the galaxy correlation function in redshift space at $z$ ($\xi^s_{\rm g}(z)$) and the matter correlation function in real space at $z$ ($\xi_{\rm M}(z)$) as
\be
b^s(z)\equiv\sqrt{\frac{\xi^s_{\rm g}(z)}{\xi_{\rm M}(z)}\Big|_{\rm LS}}\,.
\ee
The quantity $\xi^s_{\rm g}(z)$ can be obtained from the data without having to assume any information of bias or growth rate.
Furthermore, one can use the Kaiser factor $K=1+(2/3)f_{\Omega}/b+(1/5)(f_{\Omega}/b)^2$ (where $f_{\Omega}$ denotes the growth rate \citep[][]{1987MNRAS.227....1K}) to relate the galaxy correlation function in redshift space to the matter correlation function in real space, 
%\label{fig:Kfac}
$\xi_{\rm  g}^{s}(z)=K(z)\,b^2(z)\,\xi(z)$.
Combining these expressions we find a quadratic expression for $b(z)$ for each redshift $z$, with a positive solution given by \citep[see e.g.][]{2017MNRAS.467.3993A}

\begin{figure}
\includegraphics[width=8.cm]{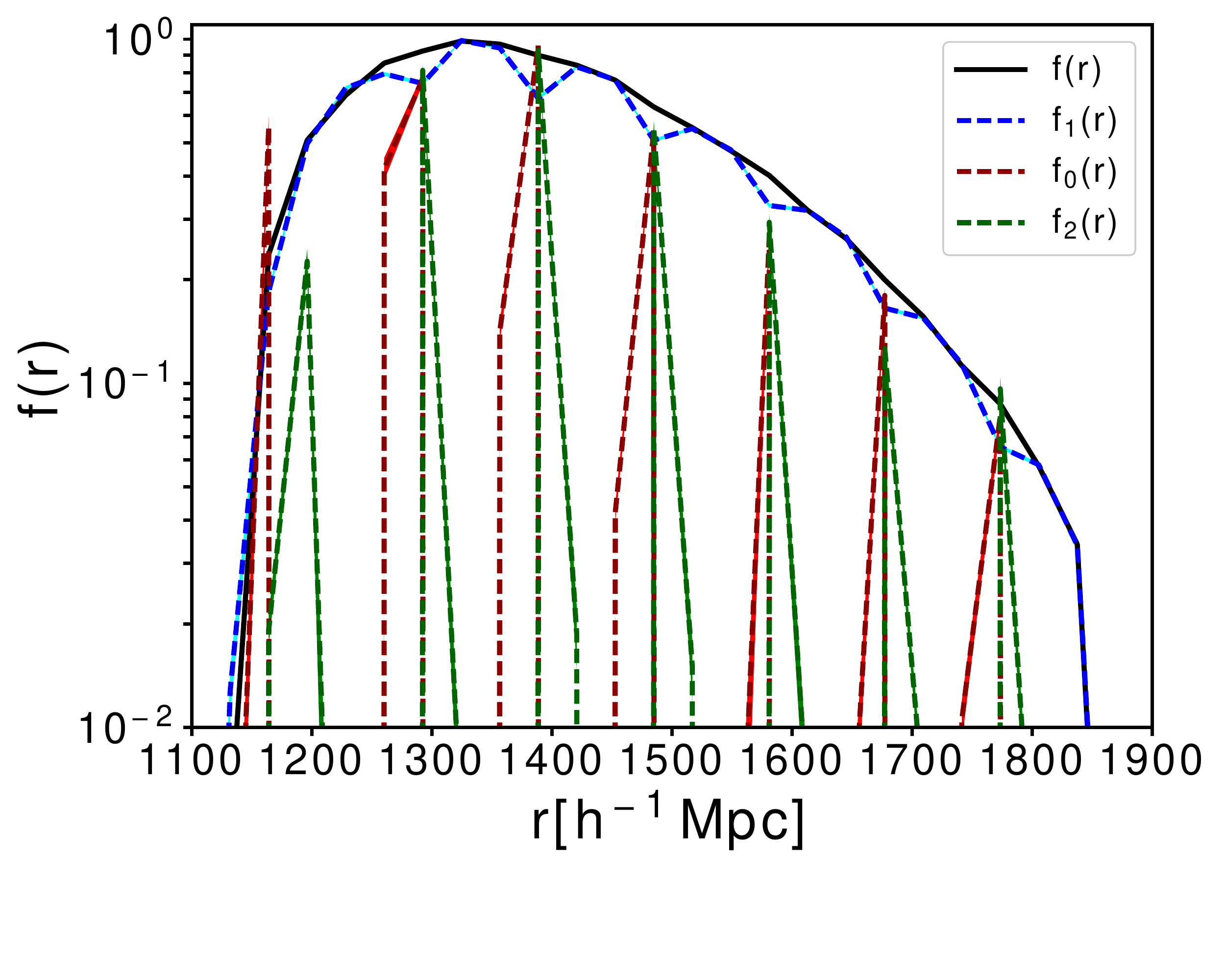}
\vspace{-1cm}
\caption{\label{fig:fsel} Radial selection function: original including all galaxies in Eulerian space (black solid line), after reconstruction in Lagrangian space corresponding to galaxies i) staying in their Eulerian redshift bin (subscript 1 and dashed blue line), ii) jumping to a lower redshift (subscript 0 and dashed red line), and iii) jumping to a higher redshift bin (subscript 2 and dashed green line). }
\end{figure}

%\be
%b^2(z)+\frac{2}{3}f_\Omega(z)b(z)+\frac{1}{5}f_\Omega^2(z)-(b^{s}(z))^2=0\,,
%\ee
\be
\label{eq:linearbias}
b(z)=-\frac{1}{3}f_{\Omega}(z)+\sqrt{-\frac{4}{45}f_\Omega(z)^2+(b^{s}(z))^2}\,.
\ee
In our case study using the light-cone mock galaxy catalog for CMASS galaxies we find a bias as a function of redshift as illustrated in Fig.~\ref{fig:bias}.

\subsubsection{Lagrangian large scale bias}
Once we have the Eulerian large-scale bias given by Eq.~(\ref{eq:linearbias}), we can translate it to any higher redshift.
%Let us briefly recap the derivation of this, as it is one key ingredient in the reconstruction process.
%According to the linear growth of perturbations the dark matter density at two different redshifts, an initial one $z_q$, and a final one $z$ is related through the following equation on large scales
%\be 
%\delta(z_q) = \frac{D(z_q)}{D(z)} \delta(z) \,, 
%\ee
%where we have transformed a field at low to high redshift.
%If we now assume that the dynamics of galaxies is governed by the dark matter field, we get following pair of continuity equations:
%\ba
%    \partial_t \delta(z) &=& -\nabla\cdot v(z)\,,\nonumber\\
%    \partial_t \delta_g(z) &=& -\nabla\cdot v(z)\,.
%\ea
%Considering finite differences and the linear growth of structures, the combination of above equations yield:
%\be
%   \delta_g(z_q)-\delta_g(z)=\delta(z_q)-\delta(z)=\left(\frac{D(z_q)}{D(z)}-1\right)\delta(z)\,,
%\ee
%and thus
%    \be
%\delta_g(z_q)=\delta_g(z)+\left(\frac{D(z_q)}{D(z)}-1\right) %\delta(z) \,.
%   \ee  
%   By introducing the notion of large scale linear bias (which %ensures Eq.~\ref{fig:Kfac}):
%\be
%\delta_g(z)=b(z) \delta(z)\,,
%\ee
%   we get
%   \be
%    \delta_g(z_q)=\left(b(z)+\frac{D(z_q)}{D(z)}-1\right) %\delta(z)\,.
%   \ee
%Since we are interested in a transformation of the bias from Eulerian to Lagrangian space, we write last equation as
%   \be
%    \delta_g(z_q)=\left(b(z)+\frac{D(z_q)}{D(z)}-1\right) \frac{D(z)}{D(z_q)} \delta(z_q)\,,
%   \ee
%    and hence 
To that aim, we assume that that the bias from Eq.~(\ref{eq:linearbias}) can be expressed as $\delta_{\rm  g}(z)=b(z) \delta(z)$. On top of this, given i) the conservation in the number of galaxies and, ii) that the galaxies follow the same velocity field as the underlying DMDF, one can demonstrate that the  large-scale bias can be written at any other redshift in terms of the linear growth factor $D(z)$ as \cite[see e.g. ][]{1994ApJ...421L...1N,1996ApJ...461L..65F,2008MNRAS.385L..78P}. 
\be
\label{eq:biasevo}
b(z_q) = \left(b(z)-1\right)\frac{D(z)}{D(z_q)}+1\,.
\ee
   We note that we do not need to include stochasticity in this relation, as introduced by \citet[][]{1998ApJ...500L..79T}, since we are not trying to model the different galaxy populations at different redshifts including galaxy formation, but the large scale bias evolution of a given galaxy population.
In fact this has been studied in detail in \citet{2019MNRAS.483.5267B}.

\subsubsection{Lagrangian non-linear bias}

Hitherto, we have only considered the bias at large scales (i.e, in the limit of $k\rightarrow0$).
If we aim at describing the DMDF on a mesh of Mpc scales resolution, we need to use a non-linear description of DMDF. In fact, a typical bias of $\sim 2$ (e.g. for LRGs) is translated through Eq.~(\ref{eq:biasevo}) to a bias of about 60 at $z=100$.
If our cell resolution is high enough of producing over-densities larger than $|\delta|>10^{-2}$, this implies that a linear model would yield negative densities, i.e. $\delta<-1$.

One of the simplest model we can assume is a power-law bias:
\be
\label{eq:lagbias}
\rho_{\rm  g}(\mbi q)=\gamma(z_q)(1+\delta(\mbi q))^{b(z_q)\,f_b(z_q)}\,,
\ee
where $\gamma(z_q)$ is a normaliziation constant and $f_b$ is a correction factor which ensures a correct large scale bias \cite[see e.g.][]{2017MNRAS.467.3993A}.
This model does not include threshold bias (see Eq.~(\ref{eq:biasthreshold})) inherent to the peak background-split model \citep[see e.g.][and references terein]{1984ApJ...284L...9K,Schmidt:2012ys} 
. This is consistent with the picture of the proto-halos associated to halos after cosmic evolution, which are not tracers of the peaks of the initial cosmic density field, but can be tracing the whole density regime \citep[see e.g.][]{2011MNRAS.413.1961L}. 
In our framework, represented in Fig.~\ref{fig:biasmodel}, galaxies tracing halos are mapped to Lagrangian space tracing the proto-halos in the entire density field. In a natural way the resulting proto-halo regions are not spherical symmetric already effectively ensuring a non-local mapping in Eulerian space  \citep[see e.g.][]{2013PhRvD..87h3002S}. The framework presented here allows to be extended to account for complex Lagrangian bias components, if that would be required.
We note however, that recent works do not find important non-local bias contributions in Lagrangian space, except for very massive haloes \citep{Modi2017,2018JCAP...07..029A}. This implies that as long as the Lagrangian tracers used to reconstruct the density field are not very massive, we can neglect additional non-local bias terms (see Fig.~\ref{fig:biasmodel}).

It is important to note that the normalisation in Eq.~(\ref{eq:lagbias}) depends on the non-linear bias model, and is only equal to the galaxy number density $\bar{N}$ for bias unity:
\be\label{eq:lagbias2}
\gamma(z_q)=\frac{\bar{N}}{\langle(1+\delta(\mbi q))^{b(z_q)\,f_b(z_q)}\rangle}\,.
\ee
The problem associated to the model represented by Eqs.~(\ref{eq:lagbias}) and (\ref{eq:lagbias2}) is its dependency on the mesh resolution on which $\rho_{\rm  g}(\mbi q)$ and $\delta(\mbi q)$ are defined, via the factor $f_b$, and thereby  on input parameters used for the representation of the data in our code. To circumvent this situation, we obtain a connection between the power-law bias of Eq.~(\ref{eq:lagbias2}) (specifically, the parameter $f_{b}$) with the large-scale bias as predicted by the renormalised perturbation theory (RPT) \citep[see e.g.][]{2009JCAP...08..020M, 2018PhR...733....1D}. We present the derivation in appendix \ref{app:rpt}.
Given the lack of a solid analytical framework which predicts the nonlinear bias, we propose here to derive it numerically. This ansatz is inspired by RPT (see appendix \ref{app:rpt}), which encodes the non-linear dependence on the resolution in the variance of the field. We can define an effective power law bias by $b_{\rm eff}(z)=b(z)\,f_b(z)$.
The large scale bias can be obtained from the ratio of the galaxy to dark matter over-density variances:
\be 
\label{eq:biaslarge}
b(z)\equiv \sqrt{\frac{\sigma_{Kg}^2(z)}{\sigma_K^2(z)}}\,,
\ee
with the variances given by
\be
\sigma^2_K(z)=\langle (K\circ \delta(\mbi q, z))^2\rangle\,,
\ee
and
\be
\sigma^2_{Kg}(z)=\langle (K\circ \delta_{\rm  g}(\mbi q, z)[b_{\rm eff}])^2\rangle\,,
\ee
for the dark matter and the galaxy field, respectively, and $K$ being a Gaussian kernel with a smoothing scale of 50-100 $h^{-1}$ Mpc \footnote{After verifying that the results do not change for our volume using different smoothing scales, we chose a scale of 50 for our numerical tests.} The  model for the galaxy over-density is accordingly written as 
\be
\label{eq:biasit}
\delta_{\rm  g}(\mbi q, z)[b_{\rm eff}] \equiv  \bar{N}(z)\,\frac{(1+\delta(\mbi q, z))^{b_{\rm eff}}}{\langle(1+\delta(\mbi q, z))^{b_{\rm eff}}\rangle} -1\,.
\ee
Now we have all the ingredients to obtain the non-linear bias correction factor $f_b(z)$, which ensures that the large scale bias (Eq.~\ref{eq:biaslarge}) is recovered for large smoothing radii from Eq.~(\ref{eq:biasit}). We do this iteratively using a Newton-Raphson method \footnote{We choose in our calculations an accuracy of eps=$10^{-5}$, which typically is achieved after 3-5 iterations for each population of galaxies and each redshift bin}.
%
%:
%\begin{verbatim}
%
%b_eff=b
%g_b=1
%conv=false
%
%while conv==false
%    b_eff*=g_b
%    sigma_Kg^2=mean[(K o delta_g[b_eff])^2]
%    g_b=sqrt(sigma_K^2*b^2/sigma_Kg^2)
%    if (g_b-1)<eps
%        conv=true
%
%f_b=b_eff/b
%\end{verbatim}

\subsubsection{Bias mixing between redshift bins}
\label{sec:biasmix}

So far we have found a Lagrangian bias description based on some clustering measurements in Eulerian redshift space.
Since those are naturally done in redshift bins, we end up having the Lagrangian bias defined on shells in redshift distance. 
This assumes that the Eulerian to Lagrangian mapping of tracers keeps spherical shells, however those are distorted in the same (reverse) way as baryon acoustic oscillation spheres are distorted through cosmic evolution. 
If the bias is interpolated to obtain a smooth varying function in redshift, then the changes in redshift from $z_s$ to $z_q$ according to Fig.~\ref{fig:lagtoeul} are not large and it can be assumed that the Lagrangian bias of a galaxy is the same evaluated at both distances. If one decides to keep a binned bias, then a galaxy in one bin might jump to a higher or lower redshift bin as shown in Fig.~\ref{fig:lightconeper}. In such scenario one would need to associate to galaxies jumping to a lower (higher) redshift bin the bias from their original higher (lower) one. This has been actually implemented in \texttt{COSMIC  BIRTH}  and we show results for both cases below.

\begin{figure*}
 \begin{tabular}{ccc}
 \hspace{2.cm}
   \includegraphics[width=7.cm]{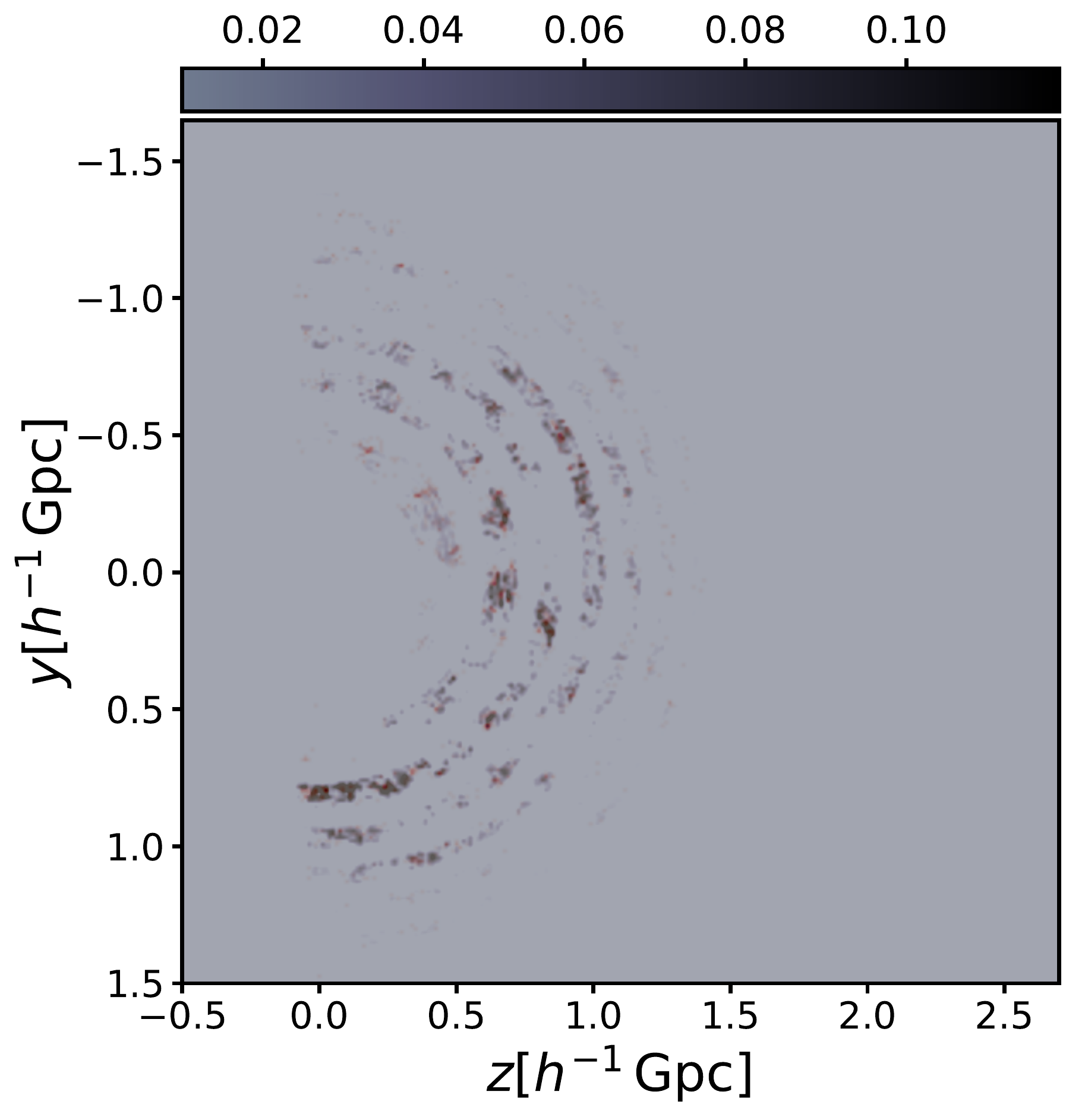}
 \hspace{-13.1cm}
   \includegraphics[width=7.cm]{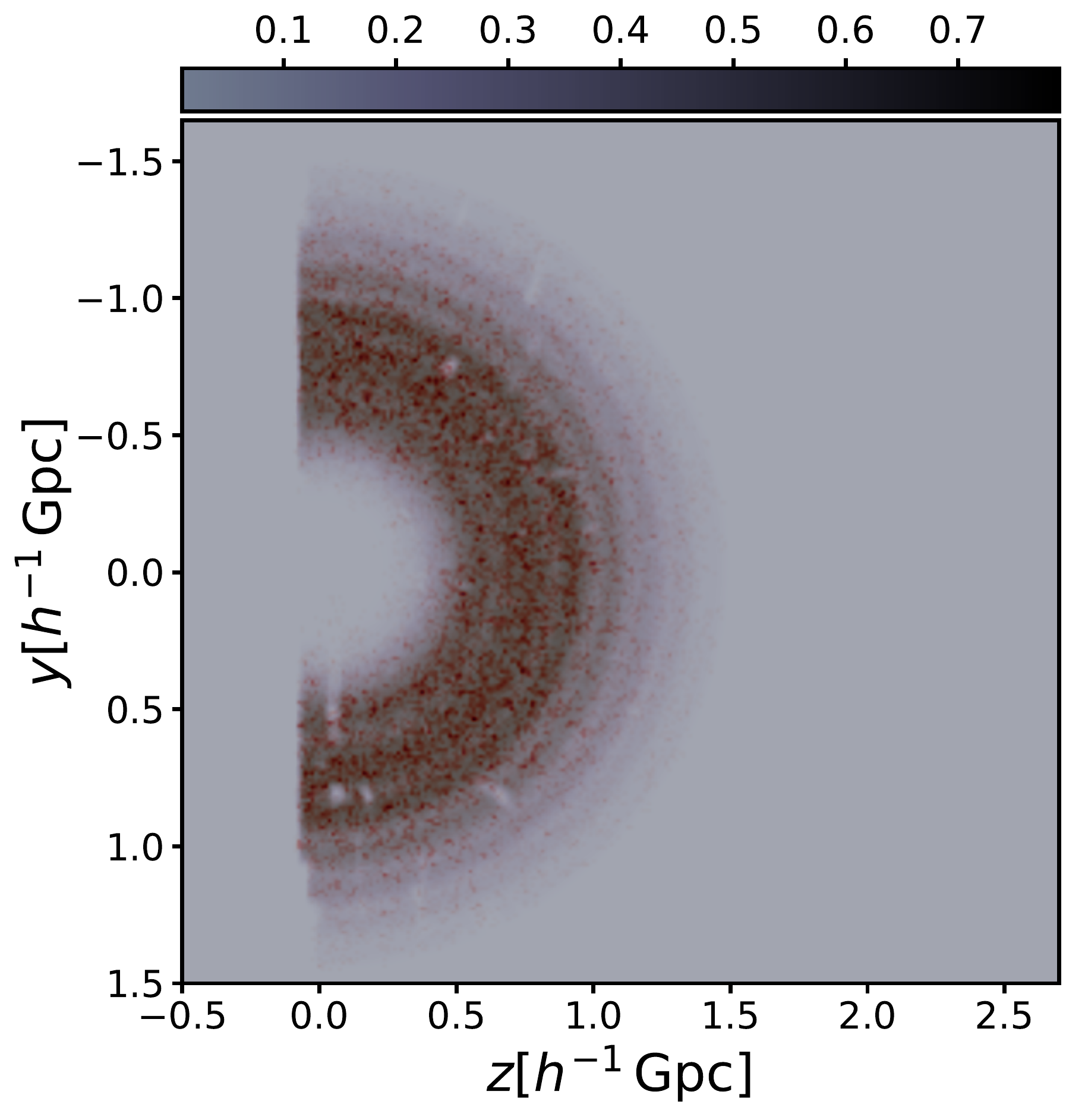}
 \hspace{-13.1cm}
  \includegraphics[width=7.cm]{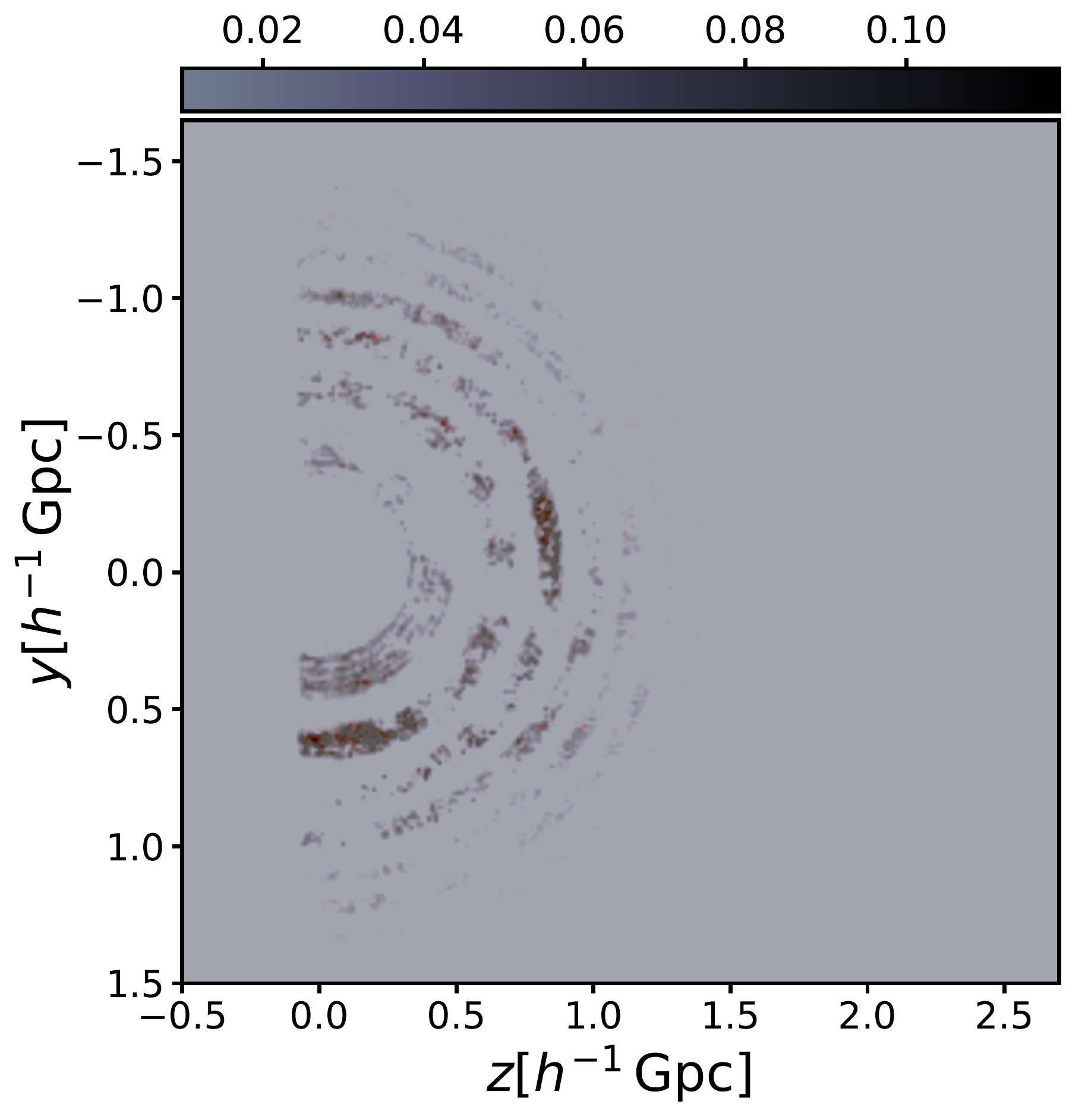}
  \end{tabular}
 \caption{\label{fig:winwg} For slices of thickness $\sim$60 $h^{-1}$ Mpc in the $z-y$  plane of  the three dimensional cubical mesh of side 3200  $h^{-1}$ Mpc and $256^3$ cells: completeness for the galaxies which jump to a lower redshift bin (left), for those which stay at the same redshift bin (middle), and for those which jump to an upper redshift bin. Galaxies are over-plotted as red dots.}
\end{figure*}

\begin{figure*}
 \begin{tabular}{ccc}
       \hspace{2.cm}
   \includegraphics[width=6.8cm]{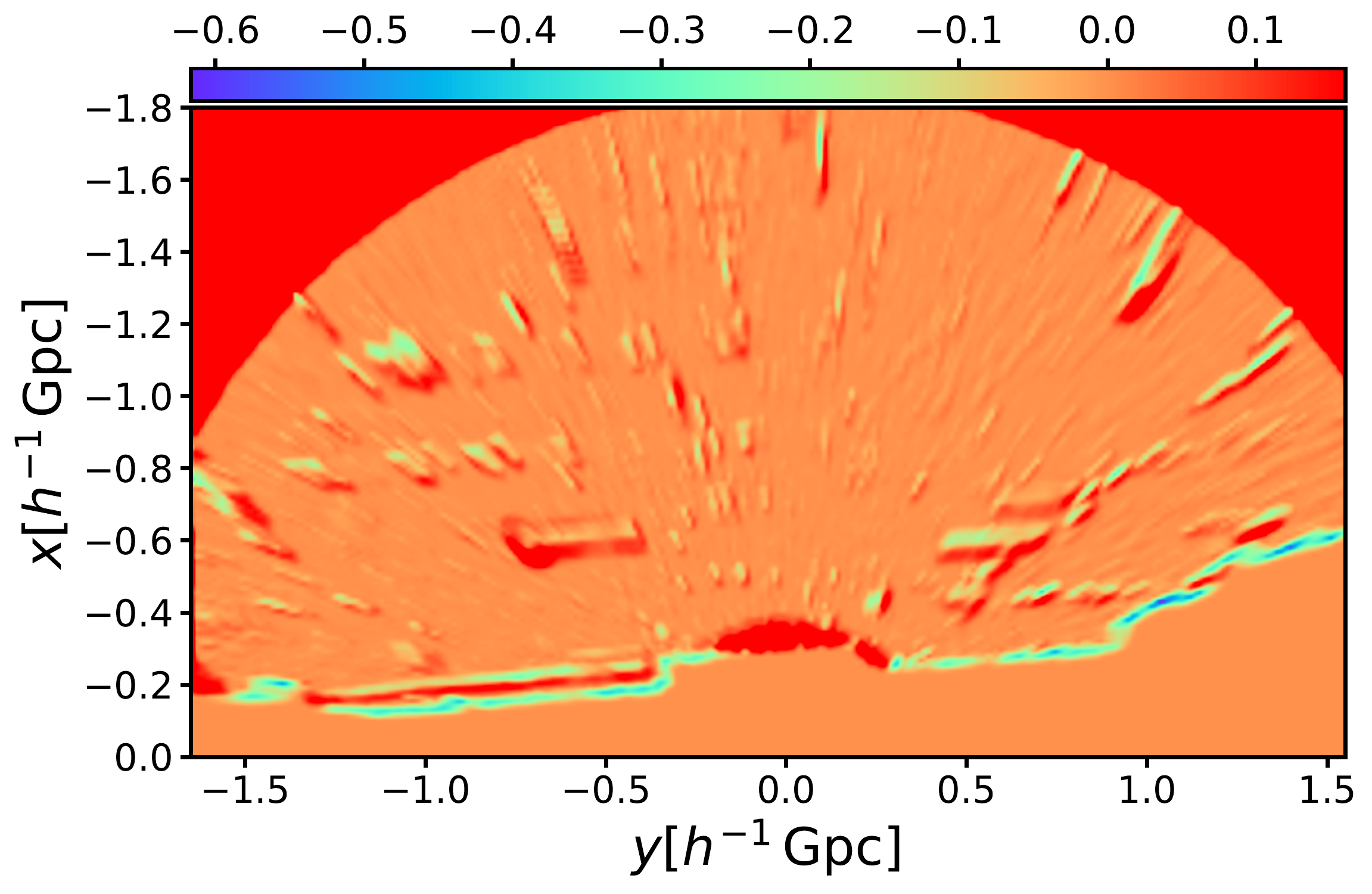}
 \hspace{-12.9cm}
   \includegraphics[width=6.8cm]{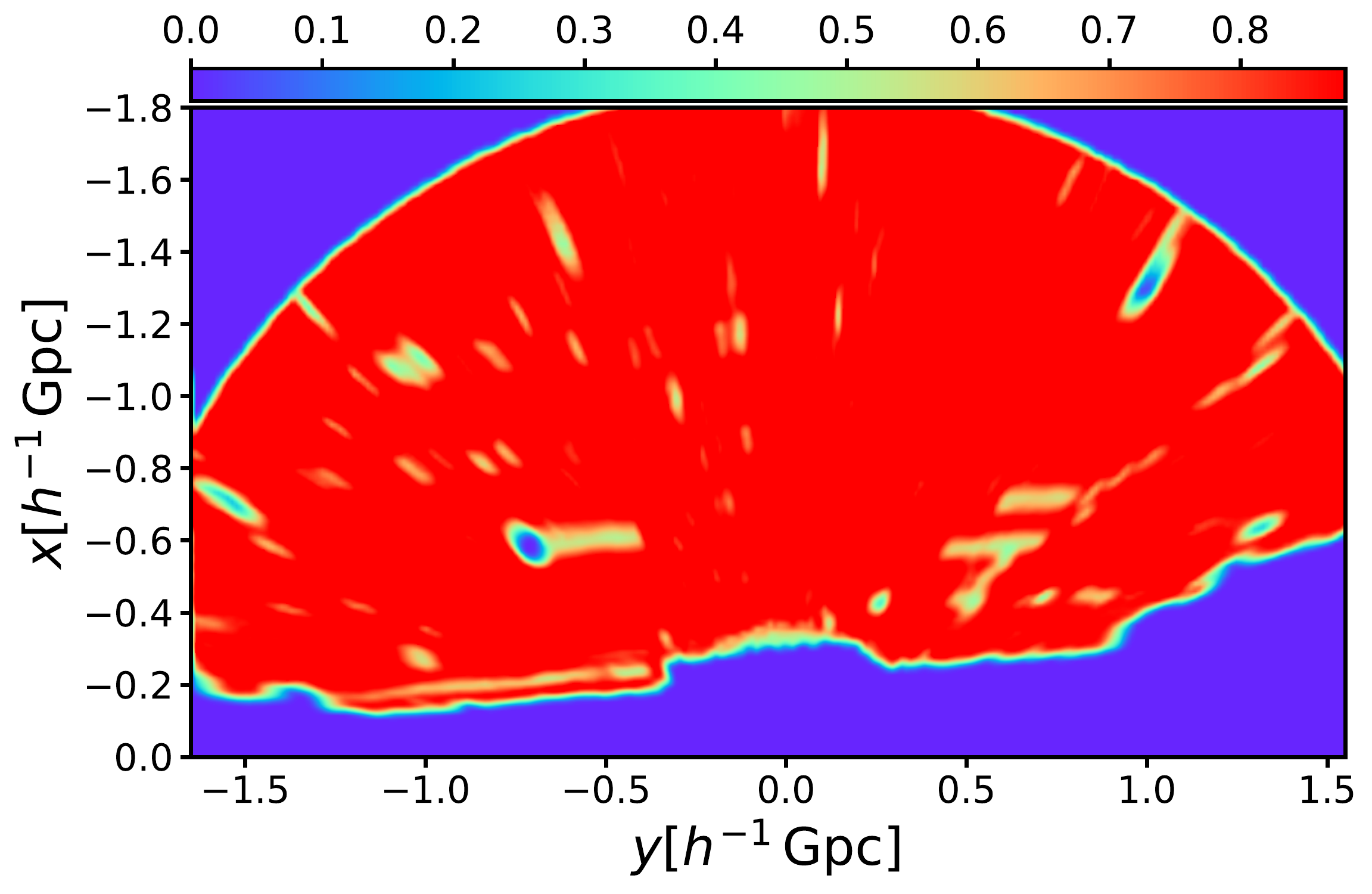}
 \hspace{-12.9cm}
  \includegraphics[width=6.8cm]{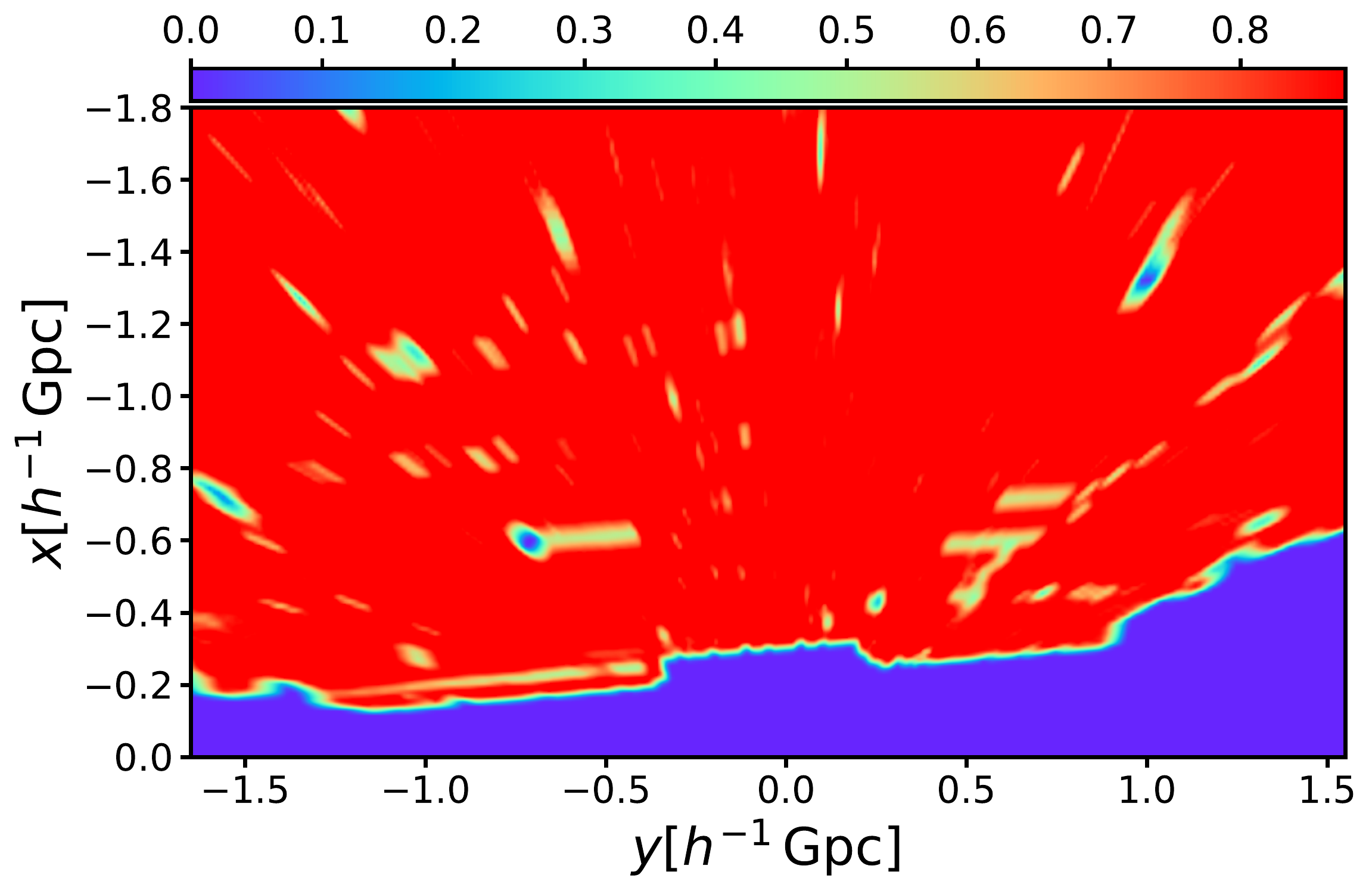}
   \end{tabular}
 \caption{\label{fig:winback} For the same cut as in Fig.~\ref{fig:winwg} but in the $x-y$ plane: 3D projected angular survey geometry, including veto mask in Eulerian space (left), Lagrangian space (middle), and the difference between both (right).}
\end{figure*}

\begin{figure*}
 \begin{tabular}{cc}
   \includegraphics[width=10.5cm]{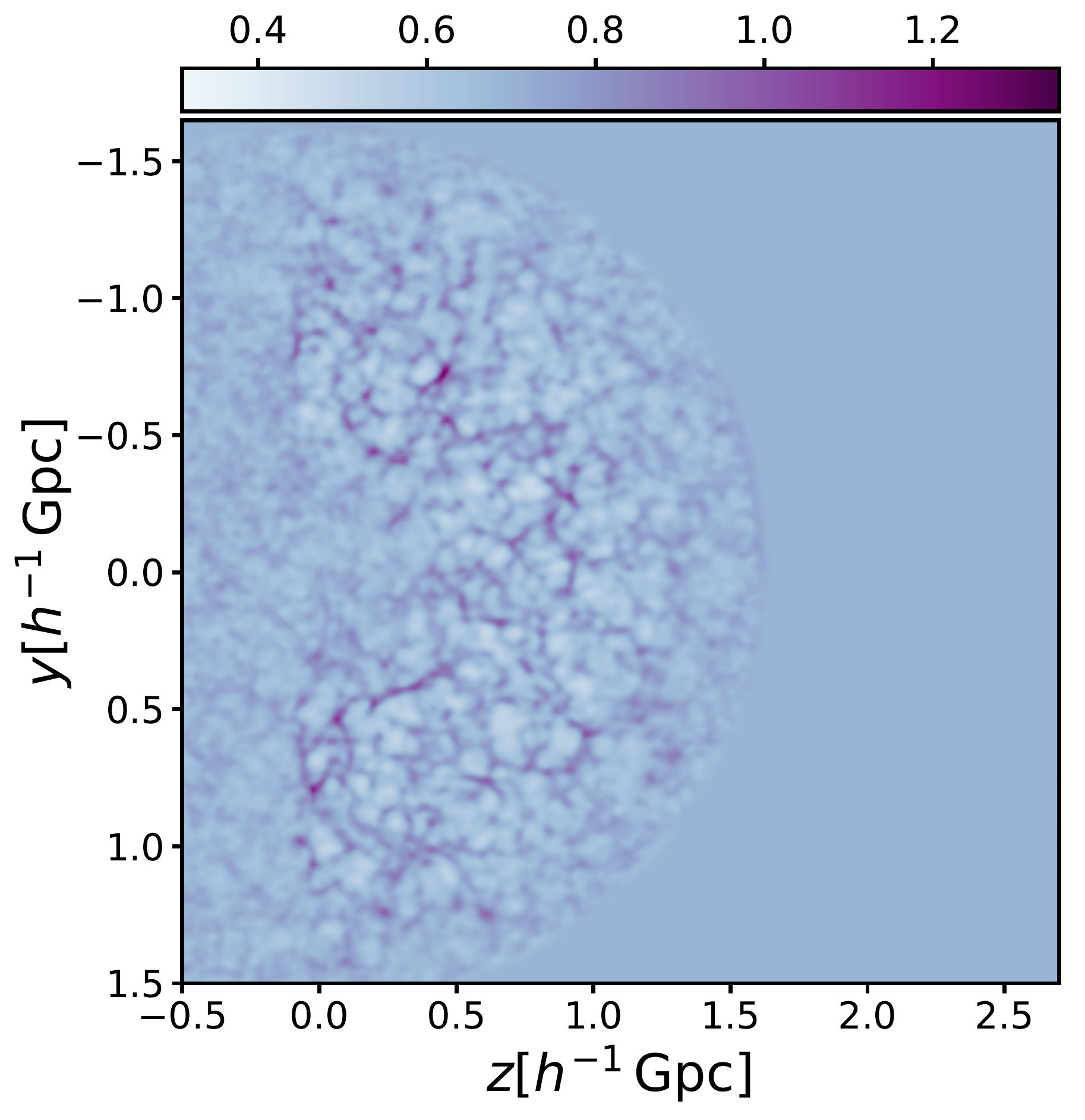}
      \hspace{-19.57cm}
   \includegraphics[width=10.5cm]{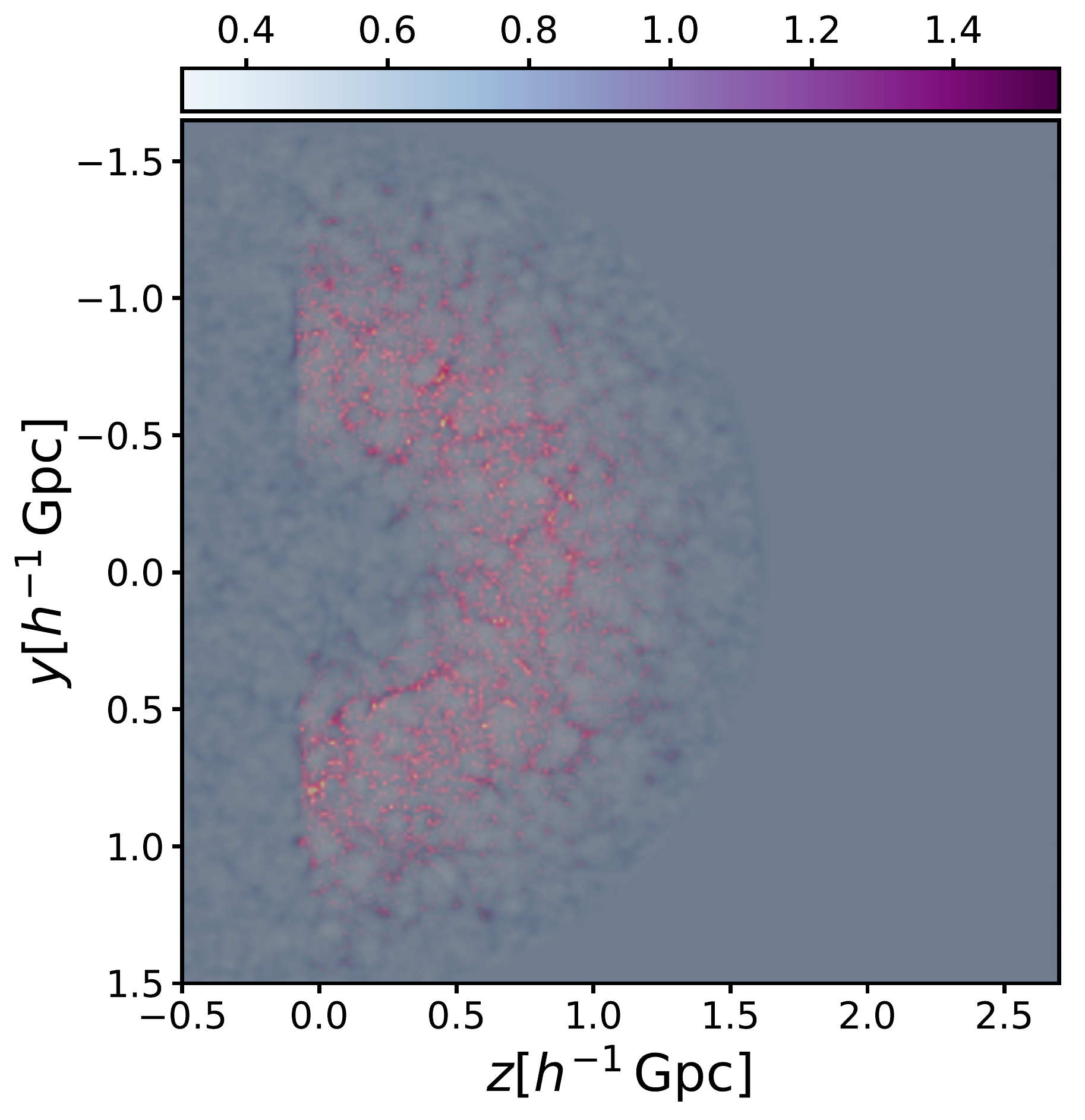}
 %  \vspace{-1.03cm}
\\
   \includegraphics[width=10.5cm]{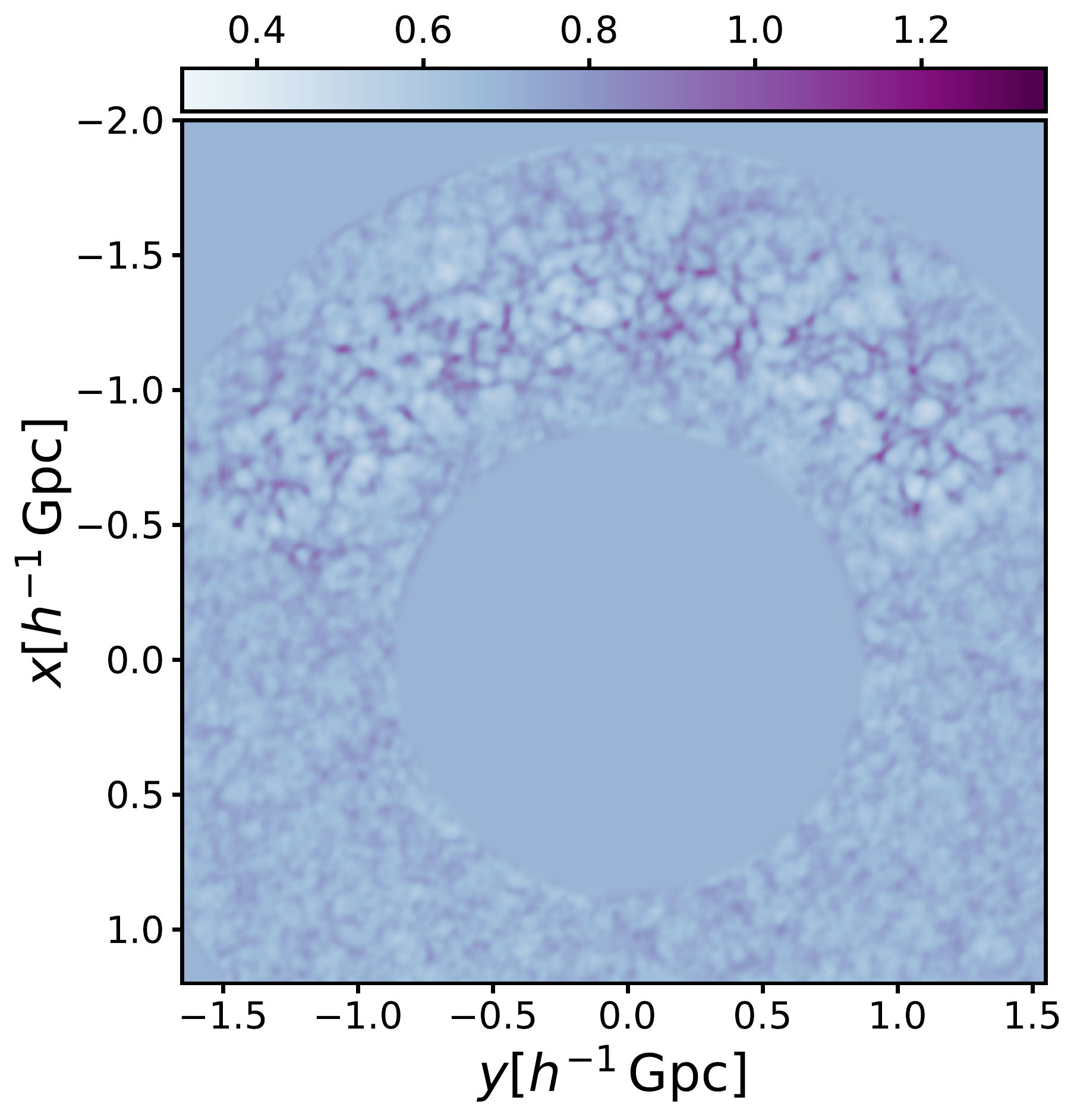}
       \hspace{-19.57cm}
   \includegraphics[width=10.5cm]{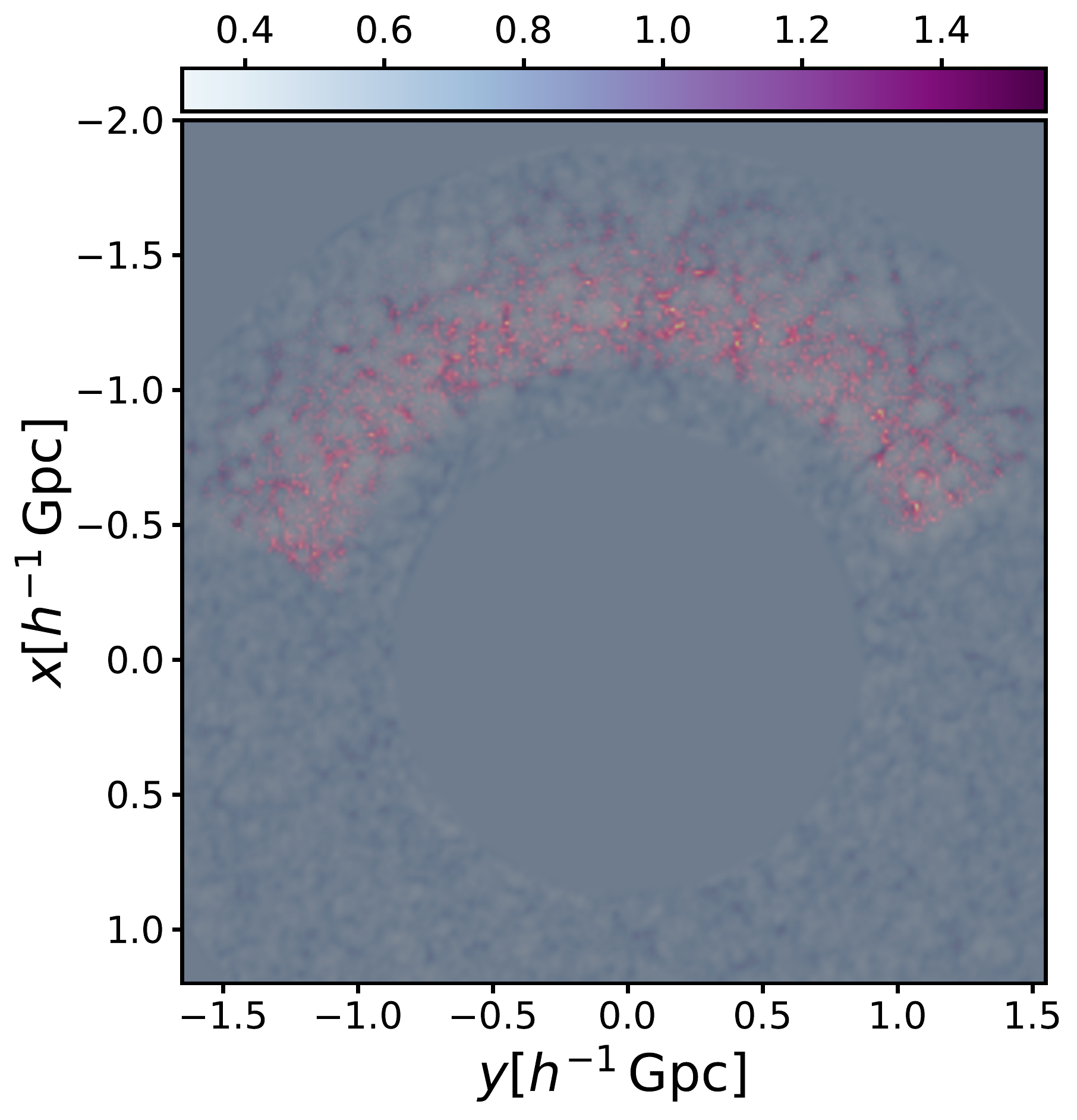}
 \end{tabular}
 \caption{\label{fig:birthzeld256}
 Slices of thickness $\sim$60 $h^{-1}$ Mpc in the $z-y$ (top) and $x-y$ (bottom) plane of  the three dimensional cubical mesh of side 3200  $h^{-1}$ Mpc and $256^3$ cells showing the averaged density field over 20 independent samples. Left panels with (right panels w/o) galaxies over-plotted. }
\end{figure*}

\section{Verification of the \texttt{COSMIC BIRTH} code}
\label{sec:application}

In this section we will describe the data used to verify the \texttt{COSMIC BIRTH} and present and discuss the results after running the \texttt{COSMIC BIRTH} code on them.

\subsection{Data used in this work}
\label{sec:mocks}

To validate the reconstruction method presented in this paper, we the Data Release DR12 of the Baryon Oscillation Spectroscopic Survey (BOSS) \citep[][]{2013AJ....145...10D}. 
The BOSS survey uses the SDSS 2.5 meter telescope at Apache Point Observatory \citep[][]{2006AJ....131.2332G} and the spectra are obtained using the double-armed BOSS spectrograph \citep[][]{2013AJ....146...32S}. The data are then reduced using the algorithms described in \citep[][]{2012AJ....144..144B}.  The target selection of the CMASS and LOWZ samples, together with the algorithms used to create large-scale structure catalogues (the \textsc{mksample} code), are presented in  \citet[][]{2016MNRAS.455.1553R}.

We restrict this analysis to the CMASS sample of luminous red galaxies (LRGs), which is a complete sample, nearly constant in mass and volume, limited between the redshifts $0.43\le z \le 0.7$ (see \citet{Andersonboss} for details of the targeting strategy). We use the $N$-body based mock galaxy catalog constructed to match the clustering bias, survey mask, selection functions, and number densities of the BOSS DR12 CMASS galaxies on the light-cone.

%This have been constructed based on the Sub-Halo Abundance Matching (SHAM) technique to connect haloes to galaxies \citep{Kravtsov04,Neyrinck04,Tasitsiomi04,Vale04,Conroy06,Kim08,Guo10,Wetzel10,Trujillo11,Nuza13}.

%At first order SHAM assumes a one-to-one correspondence between the
%luminosity or stellar and dynamical masses:
% galaxies with more stars are assigned
%to more massive haloes or subhaloes.
%The luminosity in a red-band is sometimes used instead of stellar mass.
% There should be some degree of stochasticity in the relation between stellar and dynamical
%masses due to deviations in the merger history, angular momentum,
%halo concentration, and even observational errors \citep{Tasitsiomi04,Behroozi10,Trujillo11,Leauthaud11}.
% Therefore, a scatter id included in that relation necessary to accurately  fit the clustering of the BOSS data \citep[][]{sergio15}. \aba{Is this discussion relevant for BIRTH?}

The mock galaxy catalog used in this study was presented in \citet[][]{2016MNRAS.460.1173R} and was  extracted from the BigMDPL N-body simulation\footnote{https://www.cosmosim.org/cms/simulations/bigmdpl/}, one of the Multidark simulation project, which was performed using the \texttt{GADGET-2} code \citep{2005MNRAS.364.1105S}. The BigMDPL was run with $3,840^3$ particles on a volume of $(2.5\,h^{-1}{\rm Gpc}$ $)^3$ assuming $\Lambda$CDM Planck cosmology with \{$\Omega_\Lambda =0.6928, \Omega_{\rm M}=0.307, \Omega_{b}=0.0482,\sigma_8=0.828,n_s=0.961$\}, and a Hubble constant ($H_0=100\,h\,\kmsmpc$) given by  $h=0.677$.  Halos and subhalos were identified  using the \texttt{ROCKSTAR} halo finder \citep[][]{2013ApJ...762..109B}. The DMDF on the light-cone has been constructed with the redshift snapshots between $z=0.43$ and $z=0.7$ using the stored data from the BigMDPL simulation, i.e. $0.5\%$ of the particles. As a further preparation of the data, we computed the response function following the description in \S \ref{sec:response}. In particular, the angular mask was calculated using the \texttt{MANGLE} software package \citep[][]{2004MNRAS.349..115H,2008MNRAS.387.1391S}.  For the time being we will assume the power spectrum to be known with the exact cosmology (i.e, that used to construct the mock galaxy catalog). We note that the large-scale bias on redshift bins comes as an input computed as shown in Fig.~\ref{fig:biasmodel}. 

\subsection{Results}
\label{sec:results}

We consider in our analysis cubical volumes of $L=3200$ $h^{-1}$ Mpc side length with $256^3$ cells, i.e., a cell resolution of 12.5 $h^{-1}$ Mpc\footnote{For visualisation purposes we also show results with resolutions of 6.25 $h^{-1}$ Mpc, which will be analysed in detail in forthcoming publications}.
This setting is identical to the one in \citet[][and we refer the reader to this work for further details]{2017MNRAS.467.3993A}.
We have performed a series of runs with a variation of settings. 

From now on the reference calculation is dubbed \emph{run} {\bf A}. This run includes 20 redshift snapshots in the range $0.35<z<0.8$, including {\color{black} Lagrangian tetrahedral tessellation} in the angular response function transformation to Lagrangian coordinates (see \S \ref{sec:response}), and bias interpolation (see \S \ref{sec:biasmix}).
We note that we need {\color{black}an enlarged} redshift range to ensure that we have enough redshift bins after the Eulerian to Lagrangian mapping summarised in Figs.~\ref{fig:flowchart} and \ref{fig:phasespace}. 
We have verified in an additional run {\bf A$_{\Delta}$} using 10 redshift bins for the same redshift range  that we get the same results as considering bias interpolation. In such a case we have ensured that the redshift bins were wide enough to have only three populations of tracers, i.e., galaxies, which have not changed redshift bin after doing the  Eulerian redshift-space  to Lagrangian real-space  mapping, and galaxies coming from  higher and from  lower redshift bins (see Fig.~\ref{fig:lightconeper}). This can be seen in Fig.~\ref{fig:fsel}, where the corresponding populations of galaxies are depicted in different colors after 70 Gibbs-sampling iterations. We can find that the angular response function also has to take into account the different populations according to the displacement field as shown in Fig.~\ref{fig:winwg}. The corresponding large scale tracers are over-plotted as red dots. It is interesting to make the visual inspection and verify that the galaxies are on top of the non vanishing completeness regions, which predict where those galaxies are actually expected to be mapped to according to the same set of displacement fields on the light-cone  for a given iteration (see mapping procedure described in \S \ref{sec:LtoE} and represented in Figs.~\ref{fig:flowchart} and \ref{fig:phasespace}).
The combination of different tracers is done assuming that they are independent tracers of the large scale structure (without mixed terms), according to the Poisson likelihood as described in the appendix of \citet{2015MNRAS.446.4250A}.
This framework permits to add as many tracers of the large-scale structure as additive terms in the log-likelihood used in the posterior PDF within the Hamiltonian sampler. Each of these tracers will have its own bias and response function. We will show how to add different galaxy surveys using this formalism in a forthcoming paper applied to the COSMOS field \citep{cosmos2015} (Ata et al., in prep.) and to the Local Universe (Kitaura et al. in prep.).

\begin{figure*}
\begin{tabular}{cc}
\includegraphics[width=10.5cm]{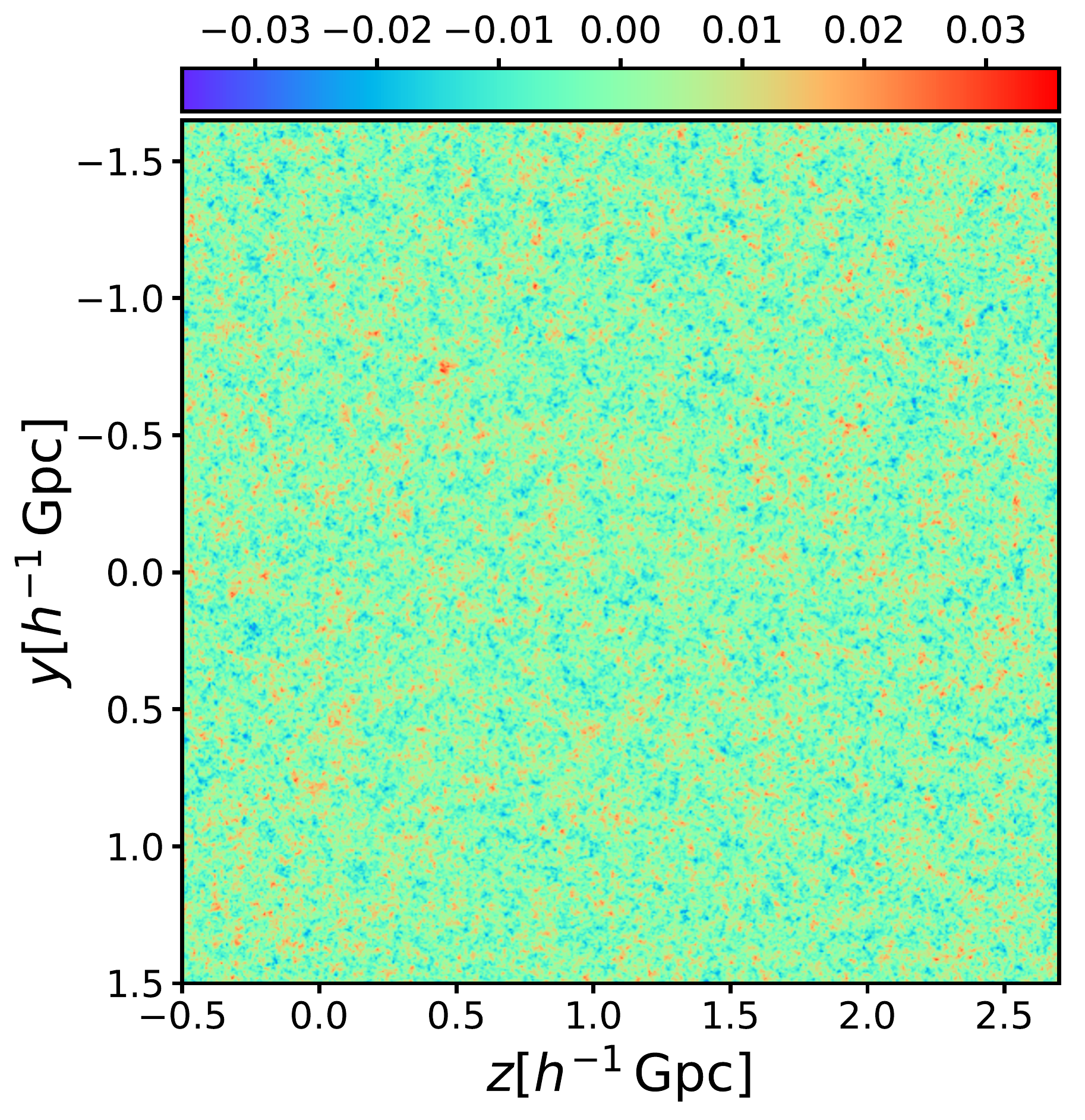}
\hspace{-19.55cm}
\includegraphics[width=10.5cm]{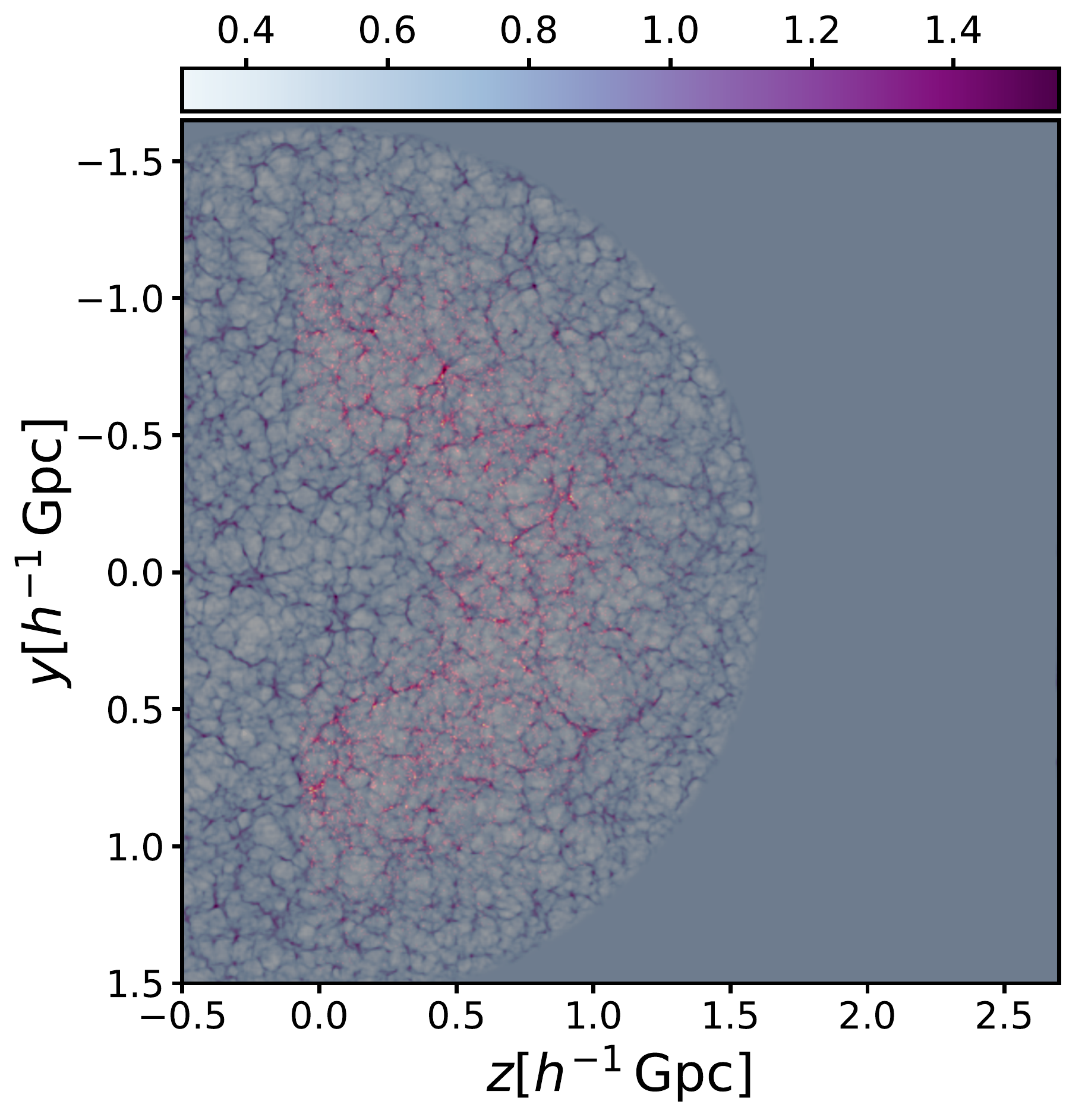}
\\
\includegraphics[width=10.5cm]{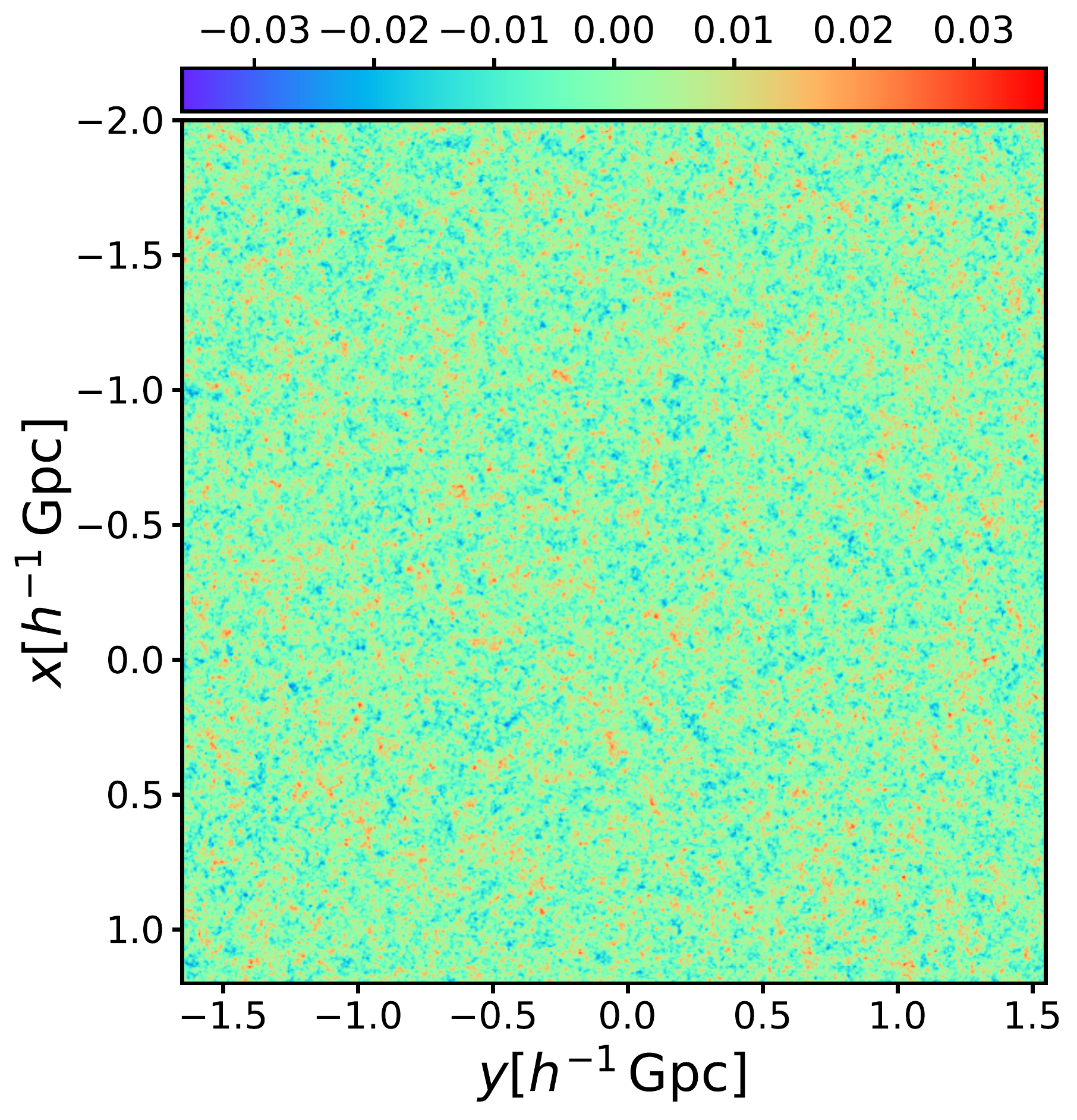}
\hspace{-19.55cm}
\includegraphics[width=10.5cm]{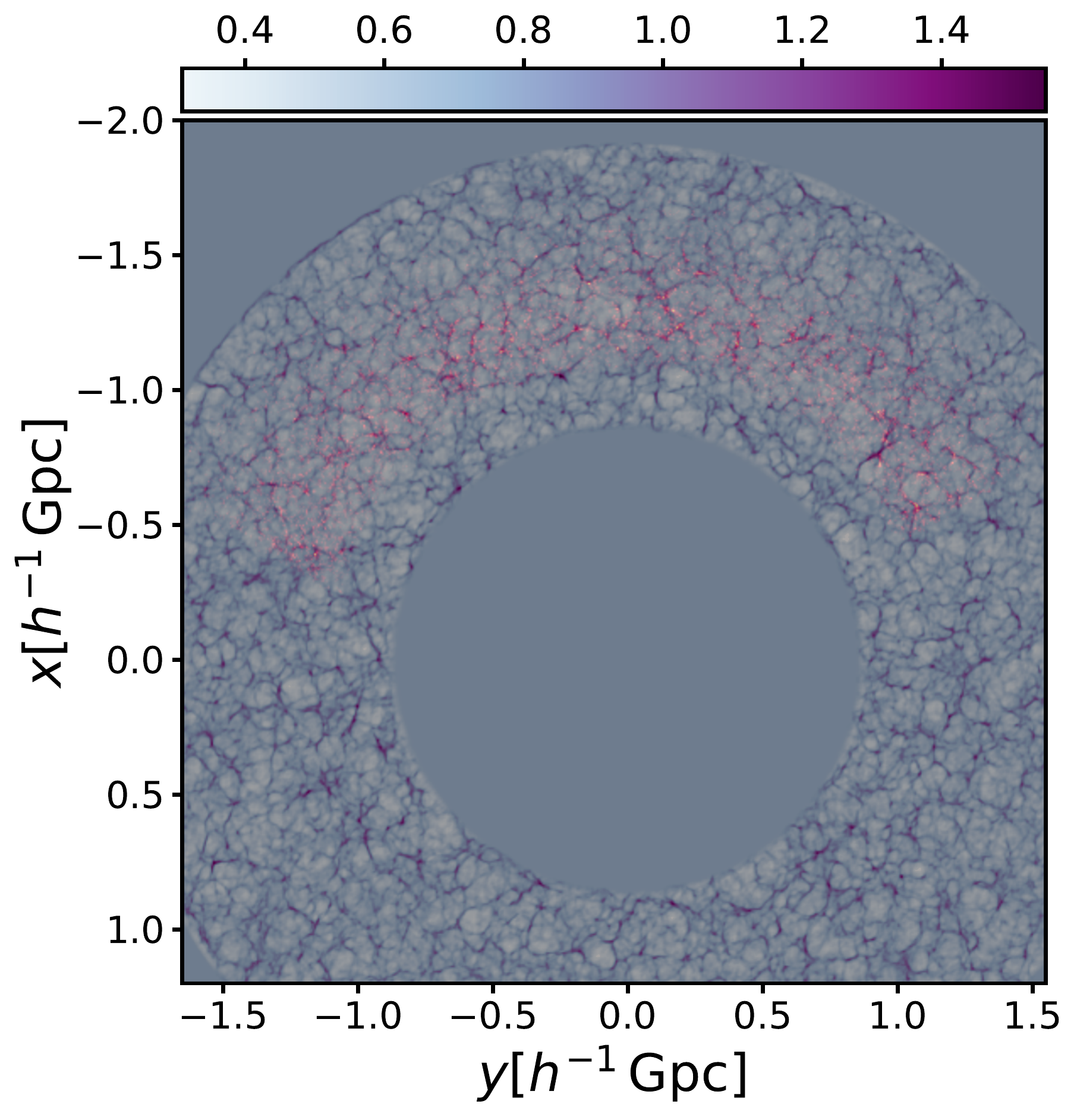}
\end{tabular}
\caption{\label{fig:birthzeld512} Left panels: same as Fig.~\ref{fig:birthzeld256} but for single reconstructions after 70 Gibbs-sampling iterations on meshes with $512^3$ cells. Right panels: corresponding initial density fields at z=100 are shown.}
\end{figure*}

\begin{figure*}
  \vspace{-.4cm}
 \begin{tabular}{cc}
    \hspace{0.4cm}
   \includegraphics[width=10.15cm]{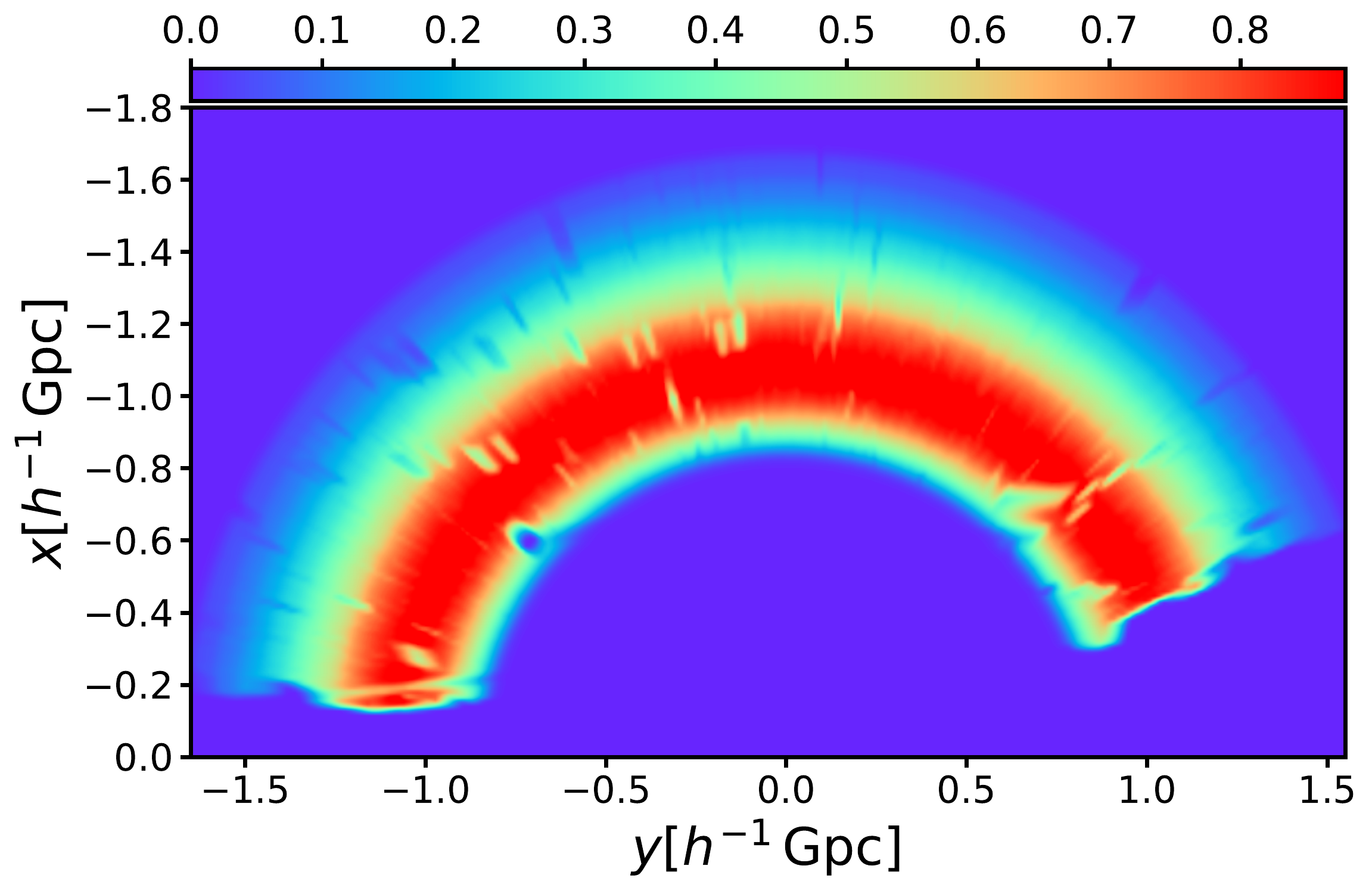}
      \hspace{-19.15cm}
   \includegraphics[width=10.15cm]{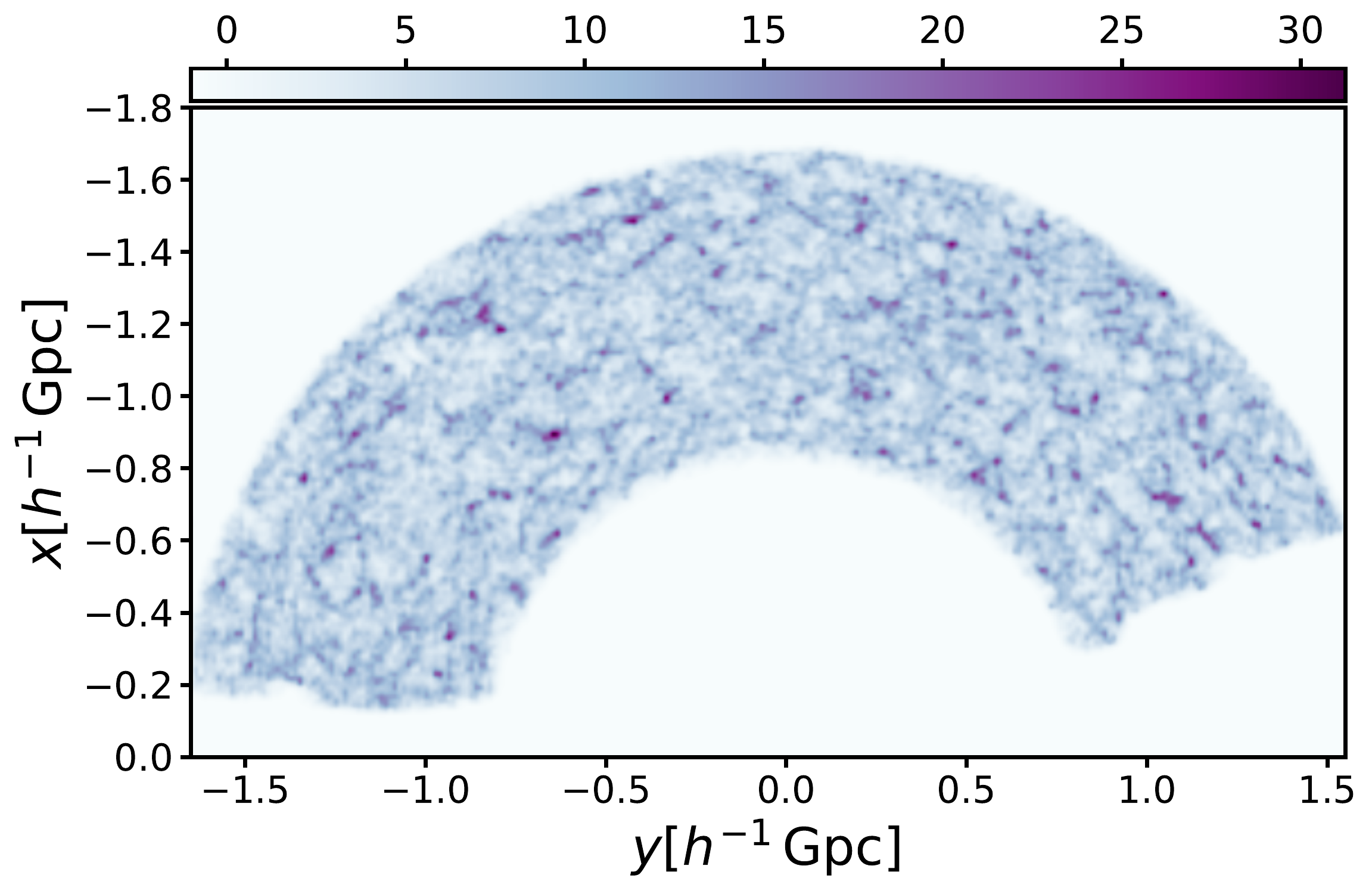}
   \vspace{-1.05cm}
\\
    \hspace{0.4cm}
   \includegraphics[width=10.15cm]{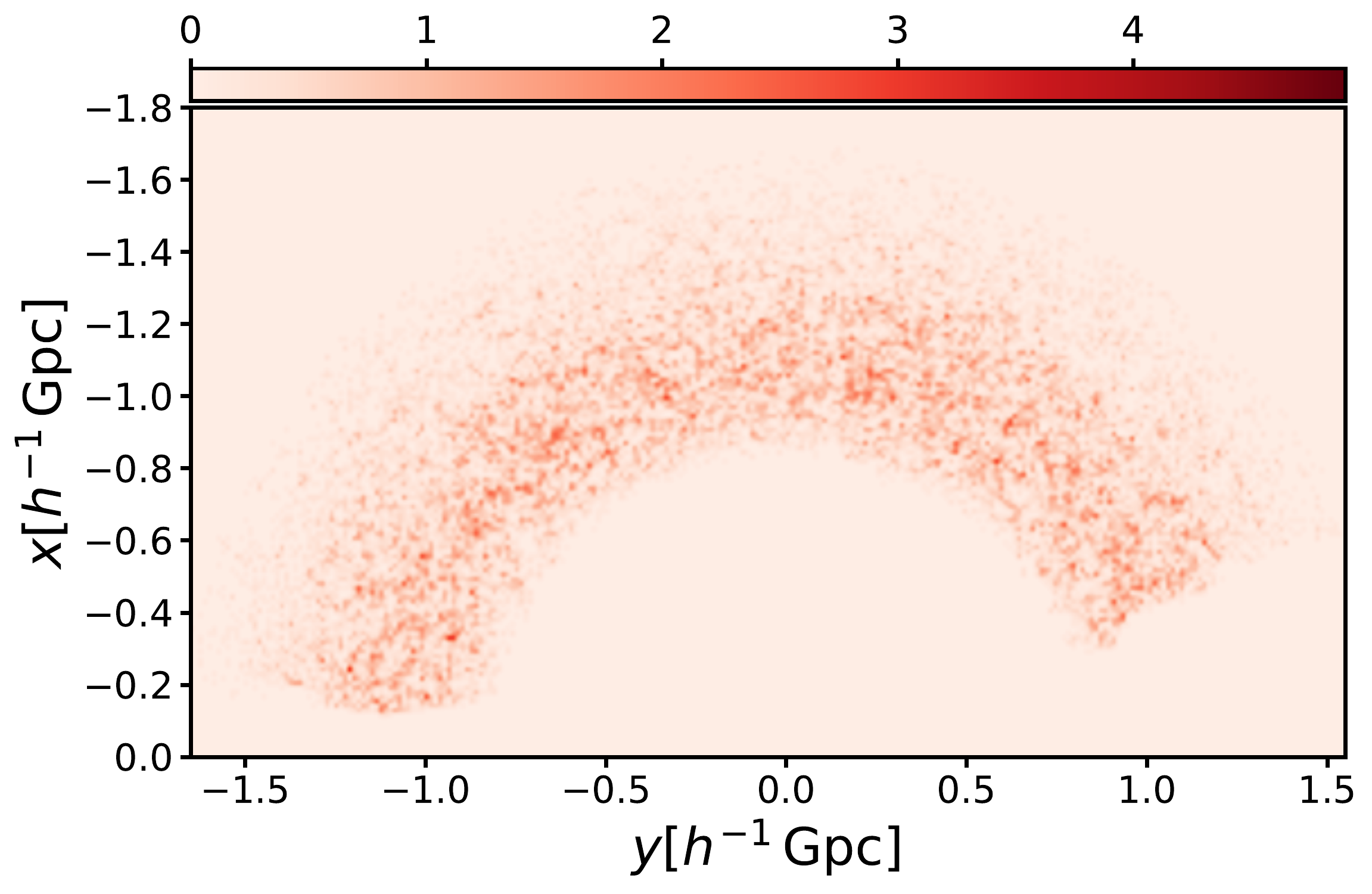}
       \hspace{-19.15cm}
   \includegraphics[width=10.15cm]{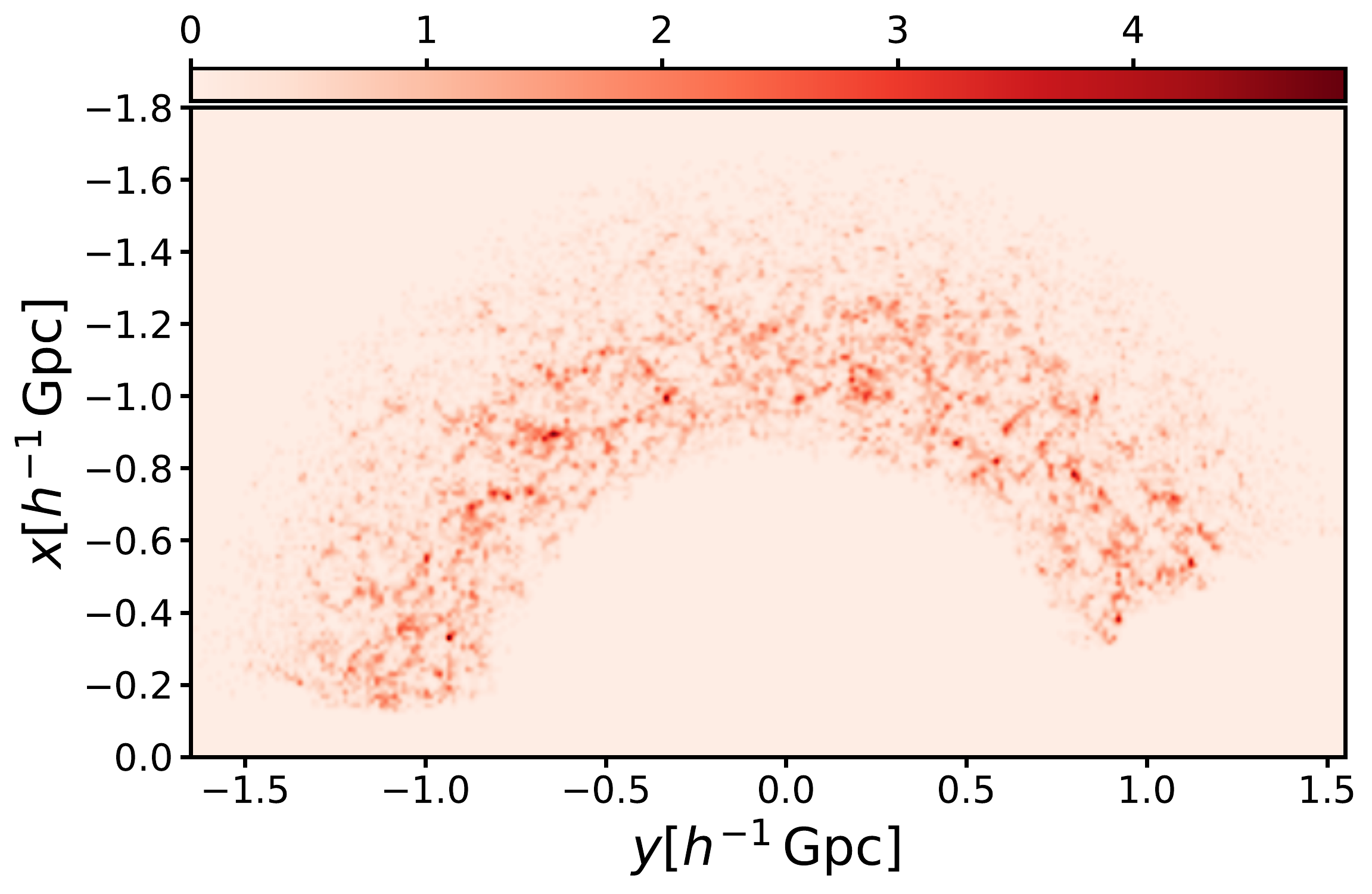}
    \vspace{-1.05cm}
    \\
  \hspace{0.4cm}
   \includegraphics[width=10.15cm]{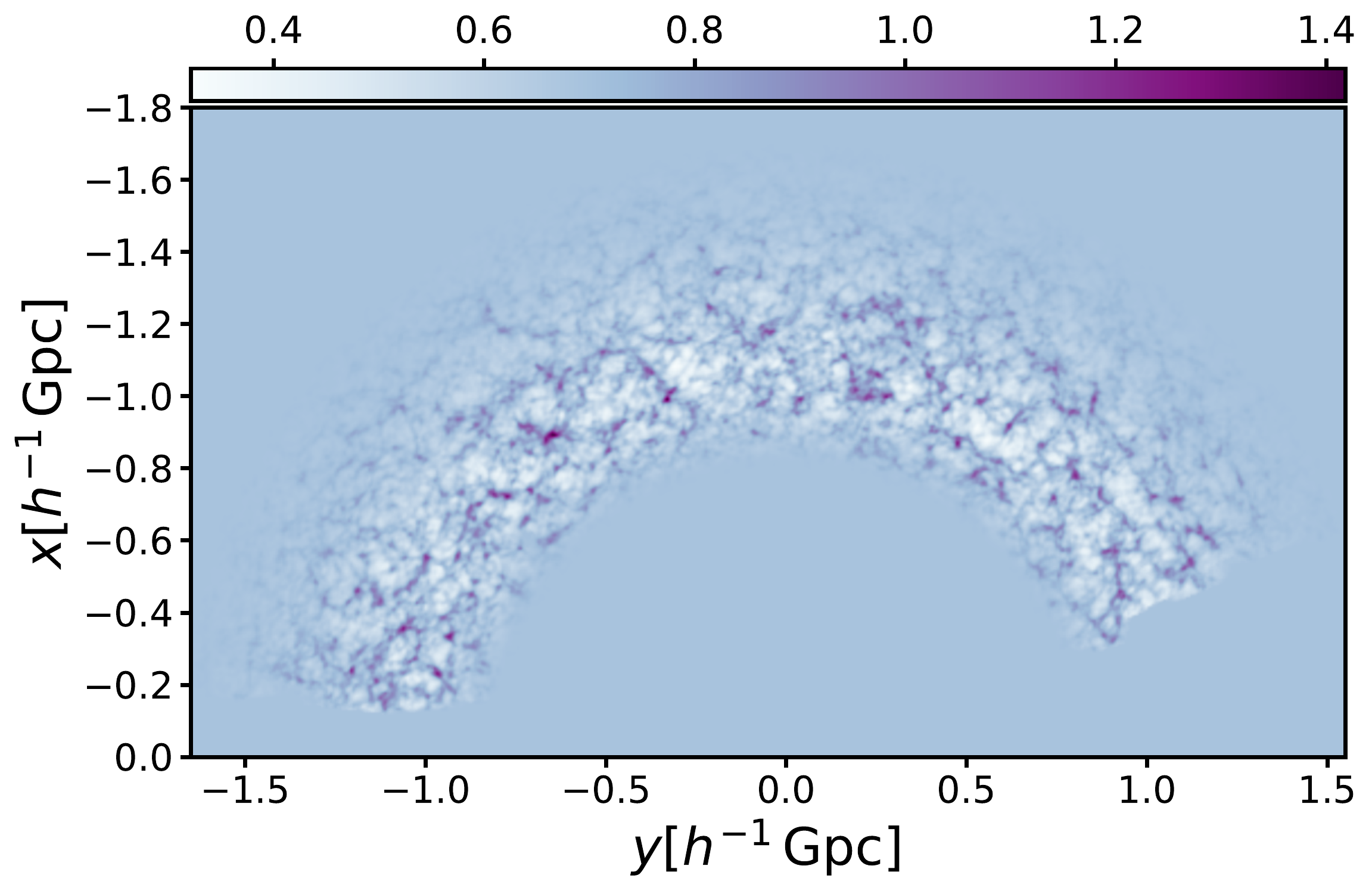}
       \hspace{-19.15cm}
   \includegraphics[width=10.15cm]{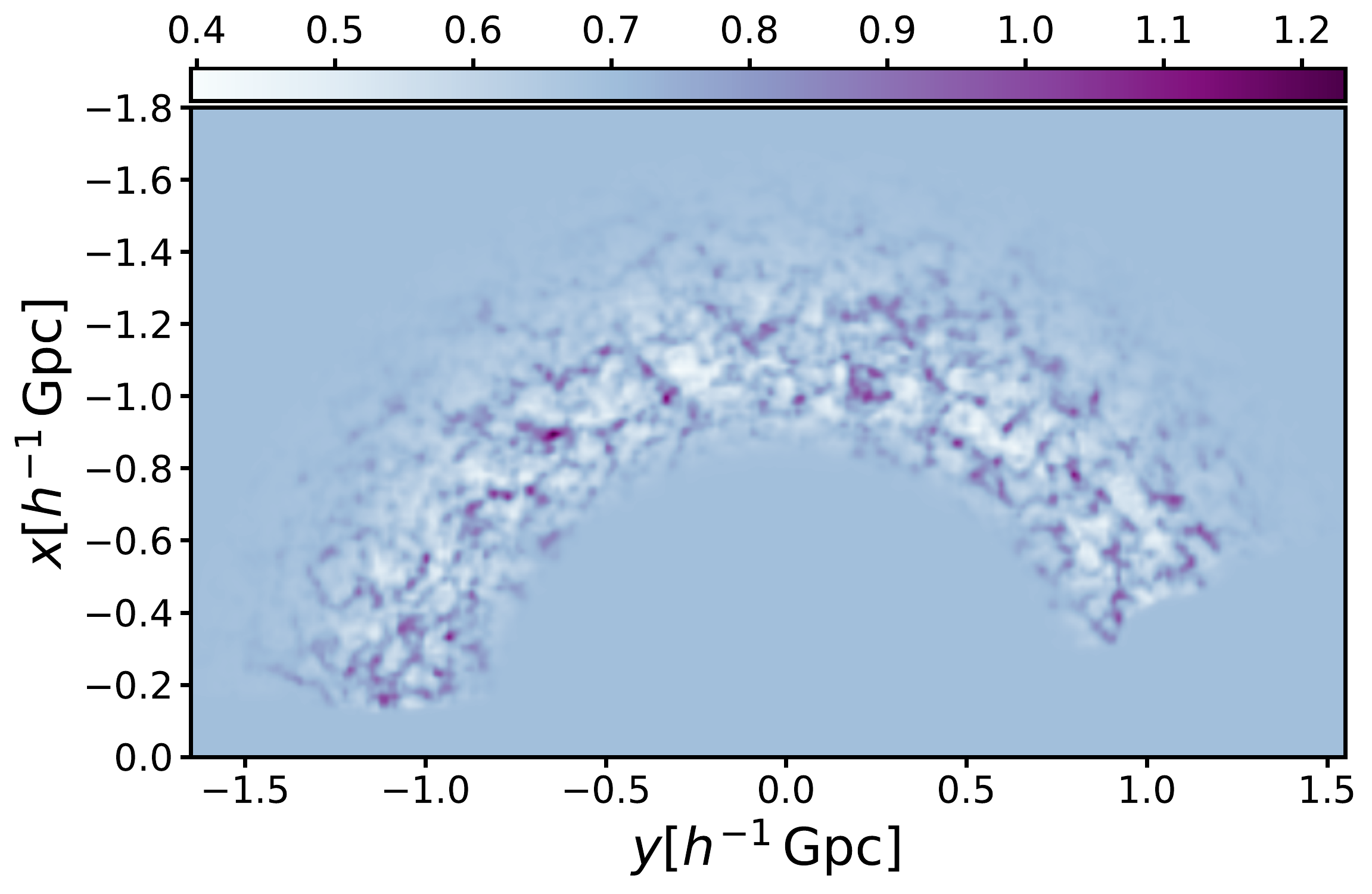}%{dmBIRTHav2000G256snaps10Naug1}
   \vspace{-1.05cm}
  \\
  \hspace{0.4cm}
   \includegraphics[width=10.15cm]{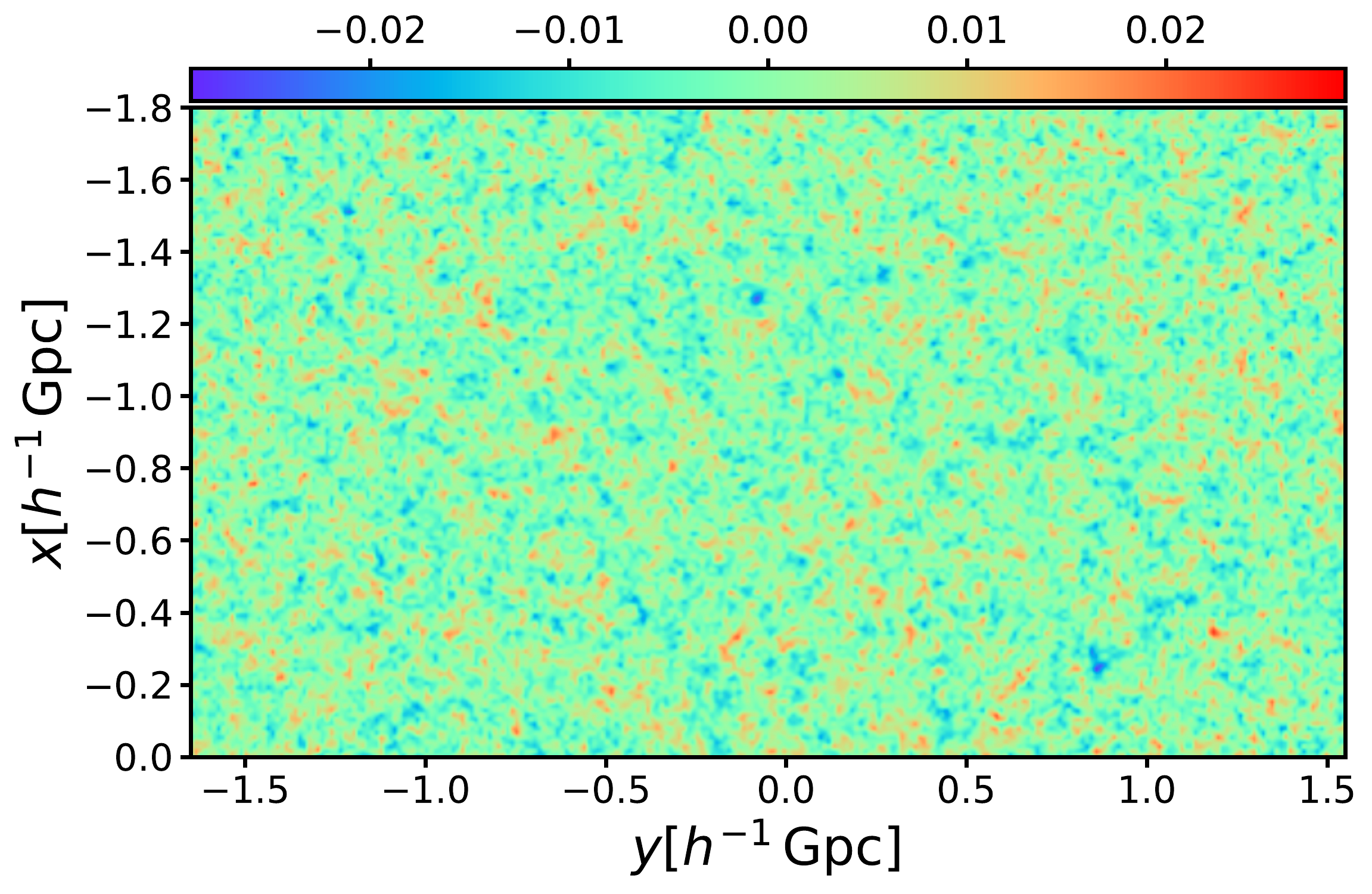}
       \hspace{-19.15cm}
   \includegraphics[width=10.15cm]{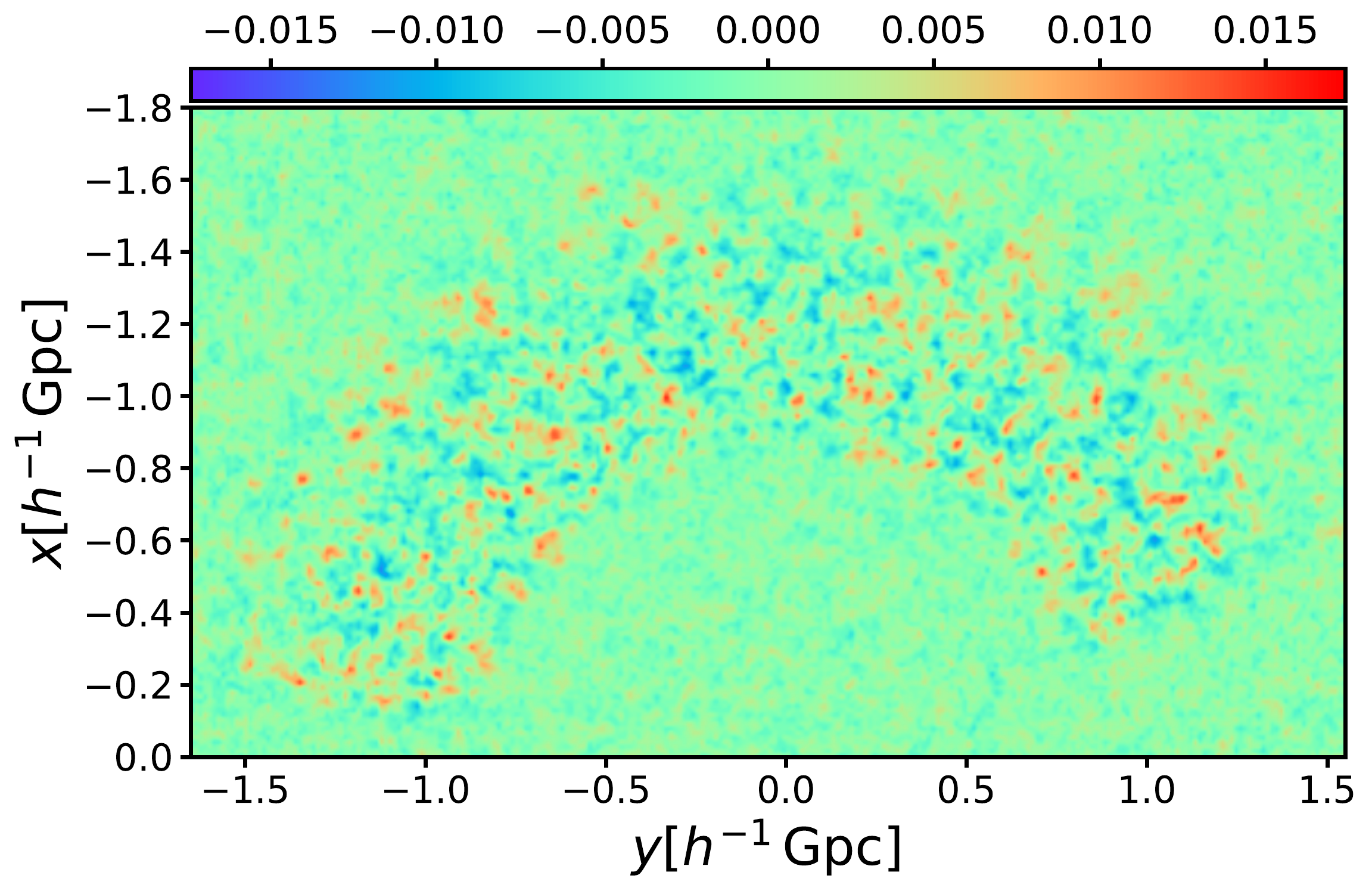}
        \vspace{-.43cm}
\end{tabular}
 \caption{\label{fig:recs} 
 For the same cut as the lower panels in Fig.~\ref{fig:birthzeld256}:
 upper left: dark matter from the BigMD simulation. Upper right: CMASS completeness. Second row: galaxy number counts based on the BigMD simulation using SHAM in Eulerian (left) and Lagrangian (right) space (reconstructed). Third row: dark matter reconstructions with ALPT on $256^3$ (left) and $512^3$ (right). Fourth row:  reconstructed density field at $z=100$, average (left), single (right).}
\end{figure*}

Since  the run {\bf A} with bias interpolation to the position of each galaxy yields numerical identical results to the classification in run  {\bf A$_{\Delta}$}, we will from now on consider only variations on run {\bf A}.
We have in this case a single angular mask for all objects, which is mapped from Eulerian redshift space to Lagrangian real space as shown in Fig.~\ref{fig:winback}. 
Here we can see that edges of the survey become less sharp due to the displacement field.
In fact a hole in the survey mask may be displaced or even filled with large-scale structure tracers when going to Lagrangian space.
The reconstructed DMDF on the light-cone are shown in Figs.~\ref{fig:birthzeld256}, \ref{fig:birthzeld512}, and \ref{fig:recs}. A first visual inspection shows a substantial correlation between the mock galaxy distribution and the underlying DMDF. The panels in Fig.~\ref{fig:birthzeld256} show the average over 2000 Gibbs-sampling iterations, which are equivalent to about 20 independent samples according to our study shown below. 
We can see how the structures cancel out in regions far away from the data (red dots). The lower panels in that figure clearly show a ''donut'' structure that the reconstruction was performed in a limited redshift range to save computations.
The left panels in Fig.~\ref{fig:birthzeld512} show the rich cosmic web for one reconstruction after 70 Gibbs-sampling iterations on a higher resolution of $6.25$ $h^{-1}$ Mpc. The right panels show the corresponding Gaussian field, which does not show a transition from the observed to the unobserved region.
To get an overview of the calculations done in the \texttt{COSMIC BIRTH} code and to further assess its performance we show in Fig.~\ref{fig:recs}: the DMDF from the original simulation on the light-cone, but without applying radial selection criteria (upper left panel), the corresponding total response function (upper right panel), the corresponding mock galaxy catalog in Eulerian redshift space (second row left panel) and in Lagrangian real space (second row right panel), the corresponding reconstructed DMDFs on the light-cone with low (left) and high (right) resolution  (panels in the third row), and finally the reconstructed dark matter field at a single snapshot at redshift 100: mean over 2000 iterations and one reconstructed sample after 1000 iterations.
The panels in the second row show how the galaxy distribution becomes considerably more homogeneous after reconstruction.
The lower panels in Fig.~\ref{fig:recs} show {\color{black} that no survey nor radial selection effects are present in}  the reconstruction, the data region is not distinguished in the single reconstruction (right panel), while the ensemble average clearly shows an enlarged region (right panel in the second row) of the original data region (left panel in the second row) due to the action of gravity.

\begin{figure*}
 \begin{tabular}{cc}

 \hspace{-1cm}
\begin{tabular}{c}
\begin{minipage}[c]{10cm}
\includegraphics[width=10cm]{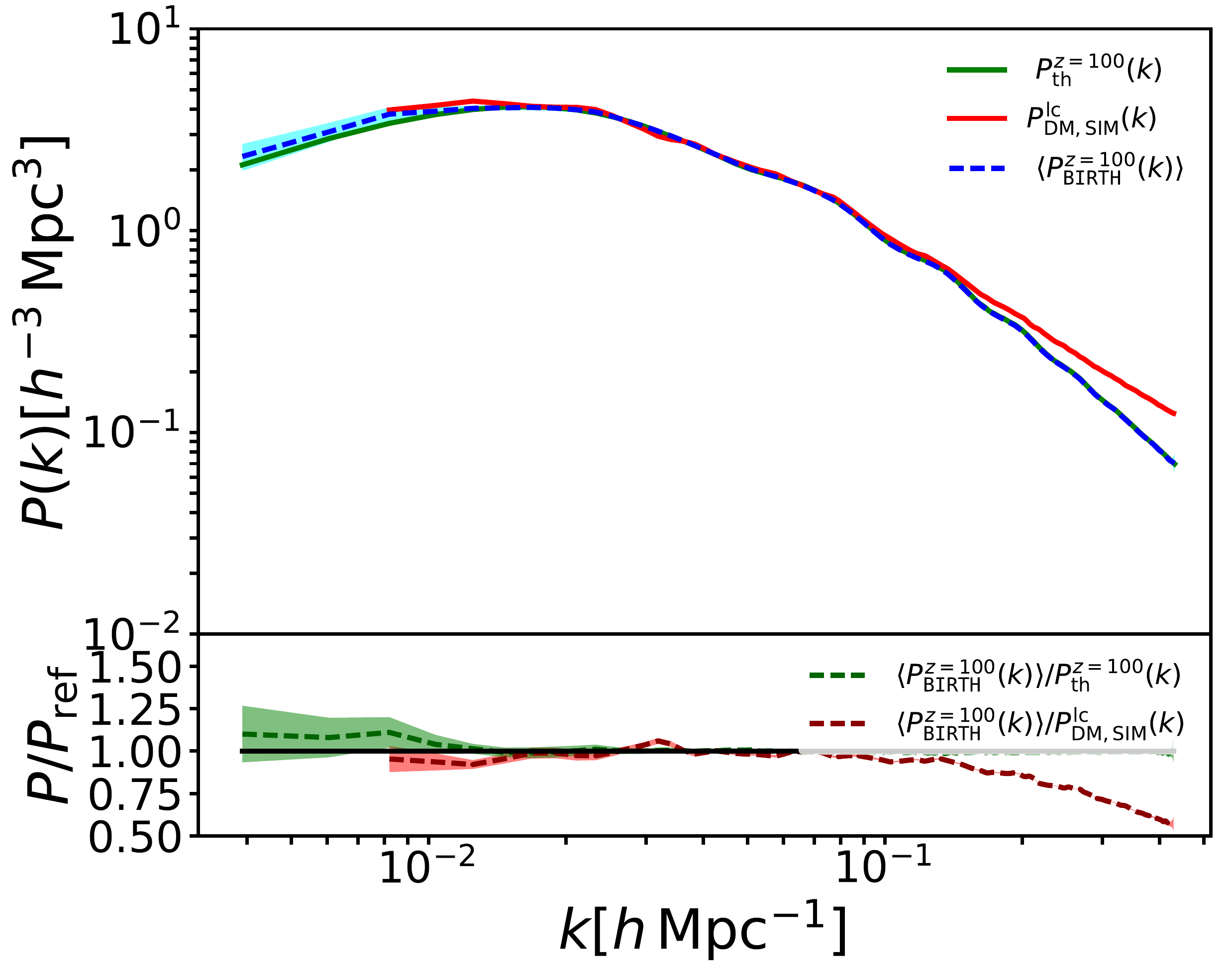}
\vspace{-6.4cm}
\end{minipage}
\\
\hspace{-2cm}
\begin{minipage}[c]{4cm}
\includegraphics[width=4cm]{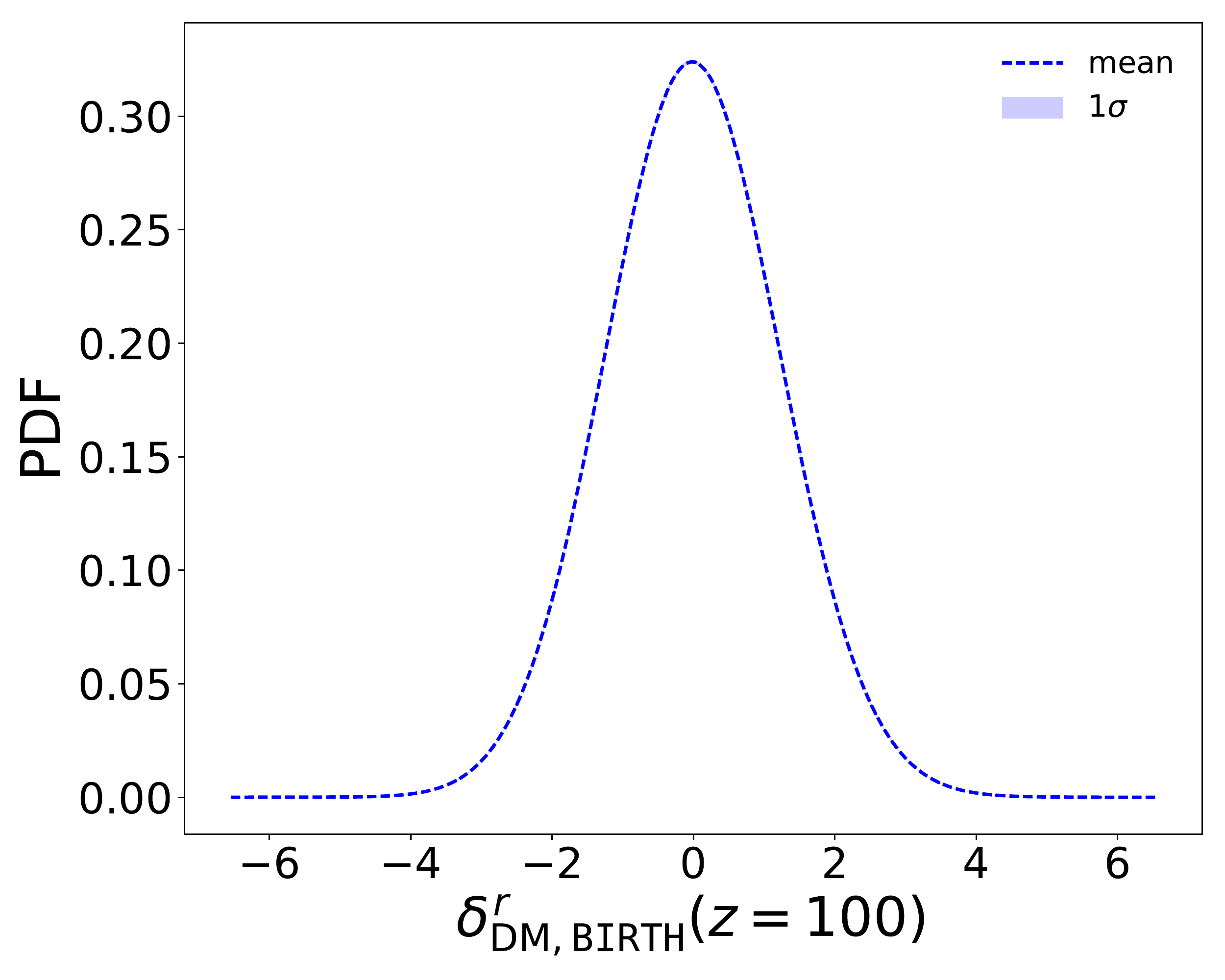}
\end{minipage}
\end{tabular}

\hspace{-19.64cm}
%hspacee-18.7cm}
 
\begin{tabular}{c}
\begin{minipage}[c]{10cm}
\includegraphics[width=10cm]{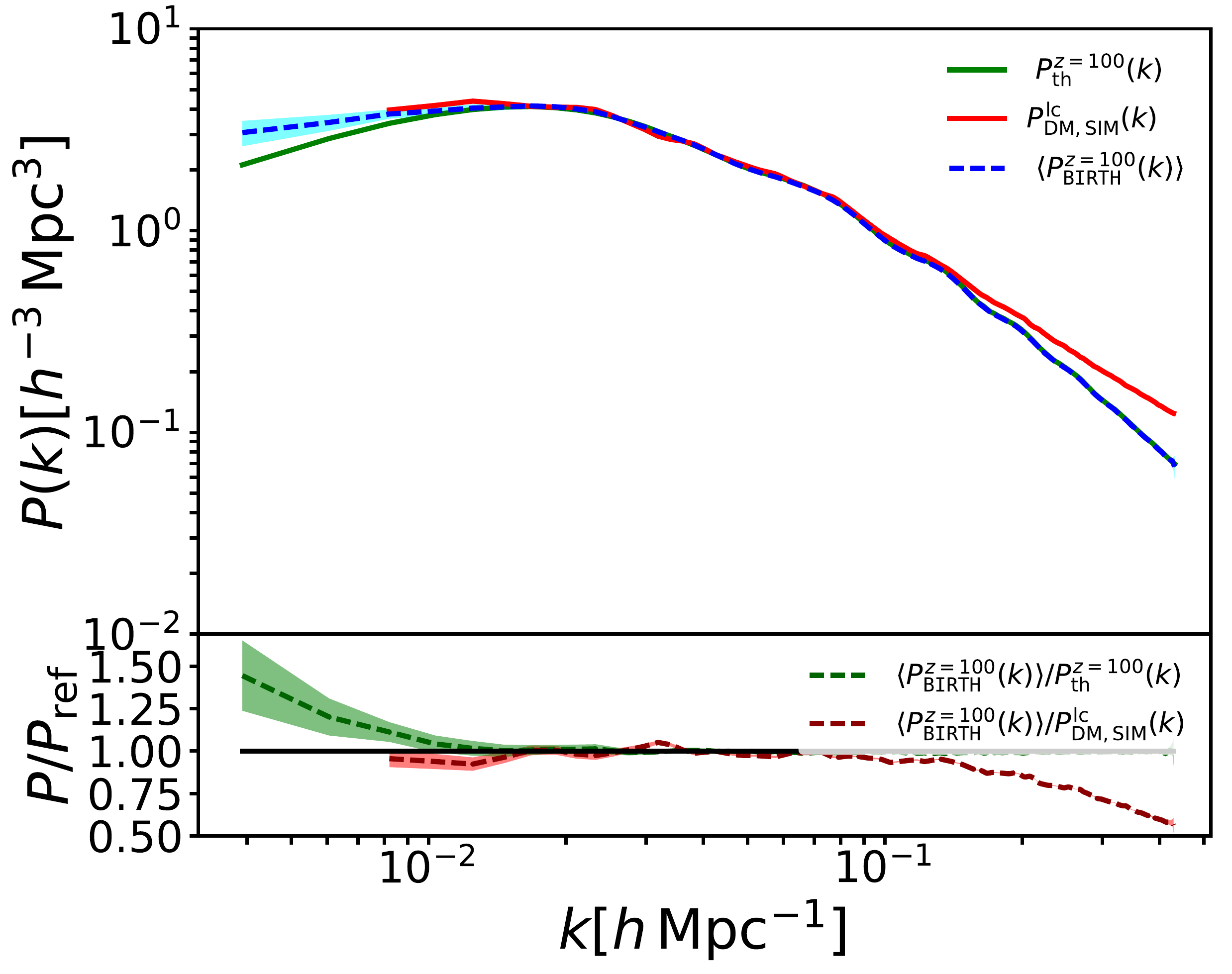}
\vspace{-6.4cm}
\end{minipage}
\\
\hspace{-2.cm}
\begin{minipage}[c]{4cm}
  \includegraphics[width=4cm]{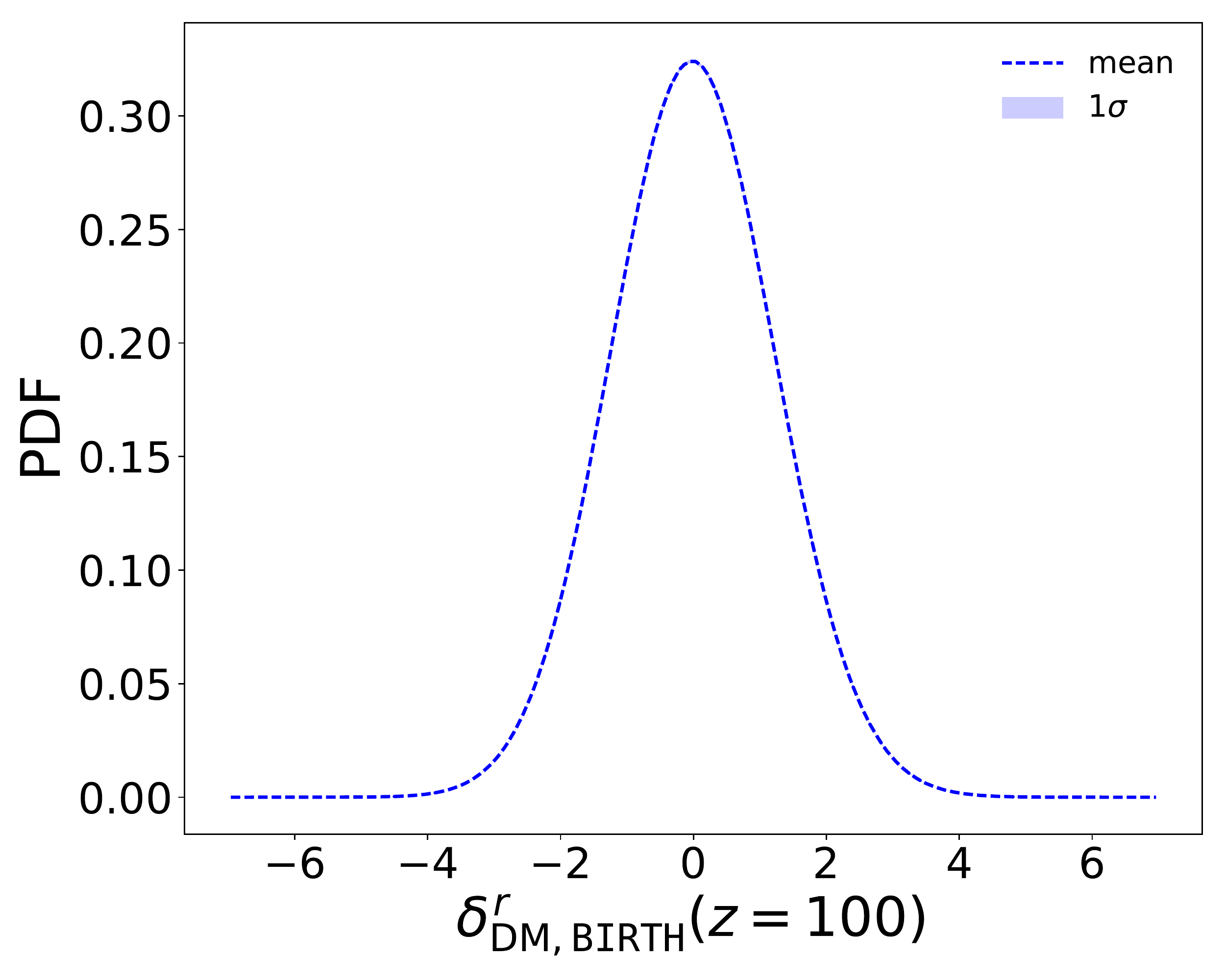}
\end{minipage}
\end{tabular}
\put(-113,-35){$d_{\rm L}=12.5\,h^{-1}$ Mpc} 
\put(-140,55){\bf C} 
\put(-113,-25){$256^3$ ALPT} 
\put(-113,-45){1 snapshot}
\put(-113,-55){with bias interpolation}
\put(-113,-65){with Lag. tetra. tess. } 
\put(-145,-67){\footnotesize{$\times100$}}
\put(130,-35){$d_{\rm L}=12.5\,h^{-1}$ Mpc} 
\put(95,55){\bf A}
\put(130,-25){$256^3$ ALPT} 
\put(130,-45){20 snapshots}
\put(130,-55){with bias interpolation}
\put(130,-65){with Lag. tetra. tess. } 
\put(100,-67){\footnotesize{\small $\times100$}}
\vspace{3cm}

\\

\hspace{-1.1cm}

\begin{tabular}{c}
\begin{minipage}[c]{10cm}
\includegraphics[width=10.2cm]{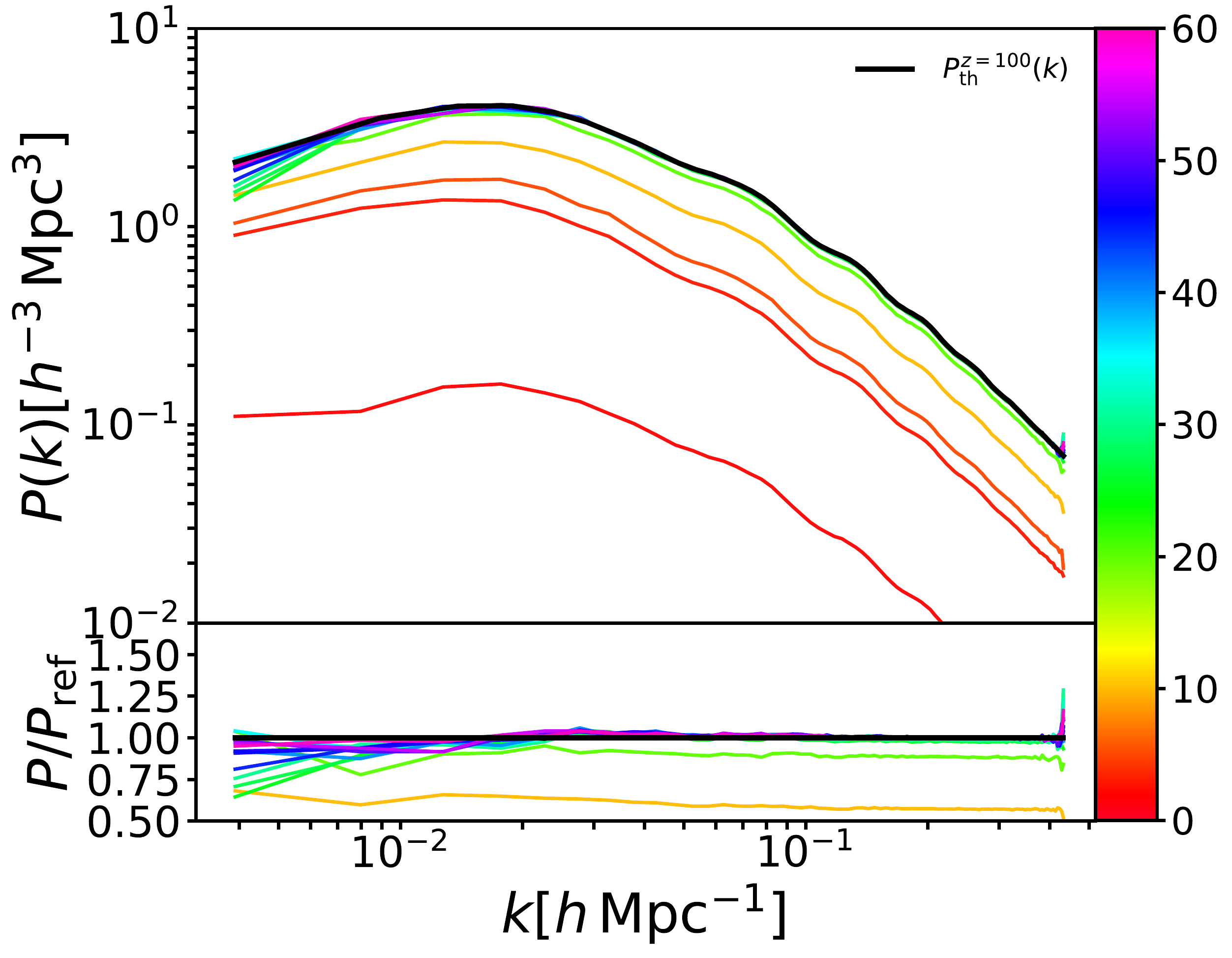}

\vspace{-3.2cm}

\end{minipage}
\\
\hspace{0.cm}
\begin{minipage}[c]{4cm}
\end{minipage}
\end{tabular}

\hspace{-19.6cm}

\begin{tabular}{c}
\begin{minipage}[c]{10cm}
\includegraphics[width=10cm]{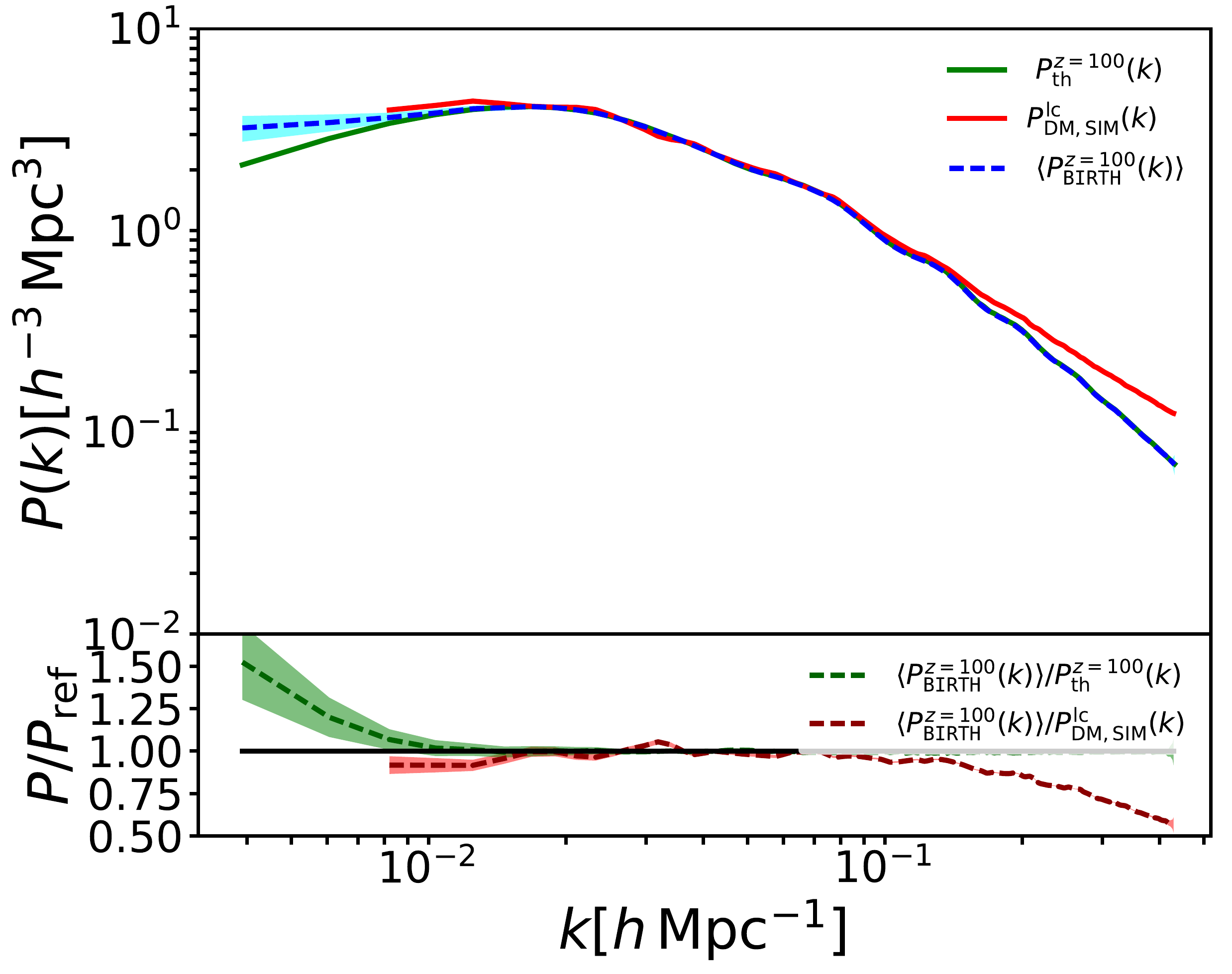}
\vspace{-6.4cm}
\end{minipage}
\\
\hspace{-1.7cm}
\begin{minipage}[c]{4cm}
\includegraphics[width=4cm]{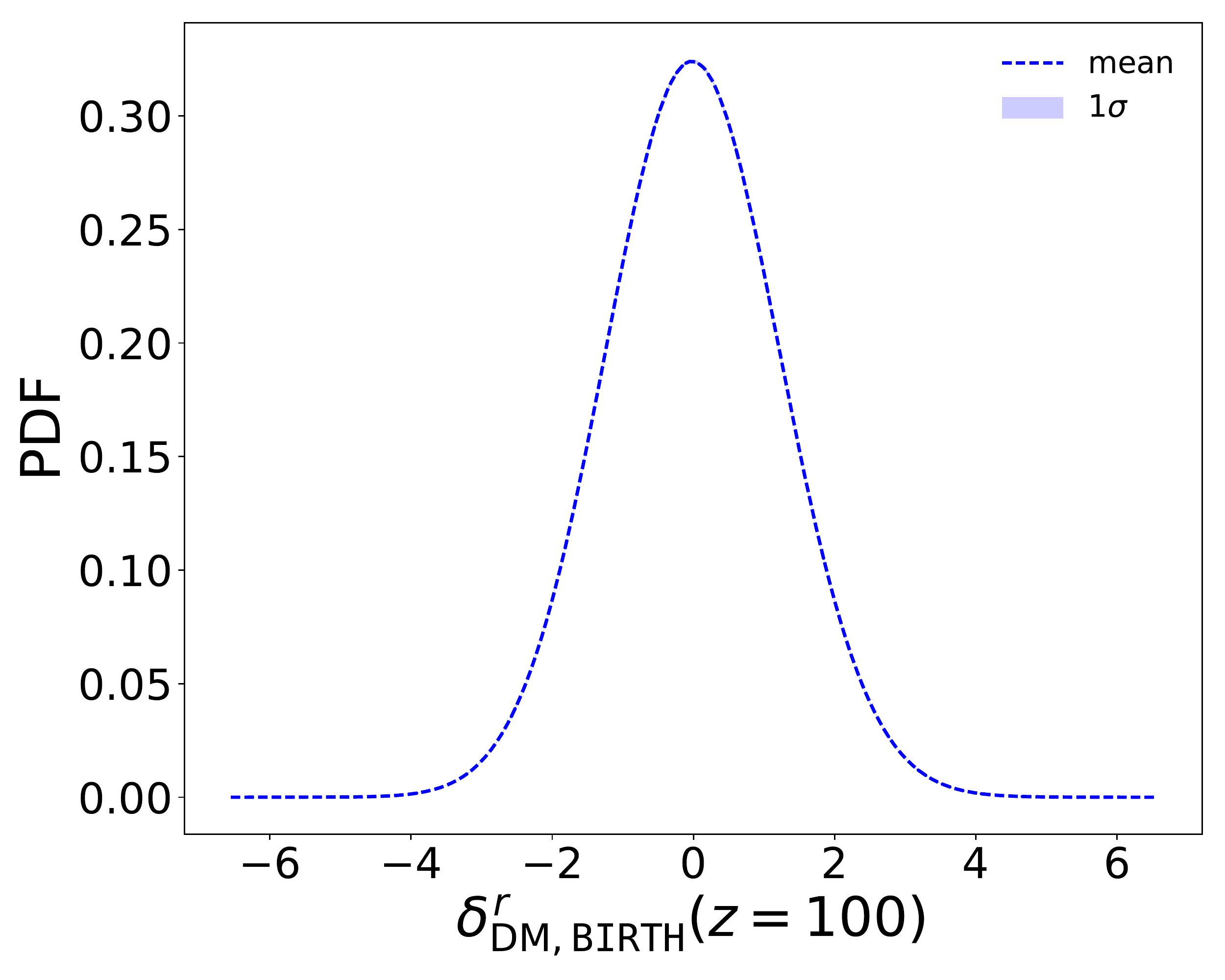}
\end{minipage}
\end{tabular}
\put(-110,-35){$d_{\rm L}=12.5\,h^{-1}$ Mpc} 
\put(-140,55){\bf B}
\put(-110,-25){$256^3$ ALPT} 
\put(-110,-45){20 snapshots} 
\put(-110,-55){with bias interpolation}
\put(-110,-65){w/o Lag. tetra. tess.} 
\put(-137,-67){\footnotesize{$\times100$}}
\put(5,-35){$d_{\rm L}=12.5\,h^{-1}$ Mpc}
\put(95,55){\bf A}
\put(5,-15){$256^3$ ALPT} 
\put(5,-45){20 snapshots} 
\put(5,-55){with bias interpolation}
\put(5,-65){with Lag. tetra. tess.}
\put(210,-130){iteration}
\vspace{.8cm}
 \end{tabular}
 \caption{\label{fig:ps} Statistics of the reconstructed density fields at $z=100$ from Bayesian posterior sampling with a lognormal-Poisson model for Lagrangian tracers. 
 The upper right panel shows results with 20 redshift snapshots (run {\bf A}), the left one with only one (run {\bf C}), both using ALPT. The lower left panel corresponds to a run with 20 redshift snapshots using ALPT without {\color{black} Lagrangian tetrahedral tessellation (Lag. tetra. tess.)} (run {\bf B}), while the lower right panel shows the convergence of the run {\bf A} (the color bar indicates the iteration number).
 The mean is represented by the blue dashed curve with the corresponding 1-$\sigma$ region in cyan, both for the power spectrum and the matter PDF (skewness and kurtosis are of the order of $10^{-2}$, and $10^{-4}$, respectively). The theoretical mean power spectrum is represented by the solid green line. The measured power spectrum from the light-cone DMDF normalised to a mean redshift of $z=0.57$ is in red. The corresponding ratio power spectra are presented at the bottom of each panel. The red solid line is close to the  green one  because the BigMD simulation was run with initial conditions with a corresponding power-spectrum relatively close to the theoretical one, selected from a set of random Gaussian realisations.}
\end{figure*}

To make a quantitative assessment we compute the mean and variance of the power spectra of the reconstructed density fields at $z=100$, as shown in the right panels of  Fig.~\ref{fig:ps}. 
The upper panel shows that we obtain exquisite unbiased power spectra including tidal fields and a nonlinear small scale correction (ALPT) with a full light-cone treatment. Also the statistics of the reconstructed density fields are Gaussian. {\color{black}We find that the particular realisation of the BigMDPL simulation has a slight excess of power on large scales which is accurately reproduced in the reconstructions \citep[see][]{2016MNRAS.457.4340K}.}
The lower right panel shows that the convergence of the \texttt{COSMIC BIRTH} code is extremely fast. It requires only about 30 iterations to converge within percentage accuracy to the theoretical power spectrum and quickly gets the right shape after about 10 iterations (see also appendix \ref{app:hmc}).

We have performed two additional runs {\bf B} and {\bf C}.  In run {\bf B} shown in the lower left panel of Fig.~\ref{fig:ps}, we used a normal CIC interpolation scheme (without tetrahedral tessellation and Lagrangian tetrahedral tessellation) to construct the angular response function in Lagrangian space with the subset of the 256$^3$ cells enclosed in the considered redshift range. This inaccuracy given the low number of tracers used to do the Eulerian-Lagrangian mapping had the same effect as a radial selection function as can be seen in the abnormal excess of power on large scales. In fact, using a single redshift bin for the whole CMASS data had a similar effect as found in run {\bf C} shown in the upper left panel. 

\begin{figure*}
\begin{tabular}{cc}
\hspace{-1cm}
\begin{tabular}{c}
\begin{minipage}[c]{8cm}
\includegraphics[width=8cm]{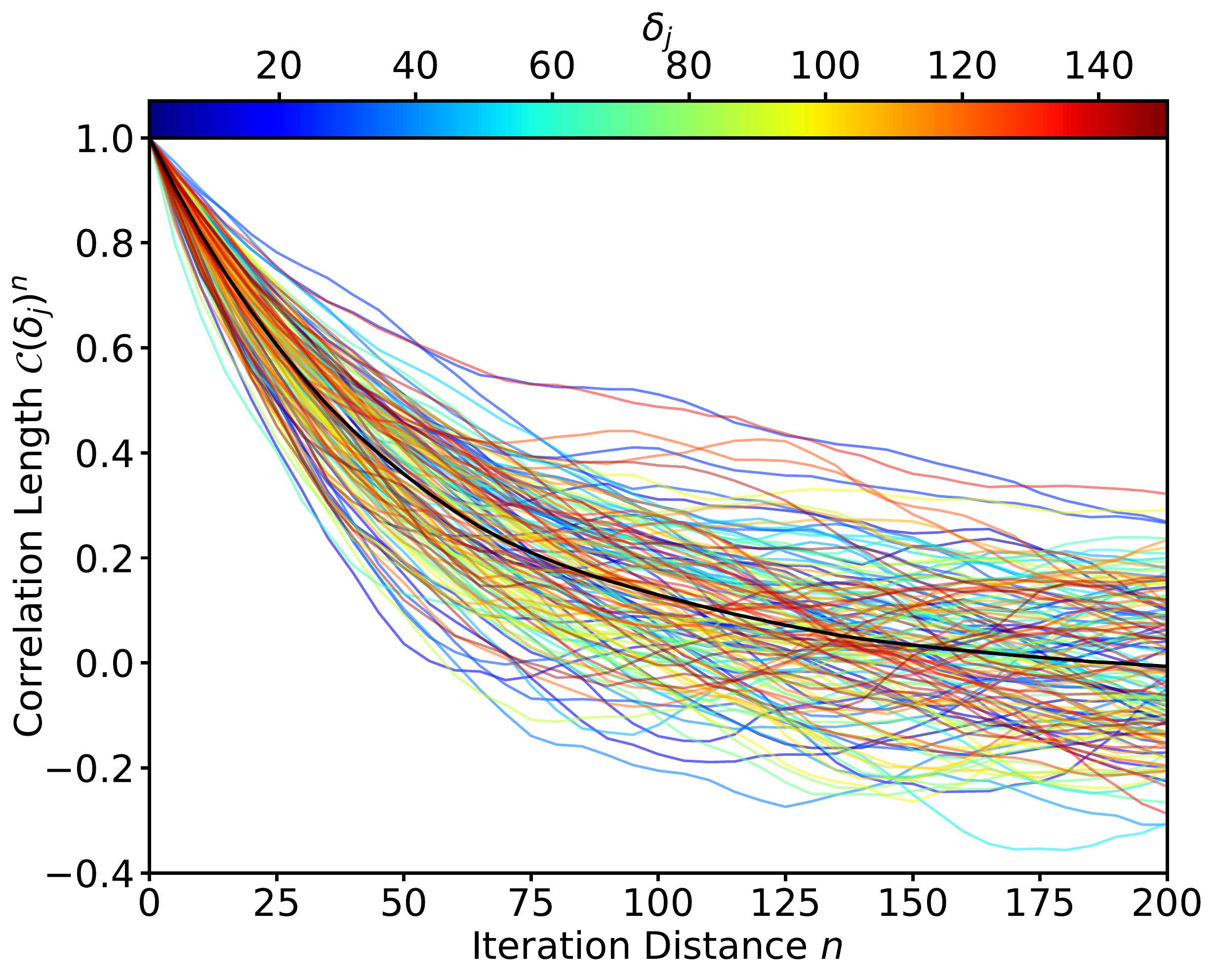} %corrG256NSnaps10it2000}
\vspace{0.5cm}
\end{minipage}
\\
\begin{minipage}[c]{8cm}
\includegraphics[width=7.5cm]{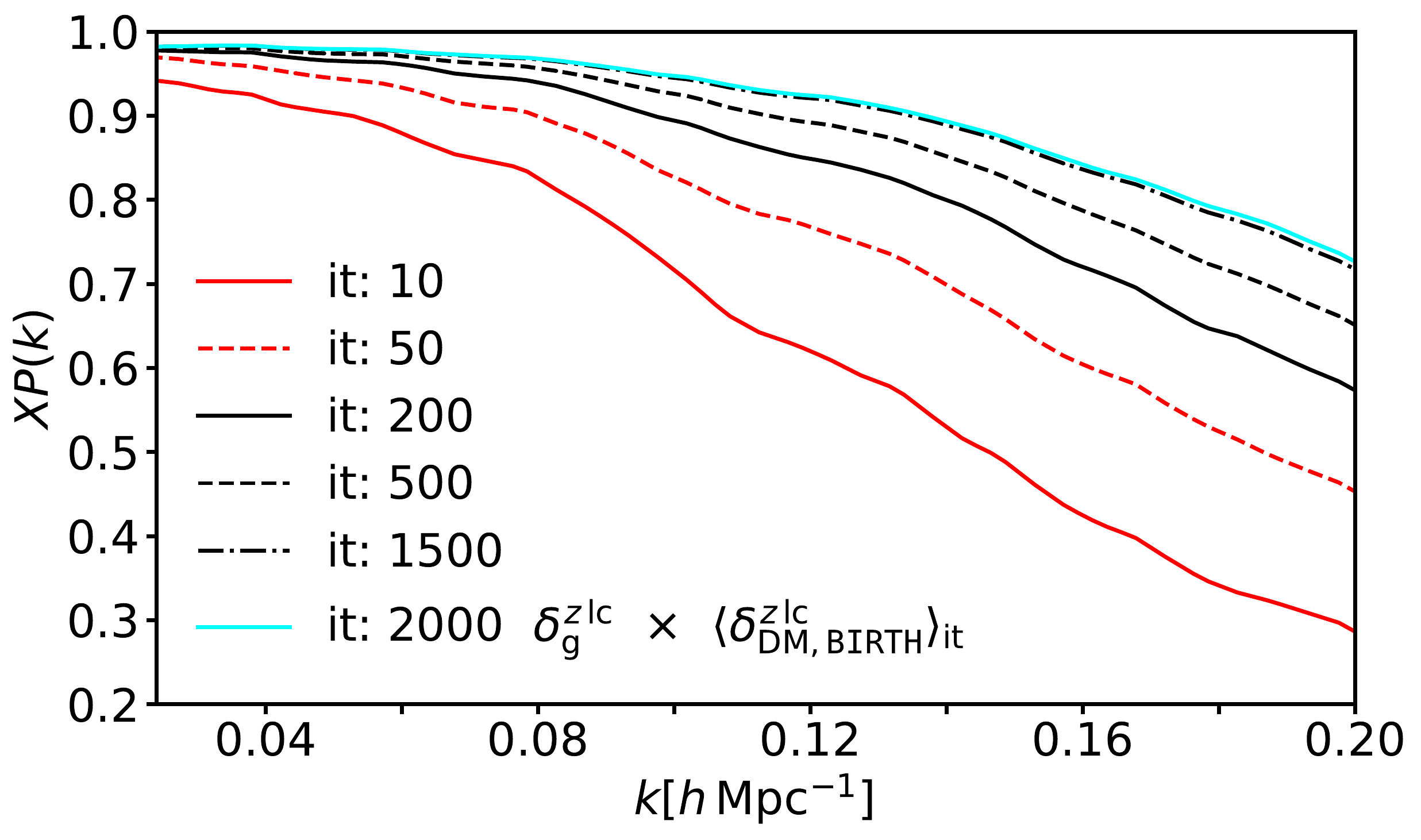}
\end{minipage}
\end{tabular}
\begin{minipage}[c]{10cm}
  \includegraphics[width=10cm]{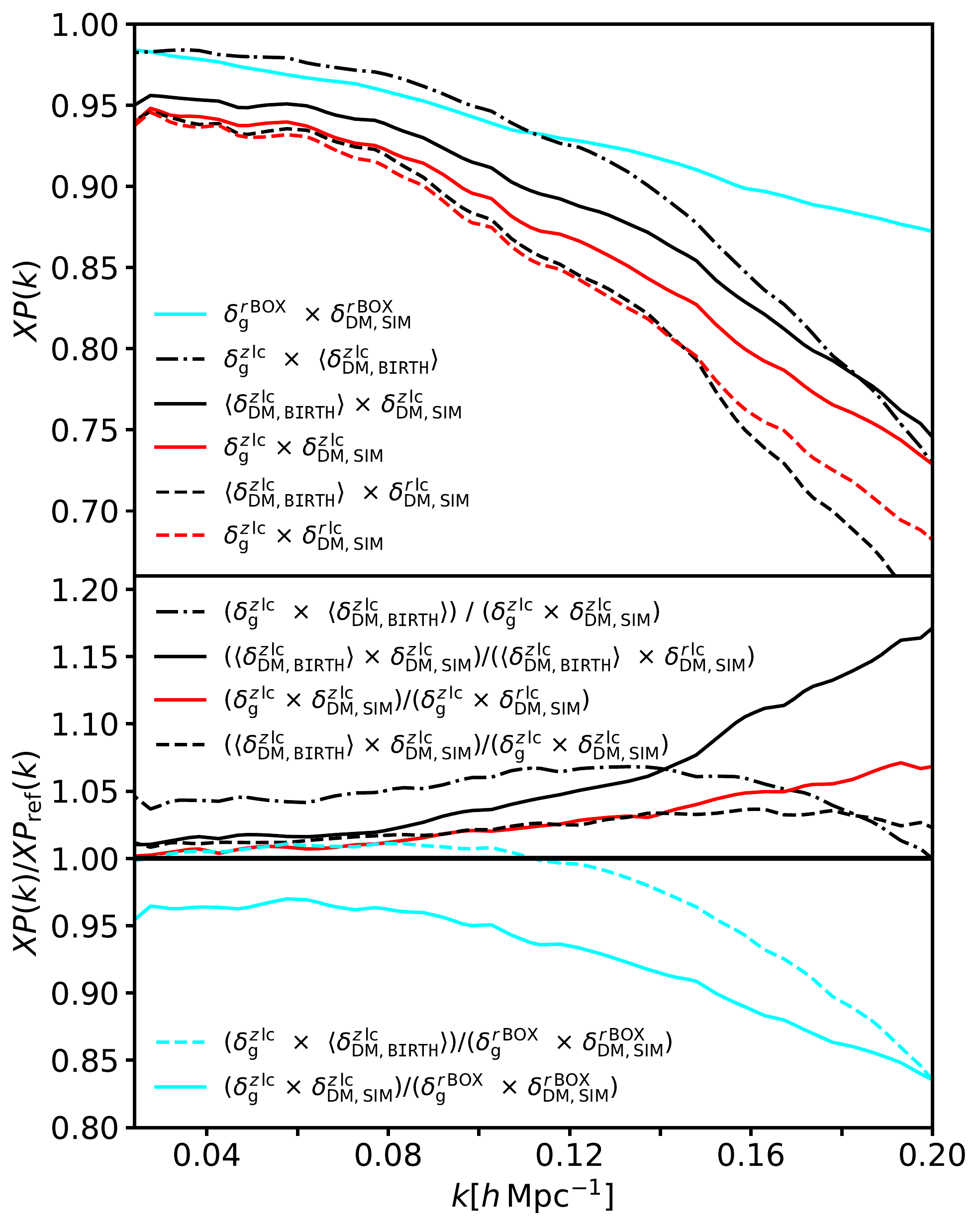}
\end{minipage}
\end{tabular}
 \caption{\label{fig:cross} Convergence behaviour and assessment of the information gain from the Bayesian posterior. Upper left panel: correlation length with the mean over all {\color{black} density bins} for each iteration represented by a solid black line. This demonstrates that independent samples are drawn each {\color{black}$\sim$100 iterations considering the density field}. 
 %Lower left panel: cross power spectra between the galaxy field in redshift space and the corresponding reconstructed dark matter fields taking the average over increasing number of iterations (10, 50, 200, 500, 1500, and 2000). The saturation of the dashed-dotted (1500 iterations) and solid cyan line (2000 iterations) shows that a fair estimation of the posterior mean is obtained after about 2000 iterations.
 Lower left panel: Growth of the the power spectrum with increasing iteration number. After $\sim$40 iterations the reconstructed power spectrum gains its target amplitude, fluctuating according to the statistical uncertainty of the individual realizations.
 The right panel presents cross power spectra (top) and the corresponding ratios (bottom) between different combinations of fields, as shown in the legend. {\color{black} The superscript "lc" stands for light-cone, if it is accompanied by $r$ or by $z$ in real-  or in redshift-space, respectively. The superscript BOX stands for data occupying the entire box without survey geometry, nor radial selection effects.} }
\end{figure*}

Another quantitative measurement of the speed of the \texttt{COSMIC BIRTH} code is presented in the upper left panel of Fig.~\ref{fig:cross}. 
Here we demonstrate that the correlation length is of about 100 iterations, meaning that we get independent samples with considerably lower number of iterations between samples than previous methods (see \S \ref{sec:intro}).

The correlation length of the density bins over the iteration distance is calculated as

\ba
C_n(\sigma_{j})=\frac{1}{N-n}\sum_{i=0}^{N-n}\frac{\left(\delta_{j}^{i}-\langle \delta_{j}\rangle\right)\left(\delta_{j}^{i+n}-\langle\delta_{j} \rangle\right)}{\sigma^{2}(\delta_{j})}\,,
\ea
where $\delta_{j}$ is the overdensity field in each iteration at voxel $j$, $N$ is the number of samples and $n$ is the distance between iterations.

We have computed a series of cross correlations following the definitions in  \citet{2012MNRAS.427L..35K,2013MNRAS.435.2065H} shown in Fig.~\ref{fig:cross}. The analysis is restricted to $k=0.2\,h\,{\rm Mpc}^{-1}$, which is about 50\% of the Nyquist frequency given the considered mesh resolution.
The lower left panel shows that after about 2000 iterations the mean over reconstructed dark matter fields on the redshift space light-cone cross correlated with the corresponding mock galaxy distribution does not improve, meaning that the ensemble average can be considered to have converged after only 2000 iterations. 
The right panel shows that we get close to the optimal cross correlation achieved between the real space dark matter field and the galaxy field on a full box without selection criteria (cyan line), as with our reconstruction (dashed dotted black line), which is a considerable information gain with respect to the red line when including the same selection criteria from the galaxy field in the dark matter field. This means that the Bayesian code is actually correcting with the given structure formation model and the the response function for incompleteness. This information gain is of course lost towards small scales and the dashed-dotted line drops towards high $k$.
It is in fact remarkable, how well the galaxy and reconstructed dark matter field correlate with each other given that a whole structure formation model displaces the large scale structure tracers on average 8 to 10 $h^{-1}$ Mpc. This is trivial when the dark matter field is simply a smooth version of the galaxy field, which is far from being the case here.
It is also interesting to verify that the cross correlation between the dark matter reconstructions and the true dark matter field are larger than the cross correlation between the mock galaxy field and the dark matter field. This means that our actual structure formation  and galaxy bias modelling is working. 
We also find a considerably higher correlation between the dark matter field reconstruction in redshift space and the true simulated dark matter field in redshift space than in real space (solid black line vs dashed black line). This implies that the redshift space distortions modelling is meaningful. 
The lower right panel shows the information gain. Here we find consistent results demonstrating that the reconstructions are adding information through the structure formation, bias, and completeness modelling.

\section{Discussion and conclusions}
\label{sec:conclusions}

In this work we have presented the {\tt COSMIC BIRTH} method. It provides a Bayesian framework to tackle the matter reconstruction problem from a distribution of galaxies. 

It is a particularly simple and efficient algorithm, which solves the Bayesian reconstruction problem including selection effects and nonlinear structure formation in the calculation of the displacements. It is important to stress that this is achieved without giving-up on accuracy or loss of generality. The strategy of splitting the approach into two reconstructions steps permits us to use a lognormal-Poisson posterior in Lagrangian space, as this model is accurate at initial cosmic times (high redshifts). The lognormal assumption ensures that the density field is positive definite converging to the Gaussian assumption for $|\delta|\ll1$. The Poisson distribution function permits us to correct for aliasing caused by describing the galaxy distribution (in Lagrangian space) as discrete number counts of large scale structure tracers on a regular mesh.
Note, that a Gaussian likelihood keeping only the two point statistics of the Poisson likelihood is not adaptive, but yields a constant mean noise covariance matrix \citep[see][]{2009MNRAS.400..183K,2010MNRAS.403..589K}. This limits very much the accuracy of Wiener filtering based on a Gaussian prior and a Gaussian likelihood \citep{1995ApJ...449..446Z}. Therefore we conclude that the simplest statistical model we can consider is the lognormal-Poisson one.
Thanks to this model (positive definite matter fields connected to a discrete number of tracers), we can include a non-linear bias description beyond the commonly used linear one in BAO reconstruction.
This Lagrangian posterior model yields the primordial Gaussian density fields defined on a regular mesh assuming a set of observed Lagrangian tracers using Hamiltonian Monte Carlo sampling.
These Lagrangian tracers in turn are connected to the observed galaxy sample on the light-cone through a forward modelling within an iterative Gibbs-sampling scheme based on an arbitrary structure formation model.
This approach dramatically simplifies the  programming structure of the code, as no gradients of structure formation models need to be computed. 
In this way, the structure formation model can be changed by any other one in this framework, and only needs to deliver information on the initial and final positions of tracers including their peculiar velocities. We will investigate in a subsequent work how a particle mesh code improves the results \citep[this was investigated to some extent with the Kigen code going from 2LPT to ALPT][]{2012MNRAS.427L..35K,2013MNRAS.435.2065H}. 
The \texttt{COSMIC BIRTH} method combines a grid based with a tracer based reconstruction, relying on Bayesian inference methods while directly yielding a set of Lagrangian tracers equivalent to BAO reconstruction.

We have introduced technical improvements to the Hamiltonian sampling scheme to gain a factor of about 20 in efficiency, yielding correlation lengths of only 40 to 50 iterations. 
This demonstrates that Bayesian methods can actually be practically used to sample posterior PDFs.
Part of these improvements are further studied in detail in a companion paper and are inspired by techniques widely used in lattice quantum field theory using higher-order discretisations of the Hamiltonian equations of motions (Hern\'andez-S\'anchez et al., in prep). Furthermore, we have introduced a strategy to efficiently deal with non-diagonal Hamiltonian mass-matrices including complex survey geometries, which speeds-up the convergence by up to $\sim$70\% with high acceptance rates of 60 to 70\%.
{\color{black} The idea is based on associating specific  uncertainties in the data augmentation modelled by the momenta depending on the completeness.  Therefore, the Hamiltonian mass needs to include the structure of the response function as derived in appendix \ref{app:hmc}. Special attention must be paid to a consistent formulation of the square root of the Hamiltonian mass, as required to generate the random fluctuations,  and its inverse to solve Hamiltonian's equations of motion.} 

This approach has the novelty of being only dependent on cosmological parameters and an arbitrary structure formation model, while solving the problem of dealing with galaxy bias on the light-cone. As we have shown, one needs in general (and in particular for the BOSS data) to consider a varying galaxy bias with redshift \citep[see Fig.~\ref{fig:bias} and][]{2016MNRAS.456.4156K}. However, one can certainly find unbiased power spectra with respect to the theoretical one in the full cubical volume with a single (wrong) bias parameter, which as we know now is inaccurate \citep[see e.g.][]{2017MNRAS.467.3993A}. These kind of crude approximations will have an impact in a detailed tomographic analysis, and will not permit to break degeneracies with, e.g. gravity or neutrino induced deviations in the power spectrum. 
We made progress to include a complete robust non-linear Lagrangian bias framework, and a {\color{black} Lagrangian tetrahedral tessellation} of the survey geometry from Eulerian to Lagrangian coordinates, as it is required in our framework. Further investigation in this direction will permit us to better understand Lagrangian bias. Also this method can be used to study Eulerian bias from the data itself and the dark matter reconstructions, without having used any Eulerian bias description in the reconstruction process.
In this work, we have found a connection between the measurable large-scale bias and the effective non-linear bias in Lagrangian space, solving the dependence on the mesh resolution. In this way non-local bias is accounted for through the displacement field.

The method has the potential to become a standard technique (particularly for BAO reconstruction, as we will show in a subsequent paper),
Our tests demonstrate that we can obtain unbiased dark matter field reconstructions on the light-cone from highly biased tracers using arbitrary structure formation models. Therefore, this method shows its great potential for the analysis of deep redshift surveys such as DESI, EUCLID, JPAS, PFS, WFIRST, 4MOST, etc.
Provided its sampling speed, other more general applications can be foreseen with this method, such as cosmological parameter estimation and growth rate sampling. 
We expect that this method contributes towards a full analysis of the large scale structure, ultimately including a full determination of the cosmological model.

\section*{Data in this article}

The data underlying this article will be shared on reasonable request to the corresponding author.  

\section*{Acknowledgments}

We thank the referee for the careful revision of the manuscript.
The authors thank Ra\'ul E. Angulo, Ginevra Favole, Mariana Vargas-Maga{\~n}a, and Cheng Zhao for discussions. FSK acknowledges financial support from the Spanish Ministry of Economy and Competitiveness (MINECO) under the Severo Ochoa program SEV-2015-0548, and for the grants RYC2015-18693 and AYA2017-89891-P. SRT thanks support from grants SEV-2015-0548 and RAYA2017-89891-P. MA thanks for the hospitality at the IAC and support from the Kavli IPMU fellowship. MHS acknowledges the {\it residente} grant at IAC. ABA acknowledges financial support from the Spanish Ministry of Economy and Competitiveness (MINECO) under the Severo Ochoa program SEV-2015-0548. 
GY acknowledges financial support  from  MINECO/FEDER under research grant AYA2015-63810-P and MICIU/FEDER PGC2018-094975-C21. FSK thanks Joaqu{\'i}n P{\'e}rez Mac{\'i}a for studying the efficiency of the non-diagonal Hamiltonian mass-matrix during his bachelor thesis.

%%%%%%%%%%%%%%%%%%%%%%%%%%%%%%%%%%%%%%%%%%%%%%%%%%

%%%%%%%%%%%%%%%%%%%% REFERENCES %%%%%%%%%%%%%%%%%%

\bibliographystyle{mnras}
\bibliography{references.bib}

%%%%%%%%%%%%%%%%%%%%%%%%%%%%%%%%%%%%%%%%%%%%%%%%%%

%%%%%%%%%%%%%%%%% APPENDICES %%%%%%%%%%%%%%%%%%%%%

%%%%%%%%%%%%%%%%%%%%%%%%%%%%%%%%%%%%%%%%%%%%%%%%%%

% Don't change these lines
\bsp	% typesetting comment
\label{lastpage}

\appendix

\section{Renormalized perturbation theory}\label{app:rpt}
Here we derive the third order equation from renormalised perturbation theory, which connects the nonlinear bias correction with the large scale bias and the dark matter variance at the cell resolution.
Let us perform a Taylor expansion in the expression to third order in the over-density field:
\ba
\label{eq:biaspow}
\lefteqn{\delta_{\rm  g}(z)\equiv\frac{\rho_{\rm  g}(z)}{\bar{N}}-1\simeq \tau(z)\,\Big[1+b(z)f_{b}(z)\delta(z)}\\
&&\hspace{-0.75cm} +\frac{1}{2}b(z)f_{b}(z)(b(z)f_{b}(z)-1)\left(\delta(z)\right)^2+\nonumber\\\
&&\hspace{-0.75cm} \left. \frac{1}{3!} b(z)f_{b}(z)(b(z)f_{b}(z)-1)(b(z)f_{b}(z)-2)\left(\delta(z)\right)^3\right]-1\nonumber\,,
\ea
with $\tau(z)\equiv\gamma(z)/\bar{N}$.
The usual expression for the perturbatively expanded over-density field to third order ignoring non-local terms is given by \citep[see e.g.][]{2018PhR...733....1D}
\be
\label{eq:biaspt}
\delta_{\rm g}(z)=c_\delta(z)\delta(z)+\frac{1}{2}c_{\delta^2}(z)(\delta^2(z)-\sigma^2(z))+\frac{1}{3!}c_{\delta^3}(z)\delta^3(z)\,.
\ee
Correspondingly, one can show that the large scale  bias is given by \citep[e.g.][]{2009JCAP...08..020M}
\ba
\label{eq:renorm}
b_\delta(z)=c_\delta(z)+\frac{34}{21}c_{\delta^2}(z)\sigma^2(z)+\frac{1}{2}c_{\delta^3}(z)\sigma^2(z)\,.
\ea
By considering that in our case the large scale bias is  given by $b(z)$ and identifying the coefficients \{$c_\delta=\tau\,f_{b}b$, $c_{\delta^2}=\tau\,f_{b}b(f_{b}b-1)$, $c_{\delta^3}=\tau\,f_{b}b(f_{b}b-1)(f_{b}b-2),\tau-1=-c_{\delta^2}\sigma^2/2\}$
from Eqs.~(\ref{eq:biaspow}) and (\ref{eq:biaspt}) one can derive the following cubic equation
 for $f_{b}$:
\ba
\lefteqn{b(z)\,f_{b}^3(z)+f_{b}(z)^2\,\left(\frac{5}{21}-b(z)\right)}\\
&&\hspace{-.5cm}+\frac{f_{b}(z)}{b(z)}\left(\frac{2}{\sigma^2(z)}-\frac{26}{21}+b(z)\right)-\frac{2}{\sigma^2(z)\,b(z)}=0 \nonumber\,.
\ea
We have verified that this model yields accurate power spectra on large scales, as long as the bias is given by the truncated Taylor expansion at third order. Although the absolute value of the over-density field is smaller than one at high redshift (say $z=100$) and resolutions of a few Mpc (say about 5 $h^{-1}$ Mpc), the Lagrangian large scale bias is so high that higher order terms in the Taylor expansion are still relevant.
Thus the validity range of this framework is thus restricted to special cases, such as low bias tracers.

\section{Efficient non-diagonal Hamiltonian mass}
\label{app:hmc}

The Hamiltonian Monte Carlo (HMC) method \citep[][]{DUANE1987216} requires a nuisance variable to sample the posterior distribution function, which is called the momenta $\mbi p$. According to the mechanical analogy the kinetic energy is given by
\be
K(\mbi p|\mat M)=\frac{1}{2}\mbi p^{\rm t} \mat M^{-1}\mbi p \,,
\ee
where $\mat M$ is the Hamiltonian mass, which acts as a pre-conditioner of the HMC sampler, and can considerably speed up the HMC \citep{2012arXiv1206.1901N}.
This mass can be interpreted as the covariance matrix of the momenta. The kinetic term can be connected to a multivariate Gaussian distribution proportional to $\exp{[-K]}$.
This implies that the generation of the momenta is equivalent to the generation of a Gaussian field with an appropriate covariance matrix $\mat M$.
Ideally, this mass should have the structure of the prior and of the likelihood \citep{2010MNRAS.407...29J}, i.e., a term related to the matter field covariance matrix, say $\mat C$, and a term related to the response function $\mat R$, which can have a structure like this:
\be
\label{eq:mass} 
\mat M=\mat C^{-1}+\beta\mat R\,,
\ee
where $\beta$ is a constant which will depend on the number density and maybe other quantities.
This mass represents a non-diagonal matrix in neither Fourier, nor configuration space, as $\mat C$ is diagonal in Fourier space, but $\mat R$ is diagonal in real space, being the three-dimensional completeness. 
For an analysis without selection function, nor angular completeness, i.e., considering a full complete volume, the second term in Eq.~(\ref{eq:mass}) can be neglected \citep[see][]{2008MNRAS.389.1284T}.
However, it is clear that an efficient sampler needs information on the completeness of the volume in a realistic case, as the uncertainty in our reconstruction is not the same in a well sampled area, as in unobserved one.
Here we face two different problems.  One problem is that we need the inverse of the mass-matrix, as we need to numerically solve the Hamiltonian equations of motion to perform HMC sampling. In particular we need to solve this equation involving the m ass matrix:
\be
  \label{eq:EoM1}
  \frac{d\mbi x}{dt} = \frac{\partial \mathcal{H}}{\partial \mbi p}=\mat M^{-1} \mbi p\, ,
\ee 
where $x$ are the positions (in our case the matter density field at initial cosmic times), t cosmic time, and $H$ the Hamiltonian.
One could consider applying efficient inversion schemes based on conjugate gradients \citep[see][and references therein]{2008MNRAS.389..497K}, but is clear that this will. lower the efficiency of the HMC sampler. But, this is not even so trivial, as we have 
a second problem, since  we  also need the square root of the mass-matrix $\sqrt{\mat M}$ to efficiently generate the Gaussian field of momenta in Fourier space \citep[see e.g.][]{2005astro.ph..6540M}.
For that reason let us  consider a different mass-matrix factorizing the term which is diagonal in Fourier sapce with the one diagonal in configuration space 
\ba
\label{eq:mass2} 
\mat M&=&\mat C^{-1}(\mathbb{1}+\beta\mat C\mat R)\,,\\
\mat M&\simeq&\mat C^{-1}(\mathbb{1}+c\mat R)\,,
\ea
approximating $\mat C$ by a constant inside the parenthesis, which multiplied by $\beta$ yields an effective constant of $c$.
Computing the  inverse of such matrix is trivial now, however, writing its square root is not. 
Since the mass-matrix is a free quantity, let us consider the naive expression for the square root as valid:
\be
\label{eq:mass3} 
\sqrt{\mat M}=\mat C^{-\frac{1}{2}}(\mathbb{1}+c\mat R)^{\frac{1}{2}}\,.
\ee
We can now go the other way round and derive the corresponding mass-matrix and in particular its inverse
\ba
\label{eq:hamilmass}
{\mat M}  =  \mat C^{-\frac 1 2} \cdot (\mathbb{1} + c \cdot \mat R)^{ \frac 1 2}  \cdot \mat C^{-\frac 1 2} \cdot (\mathbb{1} + c \cdot \mat R)^{ \frac 1 2} \,,  \\
{\mat M}^{-1}  = (\mathbb{1} + c \cdot \mat R)^{- \frac 1 2}  \cdot \mat C^{\frac 1 2} \cdot (\mathbb{1} + c \cdot \mat R)^{- \frac 1 2}  \cdot \mat C^{\frac 1 2} \, .
\ea
The important aspect to keep track of is that all expressions of the mass-matrix have to be consistent.
We have found in this way an expression for the Hamiltonian mass-matrix which can be efficiently applied as a series of convolutions going from configuration to Fourier space back and forth both for generating the momenta, where the square root is required, as to solve the Hamiltonian equations of motions, where the inverse is needed.
We have performed a series of numerical tests with a forth order discretisation of the Hamiltonian equations of motions to find the optimal $c$ value between 0.2 and 0.3 given our setting (number density, survey geometry, radial selection function). As an example a run with $c=0$ requiring 156 min (70 iterations)  of CPU time with 8 cores until the power spectra are within 1\% compatible with the theoretical one at $k>0.1\,h\,{\rm Mpc}^{-1}$ (to avoid cosmic variance at lower $k$-values), took 48 min (28 iterations) with $c=0.2$.  
 Hence we can gain a speed up of about 70\% in the convergence of the HMC sampler.

\end{document}